\documentclass[aps,prd,reprint, superscriptaddress, nofootinbib]{revtex4-2}

\usepackage{epstopdf} 
\usepackage{ulem}
\usepackage{epsfig}
\usepackage{amsmath,amssymb,amsthm}
\usepackage{graphicx}
\usepackage{bm}
\usepackage{color}

\usepackage{relsize}
\usepackage[utf8]{inputenc}
\usepackage{slashed}

\usepackage{tikz}
\usepackage{tikz-feynman}

\usepackage{amsmath}
\usepackage{tikz}
\usetikzlibrary{arrows,patterns,decorations.pathmorphing,calc}

\allowdisplaybreaks[2]

\usepackage{graphicx}

\usepackage{mathrsfs}

\usepackage{subcaption}
\usepackage{ragged2e}
\DeclareCaptionJustification{justified}{\justifying}
\usepackage[capitalize]{cleveref}

\DeclareMathAlphabet{\mathpzc}{OT1}{pzc}{m}{it}

\begin{document}

\renewcommand{\textfraction}{0.00}
\def\mean#1{\left< #1 \right>}

\definecolor{blue}{RGB}{55,160,65}

% Useful macros:
\newcommand{\Tr}{\operatorname{Tr}}
\newcommand{\vAi}{{\cal A}_{i_1\cdots i_n}}
\newcommand{\vAim}{{\cal A}_{i_1\cdots i_{n-1}}}
\newcommand{\vAbi}{\bar{\cal A}^{i_1\cdots i_n}}
\newcommand{\vAbim}{\bar{\cal A}^{i_1\cdots i_{n-1}}}
\newcommand{\htS}{\hat{S}}
\newcommand{\htR}{\hat{R}}
\newcommand{\htB}{\hat{B}}
\newcommand{\htD}{\hat{D}}
\newcommand{\htV}{\hat{V}}
\newcommand{\cT}{{\cal T}}
\newcommand{\cM}{{\cal M}}
\newcommand{\cMs}{{\cal M}^*}
\newcommand{\vk}{\vec{\mathbf{k}}}
\newcommand{\bk}{\bm{k}}
\newcommand{\kt}{\bm{k}_\perp}
\newcommand{\kp}{k_\perp}
\newcommand{\km}{k_\mathrm{max}}
\newcommand{\vl}{\vec{\mathbf{l}}}
\newcommand{\bl}{\bm{l}}
\newcommand{\bK}{\bm{K}}
\newcommand{\bb}{\bm{b}}
\newcommand{\qm}{q_\mathrm{max}}
\newcommand{\vp}{\vec{\mathbf{p}}}
\newcommand{\bp}{\bm{p}}
\newcommand{\vq}{\vec{\mathbf{q}}}
\newcommand{\bq}{\bm{q}}
\newcommand{\qt}{\bm{q}_\perp}
\newcommand{\qp}{q_\perp}
\newcommand{\bQ}{\bm{Q}}
\newcommand{\vx}{\vec{\mathbf{x}}}
\newcommand{\bx}{\bm{x}}
\newcommand{\beq}{\begin{equation}}
\newcommand{\eeq}[1]{\label{#1} \end{equation}}
\newcommand{\ee}{\end{equation}}
\newcommand{\bea}{\begin{eqnarray}}
\newcommand{\eea}{\end{eqnarray}}
\newcommand{\beqar}{\begin{eqnarray}}
\newcommand{\eeqar}[1]{\label{#1}\end{eqnarray}}

\newcommand{\half}{{\textstyle\frac{1}{2}}}
\newcommand{\ben}{\begin{enumerate}}
\newcommand{\een}{\end{enumerate}}
\newcommand{\bit}{\begin{itemize}}
\newcommand{\eit}{\end{itemize}}
\newcommand{\ec}{\end{center}}
\newcommand{\bra}[1]{\langle {#1}|}
\newcommand{\ket}[1]{|{#1}\rangle}
\newcommand{\norm}[2]{\langle{#1}|{#2}\rangle}
\newcommand{\brac}[3]{\langle{#1}|{#2}|{#3}\rangle}
\newcommand{\hilb}{{\cal H}}
\newcommand{\pleft}{\stackrel{\leftarrow}{\partial}}
\newcommand{\pright}{\stackrel{\rightarrow}{\partial}}

\newcommand{\squeezeup}{\vspace{-2.5mm}}

\title{Heavy-quark gluon radiation in transversely flowing QCD matter: First-subeikonal corrections}

\date{\today}

\author{Bojana Ilic}
\affiliation{Institute of Physics Belgrade, University of Belgrade, 11000 Belgrade, Serbia}
\author{Magdalena Djordjevic}
\affiliation{Institute of Physics Belgrade, University of Belgrade, 11000 Belgrade, Serbia}
\affiliation{Serbian Academy of Sciences and Arts, Belgrade 11000, Serbia}

\begin{abstract}
We investigate medium-induced gluon radiation from energetic heavy quarks propagating through a finite-size QCD medium with transverse collective flow. Within the DGLV framework, we calculate the radiation spectrum at first order in opacity and to first-subeikonal accuracy, retaining finite heavy-quark and effective gluon masses. Subleading corrections to the longitudinal propagator poles generate mass-dependent modifications of both the residue denominators and formation phases. For rotationally symmetric transverse-momentum integration domains, all contributions linear in the transverse flow velocity vanish after full azimuthal integration, while a nonzero static first-subeikonal correction remains. For representative medium parameters, this correction increases the integrated fractional radiative energy loss by approximately $10$--$20\%$, depending on the quark mass, initial energy, and path length. At the differential level, transverse flow produces a dipolar angular anisotropy, enhancing radiation along the flow direction and suppressing it in the opposite direction.
\end{abstract}
\maketitle

\section{\label{sec:Intro}Introduction}

Ultrarelativistic heavy-ion collisions at RHIC and the LHC produce a deconfined state of strongly interacting matter, the quark-gluon plasma (QGP), whose properties can be probed through the modification of energetic partons produced in the earliest stages of the collision~\cite{GyulassyMcLerran2005,MajumderVanLeeuwen2011,QinWang2015,ApolinarioLeeWinn2022}. These jet-quenching phenomena, characterized by the nuclear modification factor $R_{AA}(p_\perp)$ and the azimuthal anisotropy coefficients $v_n(p_\perp)$ of high-$p_\perp$ hadrons and jets, provide insight into the mechanisms of parton energy loss and the space-time evolution of the QGP~\cite{Adcox2002,Acharya,Aad}. Heavy-flavor observables offer additional sensitivity to the quark-mass dependence of parton--medium interactions, with recent charm- and bottom-flavor measurements providing increasingly stringent constraints on theoretical descriptions~\cite{ALICEDmesonRAA2022,ATLASCharmBottomRAA2022,ATLASCharmBottomFlow2020,CMSNonpromptDFlow2024}.

Several perturbative-QCD approaches have been developed to describe medium-induced gluon radiation. These include the multiple-soft-scattering BDMPS-Z~\cite{BDMPS1,BDMPS2,Z,Z1} and ASW~\cite{ASW,ASW0,ASW3} formalisms, the GLV opacity expansion~\cite{GLV1,GLV2}, the AMY framework~\cite{ArnoldMooreYaffe2002}, higher-twist approaches~\cite{HT,MajumderVanLeeuwen2011}, and effective-field-theory treatments based on Glauber interactions~\cite{OvanesyanVitev2011,KangRingerVitev2017}. Although their treatment of multiple scatterings, formation times, and medium properties differs, these frameworks generally rely on a high-energy kinematic hierarchy underlying the eikonal limit, in which transverse momenta and medium scales are small compared with the large longitudinal parton momenta. The leading corrections beyond this limit constitute the first-subeikonal order considered in the present work. The soft-gluon approximation is also widely used in radiative-energy-loss calculations, with its validity within the DGLV~\cite{DGLV} framework verified in Ref.~\cite{bsga}.

Finite parton masses introduce further modifications to the radiation spectrum. The suppression of small-angle gluon emission from heavy quarks is associated with the dead-cone effect~\cite{DokshitzerKharzeev2001}, while medium rescattering alters the corresponding angular distribution~\cite{ASW3}. Within the opacity expansion, the GLV formalism was extended to include a finite heavy-quark mass $M$ and an effective thermal gluon mass $m_g$~\cite{DG03}, leading to the DGLV description of heavy-quark radiative energy loss~\cite{DGLV}. Both masses enter the relevant propagator denominators and gluon formation phases. The conventional (D)GLV framework employs static, Yukawa-screened scattering centers~\cite{GW,GLV1,GLV2,DGLV}, whereas the dynamical energy-loss formalism incorporates recoil of thermal medium constituents~\cite{DjordjevicHeinz2008,Djordjevic2009} and has more recently been extended to a temperature-evolving QGP~\cite{KarmakarDjordjevic2026}. Here, we extend the finite-mass DGLV treatment to a transversely flowing medium at first-subeikonal order.

The collective expansion of the QGP modifies both the medium geometry and its interaction with energetic partons. Early radiative energy loss calculations incorporated longitudinal
Bjorken expansion~\cite{BaierExp,GyulassyFlow1} and the effect of  transverse expansion on the medium density~\cite{GyulassyFlow2}. Armesto, Salgado, and Wiedemann subsequently introduced an oriented momentum transfer in the scattering potential, demonstrating the sensitivity of medium-induced radiation and high-$p_\perp$ anisotropies to collective flow~\cite{ASWflow1,ASWflow2}. Baier, Mueller, and Schiff examined transverse hydrodynamic flow through a Lorentz-transformed jet-transport coefficient, without modifying the underlying radiation kernel~\cite{BaierFlow}. Hydrodynamic evolution generally builds up transverse flow from relatively small values at early times~\cite{Alqahtani2021,Hirano2002}, while high-$p_\perp$ observables are sensitive to the onset and early development of transverse expansion~\cite{Stojku2022}. This motivates isolating the leading linear response to transverse flow.

A more microscopic description was developed by Cheng and Wang, who introduced moving scattering potentials into radiative-energy-loss calculations and investigated flow effects on heavy-quark radiation and Landau--Pomeranchuk--Migdal (LPM) interference~\cite{ChengWang2009,ChengWangHeavy2009,ChengLiuWang2014}. Although finite quark and gluon masses were included, the analysis was restricted to the longitudinal component of the medium velocity. The transverse-flow configuration considered here was not addressed.

Sadofyev, Sievert, and Vitev subsequently incorporated moving color sources directly into the opacity expansion and identified subeikonal flow corrections for massless particles in a scalar-emission setup using a power counting that neglects velocity-dependent subeikonal sine contributions~\cite{SSVflow}. Related developments include flow-induced jet drift~\cite{JetDrift}, gradient corrections to momentum broadening and gluon radiation~\cite{BarataInhom,BarataMayoSadofyevSalgado2023}, all-order opacity resummation of transverse-flow effects on jet broadening~\cite{ADSS}, and the study of anisotropic flowing matter~\cite{KuzminFlow}.

At first order in opacity, Kuzmin and Mayo L\'opez derived the medium-induced spin-1 gluon spectrum for massless quarks in a flowing QCD medium while retaining first-subeikonal corrections~\cite{KML}. Their analysis showed that subeikonal contributions survive in static matter and that transverse flow modifies the magnitude and angular structure of the radiation spectrum. More recently, Barata \textit{et al.} investigated heavy-quark mass effects in a flowing medium within a resummed multiple-soft-scattering framework, identifying a directional modification of the dead-cone structure~\cite{BarataDeadCone}. Unlike the finite-mass DGLV approach, that calculation treats the emitted gluon as massless and focuses on the mass-dependent coupling between heavy quarks and collective flow.

We therefore consider a first-order-in-opacity treatment that combines finite quark and gluon masses with transverse collective flow at the level of the individual DGLV amplitudes. An additional distinction concerns the treatment of subleading longitudinal propagator-pole shifts. Under the adopted LPM power counting, these shifts contribute at first-subeikonal order both through residue denominators and through formation phases whenever their net contribution survives the contour integrations. Ref.~\cite{KML} retains the corresponding corrections in certain residue denominators, but omits their contributions to the formation phases. Retaining the surviving phase shifts is required at this accuracy and generates additional first-subeikonal $z\sin$ terms, which are included in our result alongside the mass-dependent residue-denominator corrections, while the leading LPM phases remain unexpanded.

We evaluate the single- and two-scattering amplitudes contributing at first order in opacity to heavy-quark gluon radiation in a finite-size, finite-temperature flowing QCD medium, retaining corrections through first-subeikonal order. The resulting spectrum contains the pole-induced terms discussed above, together with corrections from subleading Lorentz contractions, the scattering potential evaluated at flow-shifted propagator poles, and the residues of the scattering-potential poles. For rotationally symmetric transverse-momentum integration domains, we show analytically that all terms linear in the transverse flow velocity vanish upon full azimuthal integration, while a nonvanishing static first-subeikonal correction remains. We then examine, for light, charm, and bottom quarks, the gluon energy-fraction dependence of the static radiation spectrum, the initial energy and path-length dependence of the integrated fractional radiative energy loss, and the flow-induced angular anisotropy of the transverse-momentum-differential gluon spectrum.

The paper is organized as follows. Section~\ref{sec:TheoreticalFramework} introduces the theoretical framework, followed by the derivation and analysis of the medium-induced gluon spectrum and radiative energy loss in Sec.~\ref{sec:Analitika}. Numerical results are presented in Sec.~\ref{sec:Num}, followed by our conclusions. The assumptions, power counting, and detailed evaluation of the scattering amplitudes are provided in the Appendices.

\begin{figure*}[t]
\centering
\includegraphics[width=0.9\textwidth]{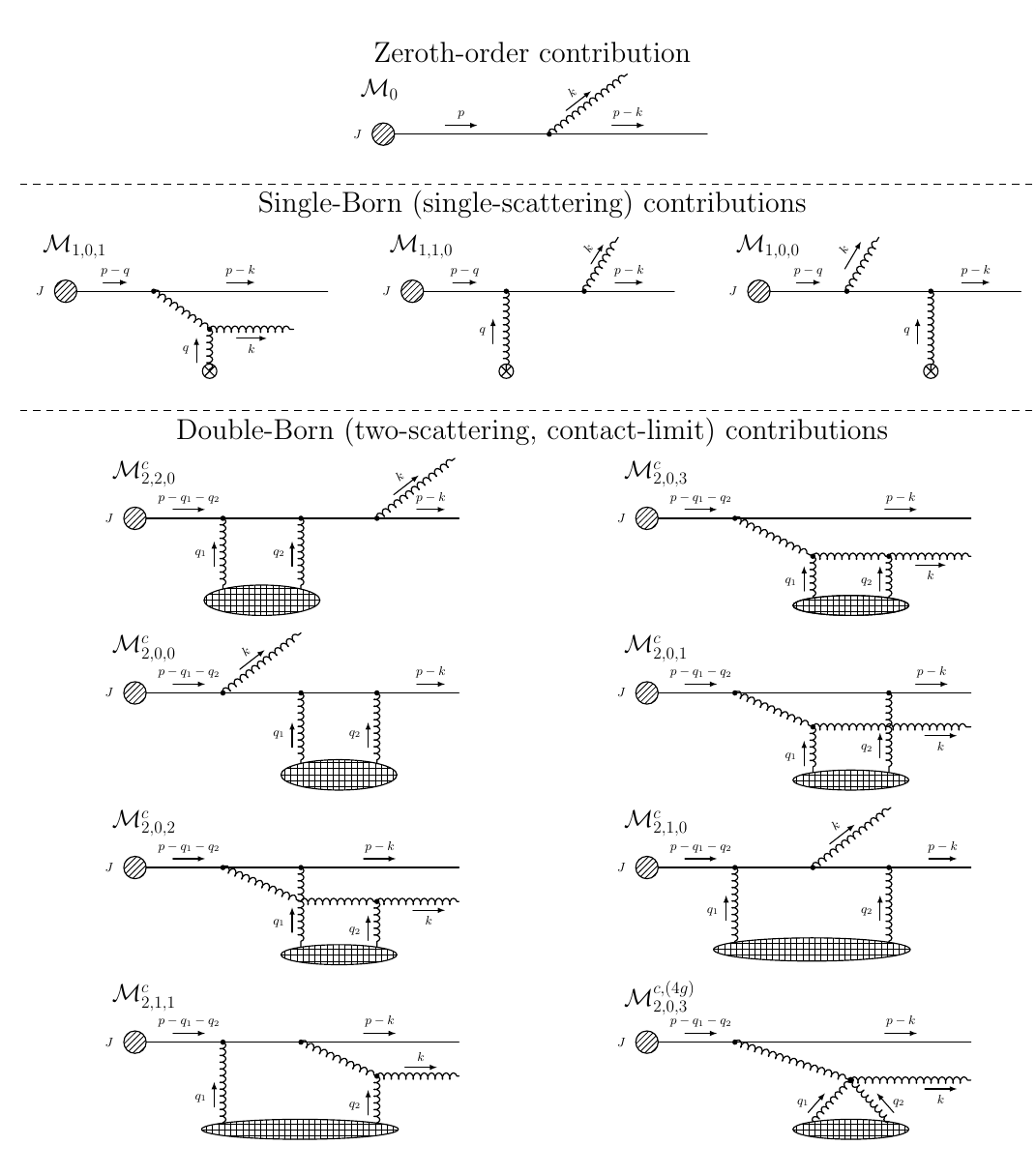}
\caption{\small Diagrams contributing to gluon radiation through first order in opacity: the zeroth-order amplitude $\mathcal{M}_0$, three single-scattering amplitudes, and eight two-scattering amplitudes shown in the contact limit, including the four-gluon vertex. }
\label{fig:Fig0}
\end{figure*}

\section{\label{sec:TheoreticalFramework}Theoretical framework}

We calculate medium-induced gluon radiation from a highly energetic heavy quark of mass $M$ propagating through a finite-size, finite-temperature, flowing QCD medium to first-subeikonal order. Starting from the DGLV opacity expansion~\cite{GLV1,GLV2,DGLV}, including the effective asymptotic mass $m_g$ of the radiated gluon~\cite{DG03}, we incorporate collective flow directly into the individual scattering amplitudes.

Following Refs.~\cite{SSVflow,KML}, we describe the interaction of the energetic heavy quark and the radiated gluon with the flowing QCD medium through an external color field generated by moving scattering centers. In momentum space, this field is given by
\begin{align}
g_s \, A_{ext}^{\mu,a}(q)
&=
\sum_i
e^{-i\left(
       \mathbf{q}\cdot\mathbf{x}_i
      +q_z z_i
     \right)}
T_i^a u_i^\mu v_i(q)\,
2\pi\,
\delta\!\left(
    q^0-\vec{u}_i\cdot\vec{q}
\right)\, .
\end{align}
Here, $u_i^\mu$ and $T_i^a$ are the four-velocity and color generator of the $i$th scattering center, $q^\mu$ is the exchanged momentum, and $(\mathbf{x}_i,z_i)$ denotes the position of that center.  The screened scattering potential of an individual scattering center is given by
\begin{align}
v_i(q)
&=
\frac{-g_s^2}
{\vec{q}^{\,2}+\mu_i^2-(\vec{u}_i\cdot\vec{q})^2}\, .
\end{align}

We work in a frame with $u_z=0$ and assume a homogeneous medium with constant transverse velocity $\mathbf{u}$, retained to linear order. This setup gives $u^\mu=(1,\mathbf{u},0)$ and $q^0=\mathbf{u}\cdot\mathbf{q}$, and is compatible with the axial-gauge choice $n^\mu=(0,\mathbf{0},1)$ used in Refs.~\cite{SSVflow,KML}, since $n\cdot u=0$.

We assume a transversely homogeneous distribution of scattering centers, described by $\rho_0(\mathbf{x},z)=N\rho(z)/A_\perp$, where $N$ is the total number of scattering centers, $A_\perp$ is the transverse area, and $\rho(z)$ is the normalized longitudinal distribution. The averaging over scattering-center positions and target color degrees of freedom follows the standard (D)GLV prescription~\cite{GLV2,DGLV}, with further details provided in Sec.~\ref{sec:Assumptions}.

The calculation is performed in the soft-gluon and soft-rescattering approximations, with the emitted gluon energy $\omega=xE$ much smaller than the parent-quark energy $E$ ($x\ll1$), and the exchanged momentum soft compared with the energetic parton energy. We adopt the high-energy, small-angle hierarchy $E \gg xE \gg |\mathbf{p}| \sim |\mathbf{k}| \sim |\mathbf{q}| \sim \mu \sim m_g \sim xM \sim \perp$, where $\mu$ is the Debye screening mass and $\perp$ denotes a generic soft transverse or mass scale, retaining corrections through first-subeikonal order in $\perp/(xE)$.

The radiated gluon carries four-momentum $k^\mu$ and polarization vector $\varepsilon^\mu(k)$, while the outgoing heavy quark has four-momentum $(p-k)^\mu$, where $p_\mu$ denotes the parent-quark momentum. The final heavy quark and emitted gluon are on shell, satisfying $(p-k)^2=M^2$ and $k^2=m_g^2$, respectively. Their four-momenta and gluon polarisation are given by
\begin{flalign}
 &(p-k)^\mu = \nonumber
\\
&\left(
(1-x)E,\,
\mathbf{p}-\mathbf{k},\,
(1-x)E-\frac{(\mathbf{p}-\mathbf{k})^2+M^2}
{2(1-x)E}
\right) \, , \nonumber \\
&{} k^\mu=
\left(
xE,\,
\mathbf{k},\,
xE-\frac{\mathbf{k}^2+m_g^2}{2xE}
\right)\, , \nonumber \\
&{} \varepsilon^\mu(k)=
\left(
\frac{\boldsymbol{\varepsilon}\cdot\mathbf{k}}{xE},\,
\boldsymbol{\varepsilon},\,
0
\right) \, ,
\label{vektori}
\end{flalign}
respectively, where \(\boldsymbol{\varepsilon}\) denotes the transverse polarization vector.

We also employ the separation condition
$ z_i \gg 1/\mu$, and the broad-source approximation
$ J(E,\mathbf{p}-\mathbf{q}_\ell) \simeq J(E,\mathbf{p})$, as in Refs.~\cite{KML,GLV2,DGLV,bsga}. These assumptions, together with the relevant Feynman rules, are described in detail in Secs.~\ref{sec:Assumptions}
and~\ref{sec:FeynmanRules}.

Following the LPM power counting of Refs.~\cite{GLV2,DGLV,KML}, we treat the leading gluon formation phases as $\phi_0\sim\perp^2 z/(xE)=\mathcal{O}(1)$ and retain them unexpanded. The first-subeikonal expansion is instead organized in powers of $\perp/(xE)$. In particular, the subleading propagator-pole corrections, $\delta Q^-\sim\perp^3/(xE)^2$, are retained wherever they contribute at this accuracy, either through pole differences entering the residue denominators or through the formation phases.
 Although $\delta Q^-$ is formally of higher order in the pole expansion, for $\phi_0=\mathcal{O}(1)$ the corresponding phase correction, $\delta\phi=\delta Q^-z=\mathcal{O}(\perp/(xE))$, is promoted to first-subeikonal order. Only this subleading correction is expanded, as specified in Eqs.~\eqref{eq:deltaphicounting} and~\eqref{eq:cosphaseexpansion}. All pole residues and Lorentz contractions are evaluated before expanding the resulting amplitudes to first-subeikonal order.
 
At first order in opacity, as considered here, the medium-induced radiation spectrum receives contributions from the squared sum of the three single-scattering (single-Born) amplitudes, $|\mathcal{M}_1|^2$, and the interference of the zeroth-order radiation amplitude, $\mathcal{M}_0$, with $\mathcal{M}_2$, representing the sum of all eight two-scattering (double-Born) amplitudes evaluated in the contact limit. The corresponding Feynman diagrams are summarized in Fig.~\ref{fig:Fig0}, comprising one zeroth-order, three single-scattering, and eight two-scattering contributions.

Unlike the leading-eikonal DGLV treatment~\cite{DGLV}, the present calculation also includes the two-scattering diagram involving a four-gluon vertex. Although suppressed at leading eikonal accuracy, this contribution enters at first-subeikonal order and must therefore be retained, as in Ref.~\cite{KML}. 

 The comprehensive diagrammatic derivations of the individual amplitudes are presented in the Appendices. More precisely, Appendix~\ref{sec:M0} addresses gluon radiation in the absence of interactions with the QGP, while Appendix subsections, Secs.~\ref{sec:M101}-~\ref{sec:M100} provide the systematic evaluation of the three amplitudes involving one interaction with the medium. The eight amplitudes involving two interactions with the medium, including the four-gluon vertex contribution, are derived separately in Appendices~\ref{sec:M220}, ~\ref{sec:M203}, ~\ref{sec:M200}, ~\ref{sec:M201}, ~\ref{sec:M210}, ~\ref{sec:M4g203}, with particular attention to their contact limits. In Appendix~\ref{sec:M1sq}, the color traces and medium averages are evaluated separately for all pairwise products of the three single-scattering amplitudes with their complex conjugates, including both diagonal and interference terms. The corresponding interference contributions between the zeroth-order and two-scattering amplitudes are calculated in Appendix~\ref{sec:M2sum}. 

 Following the (D)GLV normalization~\cite{GLV2,DGLV,bsga}, the joint gluon–heavy-quark distribution at first order in opacity reads
\begin{widetext}
\begin{align}
d^3N_g^{(1)}\,d^3N_J
={}&
\left[
\frac{1}{d_T}
\Tr\mean{|\mathcal{M}_1|^2}
+
\frac{2}{d_T}
\operatorname{Re}\Tr
\mean{\mathcal{M}_0^\dagger\mathcal{M}_2}
\right]
\frac{d^3(\vec{p}-\vec{k})}
{(2\pi)^3\,2(1-x)E}
\frac{d^3\vec{k}}
{(2\pi)^3\,2xE} \, ,
\label{spectum}
\end{align}
\end{widetext}
where $d_T$ denotes the dimension of the target color representation. 
 Finally, the combined first-order-in-opacity contributions, including the first-subeikonal corrections, are summarized in Appendix~\ref{M1M2NE}.

\section{\label{sec:Analitika} Medium-induced gluon radiation in flowing QCD matter}

\subsection{\label{FullSpectrumClassification}First-order-in-opacity gluon spectrum}

We now extract the medium-induced gluon spectrum from Eq.~\eqref{spectum}.
To separate the radiation spectrum from the parent-quark distribution, we employ the broad-source approximation.
 The change of variables $\vec{p}-\vec{k}\to\vec{p}$, performed at fixed emitted-gluon momentum, has unit Jacobian. Consequently, the hard-source factor combines with the quark phase space to give the parent-quark distribution $d^3N_J=d_F|J(E,\mathbf{p})|^2d^3\vec{p}/[(2\pi)^3\,2E]$.

Furthermore, using $dk_z=E[1+(\mathbf{k}^2+m_g^2)/(2x^2E^2)]dx$, the emitted-gluon phase space reduces to $dx\,d^2\mathbf{k}/[(2\pi)^3\,2x]$ through first-subeikonal order, which is derived in more detail in Appendix~\ref{sec:M0}. Here the soft-gluon approximation $1/(1-x)\simeq1$ has been applied, while the Jacobian correction starts at $\mathcal{O}(\perp^2/(xE)^2)$ and is therefore beyond first-subeikonal order. The common factor $d^3N_J$ consequently cancels, leaving the gluon radiation spectrum. For a propagating heavy quark in the fundamental representation, summing the emitted-gluon and exchanged color indices gives $\alpha/d_F=C_A C_2(F)/2$.

Combining the single- and two-scattering contributions, as explained in the previous section,  we obtain the medium-induced double-differential gluon spectrum at first order in opacity, retaining finite quark and gluon masses and corrections through first-subeikonal order, with the transverse-flow dependence kept to linear order:
\begin{widetext}
\begin{align}
x \, \frac{dN_g^{(1)}}
{dx\,d^2\mathbf{k}}
{}&=
\frac{\alpha_s}{ \pi^2}\,
C_A C_2(F)\,
\frac{C_2(T)N}{A_\perp d_A}
\int_0^L dz\,\rho(z)
\int\frac{d^2\mathbf{q}}{(2\pi)^2}\,
|v(\mathbf{q})|^2
\nonumber \\
&\times \Bigg\{2 \left[
1-
\cos\left(
\frac{(\mathbf{k}-\mathbf{q})^2+\chi}{2xE}\,z
\right)
\right]
\Bigg\{
\left[
\frac{(\mathbf{k}-\mathbf{q})^2}
{\left[(\mathbf{k}-\mathbf{q})^2+\chi\right]^2}
-\frac{\mathbf{k}\cdot(\mathbf{k}-\mathbf{q})}
{(\mathbf{k}^2+\chi)\left[(\mathbf{k}-\mathbf{q})^2+\chi\right]}
\right] \nonumber \\
&\quad \times\Bigg[
1
-\frac{\mathbf{u}\cdot\mathbf{q}}{2xE} \, 
\frac{2\mathbf{k}^2-(\mathbf{k}-\mathbf{q})^2+\chi}{v^2(\mathbf{q})} \, 
\frac{\partial v^2(\mathbf{q})}{\partial\mathbf{q}^2}
\Bigg]
\nonumber \\
&\quad+
\frac{\mathbf{u}\cdot(\mathbf{k}-\mathbf{q})}{xE}
\left[
\frac{\mathbf{k}^2}
{(\mathbf{k}^2+\chi)\left[(\mathbf{k}-\mathbf{q})^2+\chi\right]}
-\frac{2(\mathbf{k}-\mathbf{q})^2}
{\left[(\mathbf{k}-\mathbf{q})^2+\chi\right]^2}
\right]
\nonumber \\
&\quad-
\frac{\mathbf{u}\cdot\mathbf{k}}{xE}
\frac{\mathbf{q}\cdot(\mathbf{k}-\mathbf{q})}
{(\mathbf{k}^2+\chi)\left[(\mathbf{k}-\mathbf{q})^2+\chi\right]}
\nonumber \\
&\quad+
\frac{\mathbf{u}\cdot\mathbf{q}}{xE}
\frac{2x^2M^2}
{(\mathbf{k}-\mathbf{q})^2+\chi}
\left[
\frac{2(\mathbf{k}-\mathbf{q})^2}
{\left[(\mathbf{k}-\mathbf{q})^2+\chi\right]^2}
-\frac{\mathbf{k}\cdot(\mathbf{k}-\mathbf{q})}
{(\mathbf{k}^2+\chi)\left[(\mathbf{k}-\mathbf{q})^2+\chi\right]}
\right]
\Bigg\}
\nonumber \\
&\quad
+
\left[
1-
\cos\left(
\frac{\mathbf{k}^2+\chi}{2xE}\,z
\right)
\right]
\frac{\mathbf{u}\cdot\mathbf{k}}{2xE}
\frac{1}{(\mathbf{k}^2+\chi)^2}
\Bigg[
4\mathbf{k}^2
+
(\mathbf{k}^2-\chi)
\frac{\mathbf{q}^2}{v^2(\mathbf{q})}
\frac{\partial v^2(\mathbf{q})}{\partial\mathbf{q}^2}
\Bigg]
\nonumber \\
&\quad
+
\sin\left(
\frac{\mathbf{k}^2+\chi}{2xE}\,z
\right) \, 
\frac{\mathbf{k}^2}{(\mathbf{k}^2+\chi)^2}
\frac{R_0}{2xE}
\Bigg[
1+
\frac{\mu^2-2\mathbf{q}^2}{4v^2(\mathbf{q})}
\frac{\partial v^2(\mathbf{q})}{\partial\mathbf{q}^2}
\Bigg]
\nonumber \\
&\quad
+
\frac{1}{(\mathbf{k}-\mathbf{q})^2+\chi}
\frac{\mathbf{u}\cdot\mathbf{q}}{xE}
\frac{
(\mathbf{k}-\mathbf{q})^2+m_g^2-x^2M^2
}{xE}\,
z
\Bigg[
\frac{(\mathbf{k}-\mathbf{q})^2}
{(\mathbf{k}-\mathbf{q})^2+\chi}
-\frac{\mathbf{k}\cdot(\mathbf{k}-\mathbf{q})}
{\mathbf{k}^2+\chi}
\Bigg]
\nonumber \\
&\qquad\times
\sin\left(
\frac{(\mathbf{k}-\mathbf{q})^2+\chi}{2xE}\,z
\right)
\Bigg\} \, .
\label{spectrum}
\end{align}
\end{widetext}
Here $\alpha_s\equiv g_s^2/(4\pi)$ is the strong coupling constant, while $C_2(F)$ and $C_2(T)$ denote the Casimir operators of the heavy-quark fundamental  and scattering-centers color representations, respectively, with $C_2(F)=(N_c^2-1)/(2N_c)$, $C_A\equiv C_2(A)=N_c$, and $d_A=N_c^2-1$. Throughout this paper, we use the mass combination $\chi=m_g^2+x^2M^2$, together with $R^2_0=\mathbf{q}^2+\mu^2$. The screened scattering potential reads $v(\mathbf{q})=-g^2_s/(\mathbf{q}^2 +\mu^2)$.

The first-subeikonal contributions can be classified according to their origin:
\begin{itemize}

\item \textit{Subleading Lorentz-contraction corrections.}

The kinematic corrections arise from the systematic expansion of the Lorentz-contracted structures entering the diagrammatic amplitudes (including interaction vertices, polarization vectors, external-field four-velocities, and, where applicable, gluon-propagator projectors), retaining higher-order terms where required by the adopted power counting. Nevertheless, the flow-dependent corrections retained in the final expressions are linear in transverse-flow projections and of relative order $\mathcal{O}(|\mathbf{u}|\perp/(xE))$.

\item \textit{Flow-induced potential-derivative corrections.}
A separate class of flow-dependent terms emerges after evaluating the
scattering potential at the flow-shifted longitudinal propagator poles.
Expanding the resulting pole-evaluated potential to the working
accuracy produces contributions proportional to
$v^{-2}(\mathbf{q})\,\partial v^2(\mathbf{q})/\partial\mathbf{q}^2$,
with coefficients determined by the propagator
pole structure rather than by the scattering-potential pole residues.

\item \textit{Scattering-potential-pole sine contribution.}

The term proportional to
\[
\sin\left(\frac{\mathbf{k}^2+\chi}{2xE}\,z\right)
\]
originates from the scattering-potential-pole residues in the two-scattering longitudinal integrations and survives in 
the static limit.
These
residues also generate the associated derivative-of-potential
structures proportional to
$v^{-2}(\mathbf{q})\,\partial v^2(\mathbf{q})/\partial\mathbf{q}^2$.

\item \textit{Mass-dependent residue-denominator correction.}

Finite heavy-quark mass generates, in addition, the residue correction proportional to
\begin{align}
 \frac{\mathbf{u}\cdot\mathbf{q}}{xE}\, \frac{2x^2M^2} {(\mathbf{k}-\mathbf{q})^2+\chi}\,.
\label{eq:flow_massresidue}
\end{align}
This contribution originates from the flow-dependent $\mathcal{O}\!\left(\perp^3/(xE)^2\right)$ mismatch between gluon and heavy-quark propagator poles. When their common $\mathcal{O}(\perp)$ terms cancel, the remaining separation is of order $\mathcal{O}(\perp^2/(xE))$, allowing the subleading mismatch to enter the residue denominators at relative first-subeikonal accuracy. In the two-scattering sector, the analogous gluon-heavy-quark pole difference emerges effectively after the first longitudinal integration.

In $|\mathcal{M}_{1,0,1}|^2$, the contributions from the amplitude and its conjugate double the coefficient in Eq.~\eqref{eq:flow_massresidue}. The mass-dependent term also enters the interference contributions involving $\mathcal{M}_{1,0,1}$ and the remaining single-scattering amplitudes, with analogous corrections arising in the interferences of $\mathcal{M}_{2,0,1}^{c}$ and $\mathcal{M}_{2,0,2}^{c}$ with $\mathcal{M}_0^\dagger$. In contrast, for $\mathcal{M}_{2,1,1}^{c}$, the subleading pole-difference corrections enter the individual propagator residue denominators but cancel upon summing the corresponding residues.

\item \textit{Phase-induced $z\sin$ correction.}

The same subleading gluon-heavy-quark pole mismatch can also modify the relative formation phase. When the corresponding correction survives in the longitudinal pole combination determining the phase argument, it takes the form
\begin{align}
\delta\phi
&=
\frac{\mathbf{u}\cdot\mathbf{q}}{xE}
\frac{
(\mathbf{k}-\mathbf{q})^2+m_g^2-x^2M^2} {2 x E} \, z\, .
\label{eq:flow_deltaphi}
\end{align}
Expanding the relative formation phase according to Eq.~\eqref{eq:cosphaseexpansion} in Appendix~\ref{sec:Assumptions}, this shift generates a first-subeikonal $z\sin$ contribution. In the single-scattering sector, nonvanishing phase-induced corrections arise within $|\mathcal{M}_{1,0,1}|^2$ and in its interference with the remaining amplitudes, where the relative phase involves contributions from both amplitudes.

For the two-scattering sector, the relevant longitudinal combination involves $q_{1z}+q_{2z}$. In $\mathcal{M}_{2,0,1}^{c}$ and $\mathcal{M}_{2,0,2}^{c}$, the effective gluon-heavy-quark pole mismatch responsible for the residue-denominator correction also determines the surviving subleading phase shift. However, the cancellation of individual pole corrections in the longitudinal phase depends on the selected residues. Consequently, phase-induced contributions survive only for specific residues in their interferences with $\mathcal{M}_0^\dagger$.

Unlike these phase-induced $z\sin$ terms, ordinary sine contributions originate from scattering-potential-pole residues and appear only in the two-scattering sector.

\end{itemize}

In the massless limit, $M=m_g=0$, the Lorentz-contraction, potential-derivative and ordinary sine  terms reduce, at linear order in $\mathbf{u}$, to the corresponding terms in Eq.~(65) of Ref.~\cite{KML}. The subleading pole shift in the formation phase also generates an additional first-subeikonal $z\sin$ term in Eq.~\eqref{spectrum}, with finite quark and gluon masses introducing the corresponding mass-dependent corrections.

\subsection{\label{KinInter}Kinematic interpretation of the mass-dependent and phase-induced terms}

The following discussion provides a compact kinematic interpretation of the mass-dependent and phase-induced structures associated with $\mathcal{M}_{1,0,1}$, $\mathcal{M}_{2,0,1}$, and $\mathcal{M}_{2,0,2}$.

In a flowing medium, the exchanged momentum carries a nonvanishing energy
$q^0=\mathbf{u}\cdot\mathbf{q}$,
so that the energy flowing through an intermediate state
differs from its static-medium (DGLV) value.
The relevant pole differences contain corrections of $\mathcal{O}\left(\perp^3/(xE)^2\right)$, which contribute at first-subeikonal order through the residue denominators or relative formation phases; see Appendix~\ref{sec:Assumptions}.

For the relevant intermediate state in
$\mathcal{M}_{1,0,1}$, this mechanism can be represented as a shift in the gluon
energy fraction,
\begin{equation}
x\rightarrow x-\delta x\,,
\qquad
\frac{\delta x}{x}
=
\frac{\mathbf{u}\cdot\mathbf{q}}{xE}\,,
\label{eq:flow_deltax}
\end{equation}
where the minus sign follows from the intermediate momentum $k-q$  ($(k-q)^0=xE -\mathbf{u} \cdot \mathbf{q}$). This replacement provides a kinematic representation of the pole shift in Appendix subsection~\ref{sec:M101}. The signs for $\mathcal{M}_{2,0,1}$ and $\mathcal{M}_{2,0,2}$ are fixed by their respective momentum flow. Detailed derivations are presented in Appendix~\ref{sec:M201}.

For
\begin{align}
D(x)
&=
(\mathbf{k}-\mathbf{q})^2+m_g^2+x^2M^2
\nonumber \\
&=
(\mathbf{k}-\mathbf{q})^2+\chi\,,
\label{eq:flow_D}
\end{align}
the finite-mass part of this shift gives
\begin{align}
D(x-\delta x)
&=
D(x) -
\delta x\,\frac{\partial D(x)}{\partial x}
\nonumber \\
&=
D(x)
-
\frac{\mathbf{u}\cdot\mathbf{q}}{xE}\,
2x^2M^2\,,
\label{eq:flow_Dshift}
\end{align}
and hence
\begin{align}
\frac{1}{D(x-\delta x)}
\simeq{}&
\frac{1}{D(x)}
\left[
1+
\frac{\mathbf{u}\cdot\mathbf{q}}{xE}\,
\frac{2x^2M^2}{D(x)}
\right]\,.
\label{eq:flow_deadcone}
\end{align}
The same mechanism appears in the corresponding intermediate pole
structures of $\mathcal{M}_{2,0,1}$ and $\mathcal{M}_{2,0,2}$, with
the precise momentum assignment fixed by the respective diagram.

The same subleading pole mismatch also shifts the formation frequency when it survives in the corresponding phase argument. With
\begin{equation}
\Omega(x)
=
\frac{D(x)}{2xE}\,,
\label{eq:flow_Omega}
\end{equation}
the shift associated with the intermediate state above gives
\begin{align}
\delta\Omega
&=
-\delta x\,\frac{\partial\Omega(x)}{\partial x}
\nonumber \\
&=
\frac{\mathbf{u}\cdot\mathbf{q}}{2(xE)^2}
\left[
(\mathbf{k}-\mathbf{q})^2+m_g^2-x^2M^2
\right]\,.
\label{eq:flow_deltaOmega}
\end{align}
The corresponding shift in the relative formation phase, $\delta\phi=z\,\delta\Omega$, reproduces Eq.~\eqref{eq:flow_deltaphi}. Thus, $\delta\phi=\mathcal{O}\left(\perp/(xE)\right)$, as shown in Eq.~\eqref{eq:deltaphicounting}. Expanding the
corresponding relative phases according to
Eq.~\eqref{eq:cosphaseexpansion} then produces the $z\sin$ terms. 

The mass-dependent residue-denominator and phase-induced $z\sin$ corrections thus represent two manifestations of the same subleading longitudinal-pole mismatch. Both probe the $x$ dependence of the massive radiative denominator, but through different quantities: $D(x)$ in the residue correction and $\Omega(x)$ in the formation-phase shift. Consequently, the heavy-quark dead-cone term $x^2M^2$ enters their final expressions through distinct algebraic structures.

\subsection{\label{RadELoss}Radiative energy loss}

The differential radiative energy loss is obtained by weighting the gluon spectrum $dN^{(1)}_g/(dx d^2 \mathbf{k})$ with the emitted-gluon energy $\omega=xE$ and integrating over its transverse momentum, following the standard (D)GLV prescription~\cite{GLV2, DGLV,bsga}:
\begin{equation}
\frac{dE^{(1)}}{dx}
=
E\int d^2\mathbf{k}\,
x\frac{dN_g^{(1)}}{dx\,d^2\mathbf{k}}\,.
\label{frEloss}
\end{equation}

Substituting the gluon spectrum from Eq.~\eqref{spectrum} into Eq.~\eqref{frEloss}, we obtain the explicit expression for the differential radiative energy loss:
\begin{widetext}
\begin{align}
\frac{dE^{(1)}}{dx}
{}&=E 
\frac{\alpha_s}{\pi^2}\,
C_A C_2(F)\,
\frac{C_2(T)N}{A_\perp d_A}
\int d^2\mathbf{k}\,
\int_0^L dz\,\rho(z)
\int\frac{d^2\mathbf{q}}{(2\pi)^2}\,
|v(\mathbf{q})|^2
\nonumber \\
&\times \Bigg\{2 \left[
1-
\cos\left(
\frac{(\mathbf{k}-\mathbf{q})^2+\chi}{2xE}\,z
\right)
\right]
\Bigg\{
\left[
\frac{(\mathbf{k}-\mathbf{q})^2}
{\left[(\mathbf{k}-\mathbf{q})^2+\chi\right]^2}
-\frac{\mathbf{k}\cdot(\mathbf{k}-\mathbf{q})}
{(\mathbf{k}^2+\chi)\left[(\mathbf{k}-\mathbf{q})^2+\chi\right]}
\right] \nonumber \\
&\quad \times\Bigg[
1
-\frac{\mathbf{u}\cdot\mathbf{q}}{2xE} \, 
\frac{2\mathbf{k}^2-(\mathbf{k}-\mathbf{q})^2+\chi}{v^2(\mathbf{q})} \, 
\frac{\partial v^2(\mathbf{q})}{\partial\mathbf{q}^2}
\Bigg]
\nonumber \\
&\quad+
\frac{\mathbf{u}\cdot(\mathbf{k}-\mathbf{q})}{xE}
\left[
\frac{\mathbf{k}^2}
{(\mathbf{k}^2+\chi)\left[(\mathbf{k}-\mathbf{q})^2+\chi\right]}
-\frac{2(\mathbf{k}-\mathbf{q})^2}
{\left[(\mathbf{k}-\mathbf{q})^2+\chi\right]^2}
\right]
\nonumber \\
&\quad-
\frac{\mathbf{u}\cdot\mathbf{k}}{xE}
\frac{\mathbf{q}\cdot(\mathbf{k}-\mathbf{q})}
{(\mathbf{k}^2+\chi)\left[(\mathbf{k}-\mathbf{q})^2+\chi\right]}
\nonumber \\
&\quad+
\frac{\mathbf{u}\cdot\mathbf{q}}{xE}
\frac{2x^2M^2}
{(\mathbf{k}-\mathbf{q})^2+\chi}
\left[
\frac{2(\mathbf{k}-\mathbf{q})^2}
{\left[(\mathbf{k}-\mathbf{q})^2+\chi\right]^2}
-\frac{\mathbf{k}\cdot(\mathbf{k}-\mathbf{q})}
{(\mathbf{k}^2+\chi)\left[(\mathbf{k}-\mathbf{q})^2+\chi\right]}
\right]
\Bigg\}
\nonumber \\
&\quad
+
\left[
1-
\cos\left(
\frac{\mathbf{k}^2+\chi}{2xE}\,z
\right)
\right]
\frac{\mathbf{u}\cdot\mathbf{k}}{2xE}
\frac{1}{(\mathbf{k}^2+\chi)^2}
\Bigg[
4\mathbf{k}^2
+
(\mathbf{k}^2-\chi)
\frac{\mathbf{q}^2}{v^2(\mathbf{q})}
\frac{\partial v^2(\mathbf{q})}{\partial\mathbf{q}^2}
\Bigg]
\nonumber \\
&\quad
+
\sin\left(
\frac{\mathbf{k}^2+\chi}{2xE}\,z
\right) \, 
\frac{\mathbf{k}^2}{(\mathbf{k}^2+\chi)^2}
\frac{R_0}{2xE}
\Bigg[
1+
\frac{\mu^2-2\mathbf{q}^2}{4v^2(\mathbf{q})}
\frac{\partial v^2(\mathbf{q})}{\partial\mathbf{q}^2}
\Bigg]
\nonumber \\
&\quad
+
\frac{1}{(\mathbf{k}-\mathbf{q})^2+\chi}
\frac{\mathbf{u}\cdot\mathbf{q}}{xE}
\frac{
(\mathbf{k}-\mathbf{q})^2+m_g^2-x^2M^2
}{xE}\,
z
\Bigg[
\frac{(\mathbf{k}-\mathbf{q})^2}
{(\mathbf{k}-\mathbf{q})^2+\chi}
-\frac{\mathbf{k}\cdot(\mathbf{k}-\mathbf{q})}
{\mathbf{k}^2+\chi}
\Bigg]
\nonumber \\
&\qquad\times
\sin\left(
\frac{(\mathbf{k}-\mathbf{q})^2+\chi}{2xE}\,z
\right)
\Bigg\} \, .
\label{frEloss1}
\end{align}
\end{widetext}
The total radiative energy loss at first order in opacity follows upon integration over the gluon momentum fraction $x$.

\subsection{\label{sec:AzimSymm}Azimuthal symmetry of the integrated radiation}

We now examine the angular structure of the result in
Eq.~\eqref{frEloss1}. At linear order in $\mathbf{u}$,
all flow-dependent terms are proportional to a scalar product of
$\mathbf{u}$ with $\mathbf{k}$, $\mathbf{q}$, or
$\mathbf{k}-\mathbf{q}$. All remaining factors depend on the
transverse directions through $\mathbf{k}\cdot\mathbf{q}$ only.
 The following argument assumes rotationally symmetric
transverse-momentum integration domains, as satisfied by the radial
cutoffs used in the calculation.
Choosing $\mathbf{u}=(|\mathbf{u}|,0)$ as the fixed reference direction, we denote by $\theta$ and $\Phi$ the azimuthal angles of $\mathbf{k}$ and $\mathbf{q}$, respectively, and introduce their relative angle $\psi=\theta-\Phi$. Thus, we obtain
\begin{align}
\mathbf{k}\cdot\mathbf{q}
={}&
|\mathbf{k}|\,|\mathbf{q}|\cos\psi,
\nonumber\\
\mathbf{u}\cdot\mathbf{k}
={}&
|\mathbf{u}|\,|\mathbf{k}|\cos(\psi+\Phi),
\nonumber\\
\mathbf{u}\cdot\mathbf{q}
={}&
|\mathbf{u}|\,|\mathbf{q}|\cos\Phi,
\nonumber\\
\mathbf{u}\cdot(\mathbf{k}-\mathbf{q})
={}&
|\mathbf{u}|
\left[
|\mathbf{k}|\cos(\psi+\Phi)
-|\mathbf{q}|\cos\Phi
\right].
\end{align}
The full angular integration of the contribution linear in
$\mathbf{u}$ can therefore be written as
\begin{align}
\left.
\frac{dE^{(1)}}{dx}
\right|_{\mathcal{O}(\mathbf{u})}
\propto{}&
|\mathbf{u}|
\int d\mathbf{k}^2
\int d\mathbf{q}^2
\int_0^{2\pi}d\psi
\int_0^{2\pi}d\Phi
\nonumber\\
&\times
\Bigl[
\cos(\psi+\Phi)\,
F(|\mathbf{k}|,|\mathbf{q}|,\cos\psi)
\nonumber\\
&\qquad+
\cos\Phi\,
G(|\mathbf{k}|,|\mathbf{q}|,\cos\psi)
\Bigr]
=0\, ,
\label{fullangsym}
\end{align}
where $F$ and $G$ collect the coefficients of the linear-flow terms in Eq.~\eqref{frEloss1}, with their dependence on $x$, $z$, and the remaining kinematic and medium parameters left implicit.  Both functions are independent of $\Phi$, so that
\begin{align}
\int_0^{2\pi}d\Phi\,\cos(\psi+\Phi)
={}&0,
\nonumber\\
\int_0^{2\pi}d\Phi\,\cos\Phi
={}&0.
\end{align}
Consequently, each term linear in the transverse flow vanishes upon integration over both azimuthal angles, independently of the finite quark and gluon masses. The remaining flow-independent kernel in Eq.~\eqref{frEloss1} is
\begin{widetext}
\begin{align}
\left.
\mathcal{K}\right|_{\mathrm{static}}
={}&
2\left[
1-\cos\left(
\frac{(\mathbf{k}-\mathbf{q})^2+\chi}{2xE}\,z
\right)
\right]
\Bigg[
\frac{(\mathbf{k}-\mathbf{q})^2}
{\left[(\mathbf{k}-\mathbf{q})^2+\chi\right]^2}
-
\frac{\mathbf{k}\cdot(\mathbf{k}-\mathbf{q})}
{(\mathbf{k}^2+\chi)
\left[(\mathbf{k}-\mathbf{q})^2+\chi\right]}
\Bigg]
\nonumber\\
&+
\sin\left(
\frac{\mathbf{k}^2+\chi}{2xE}\,z
\right)
\frac{\mathbf{k}^2}{(\mathbf{k}^2+\chi)^2}
\frac{R_0}{2xE}
\Bigg[
1+
\frac{\mu^2-2\mathbf{q}^2}
{4v^2(\mathbf{q})}\,
\frac{\partial v^2(\mathbf{q})}
{\partial\mathbf{q}^2}
\Bigg]\, .
\label{statkernel}
\end{align}
\end{widetext}
The first term represents the leading static DGLV contribution~\cite{DGLV}, while the second is the static first-subeikonal correction, which survives the complete azimuthal integration.

Thus, the static first-subeikonal term contributes to the azimuthally integrated spectrum and energy loss, whereas the terms linear in transverse flow contribute only to angularly differential observables.

\subsection{\label{static1stSubExp}Static first-subeikonal correction}

The cancellation established in Eq.~\eqref{fullangsym} implies that the azimuthally integrated radiation quantities are determined by the static kernel in Eq.~\eqref{statkernel}, including its first-subeikonal correction. To obtain explicit expressions for the numerical analysis, we adopt the normalized exponential longitudinal profile employed in Refs.~\cite{GLV2,DGLV, KML}, $\rho(z)=2\,e^{-2z/L}/L$. For this profile, the normalized longitudinal average is taken over $0\leq z<\infty$. 
For this exponential profile, the longitudinal averages of the formation-phase factors are evaluated explicitly in Appendix~\ref{M1M2NE}. The resulting formation-dependent denominators contain the characteristic scale $4xE/L$, which enters the expressions used in the numerical analysis.

We use the standard DGLV gluon-opacity normalization~\cite{GW,GLV2,DGLV}, $L/\lambda_g=N\sigma_g/A_\perp$, where $\lambda_g$ denotes the gluon mean free path; the corresponding normalization is specified in Appendix~\ref{M1M2NE}.

Using this normalization and the longitudinal averages above, the azimuthally integrated static gluon spectrum, i.e., fractional differential radiative energy loss, $x\,dN_g^{(1)}/dx \equiv E^{-1}dE^{(1)}/dx$, takes the form
\begin{widetext}
\begin{align}
\left.
x\frac{dN_g^{(1)}}{dx}
\right|_{\mathrm{static}}
{}&=
\frac{C_2(F)\alpha_s}{\pi}\,
\frac{L}{\lambda_g}
\int
\frac{d\mathbf{k}^2}{\mathbf{k}^2+\chi}
\int
\frac{d^2\mathbf{q}}{\pi}\,
\frac{\mu^2}{\left(\mathbf{q}^2+\mu^2\right)^2}
\Bigg\{
2\,
\frac{
\left(\mathbf{k}\cdot\mathbf{q}\right)
\left(\mathbf{k}-\mathbf{q}\right)^2
+\chi\,\mathbf{q}\cdot\left(\mathbf{q}-\mathbf{k}\right)
}{
\left[\left(\mathbf{k}-\mathbf{q}\right)^2+\chi\right]^2
+\left(\frac{4xE}{L}\right)^2
}
\nonumber \\
&\hspace{0.8cm}
+
\frac{
\mathbf{k}^2\left(4\mathbf{q}^2+\mu^2\right)
}{
L\sqrt{\mathbf{q}^2+\mu^2}\,
\left[
\left(\mathbf{k}^2+\chi\right)^2
+\left(\frac{4xE}{L}\right)^2
\right]
}
\Bigg\} \, .
\label{spectExpDistr}
\end{align}
\end{widetext}

\begin{widetext}
The corresponding fractional radiative energy loss reads
\begin{align}
\left.
\frac{\Delta E^{(1)}}{E}
\right|_\mathrm{{static}}
={}&
\frac{C_2(F)\alpha_s}{\pi}\,
\frac{L}{\lambda_g}
\int dx
\int
\frac{d\mathbf{k}^2}{\mathbf{k}^2+\chi}
\int
\frac{d^2\mathbf{q}}{\pi}\,
\frac{\mu^2}{\left(\mathbf{q}^2+\mu^2\right)^2}
\Bigg\{
2\,
\frac{
\left(\mathbf{k}\cdot\mathbf{q}\right)
\left(\mathbf{k}-\mathbf{q}\right)^2
+\chi\,\mathbf{q}\cdot\left(\mathbf{q}-\mathbf{k}\right)
}{
\left[\left(\mathbf{k}-\mathbf{q}\right)^2+\chi\right]^2
+\left(\frac{4xE}{L}\right)^2
}
\nonumber \\
&+
\frac{
\mathbf{k}^2\left(4\mathbf{q}^2+\mu^2\right)
}{
L\sqrt{\mathbf{q}^2+\mu^2}\,
\left[
\left(\mathbf{k}^2+\chi\right)^2
+\left(\frac{4xE}{L}\right)^2
\right]
}
\Bigg\}\, .
\label{ElossExpDstatic}
\end{align}
\end{widetext}

In both expressions, the first term reproduces the leading static DGLV contribution~\cite{DGLV}, while the second represents the static first-subeikonal correction generated by the scattering-potential pole residues. In the massless limit, $M=m_g=0$, the resulting static spectrum agrees with Eq.~(68) of Ref.~\cite{KML}, upon integration over the emitted-gluon transverse momentum. The second term introduces an additional formation-dependent denominator and remains finite after complete azimuthal integration. Equations~\eqref{spectExpDistr} and~\eqref{ElossExpDstatic} therefore allow the static first-subeikonal effects on the fractional differential radiative energy loss and fractional radiative energy loss to be quantified directly relative to the leading DGLV limit.

\subsection{\label{Flow1stSubExp} Flow-dependent differential radiation}

The angular cancellation above does not apply to a spectrum resolved
in the emission direction. Keeping the linear transverse-flow terms
and performing the same exponential longitudinal average and using the same normalization gives
\begin{widetext}
\begin{align}
x \, \frac{dN_g^{(1)}}{dx\,d^2\mathbf{k}}
={}&
\frac{C_2(F)\alpha_s}{\pi^2}\,
\frac{L}{\lambda_g}
\int\frac{d^2\mathbf{q}}{\pi}\,
\frac{\mu^2}{\left(\mathbf{q}^2+\mu^2\right)^2}
\Bigg\{
\frac{2}
{\left[(\mathbf{k}-\mathbf{q})^2+\chi\right]^2
+\left(\frac{4xE}{L}\right)^2}
\Bigg\{
\left[
(\mathbf{k}-\mathbf{q})^2
-\mathbf{k}\cdot(\mathbf{k}-\mathbf{q}) \, \frac{(\mathbf{k}-\mathbf{q})^2+\chi
}
{\mathbf{k}^2+\chi}
\right]
\nonumber\\
&\times
\left[
1+
\frac{\mathbf{u}\cdot\mathbf{q}}{xE}\,
\frac{
2\mathbf{k}^2-(\mathbf{k}-\mathbf{q})^2+\chi
}
{\mathbf{q}^2+\mu^2}
\right]
+
\frac{\mathbf{u}\cdot(\mathbf{k}-\mathbf{q})}{xE}
\left[
\frac{
\mathbf{k}^2
\left[(\mathbf{k}-\mathbf{q})^2+\chi\right]
}
{\mathbf{k}^2+\chi}
-2(\mathbf{k}-\mathbf{q})^2
\right]
\nonumber\\
&-
\frac{\mathbf{u}\cdot\mathbf{k}}{xE}\,
\mathbf{q}\cdot(\mathbf{k}-\mathbf{q}) \, \frac{
(\mathbf{k}-\mathbf{q})^2+\chi
}
{\mathbf{k}^2+\chi}
+
\frac{\mathbf{u}\cdot\mathbf{q}}{xE}\,
2x^2M^2
\left[
\frac{2(\mathbf{k}-\mathbf{q})^2}
{(\mathbf{k}-\mathbf{q})^2+\chi}
-\frac{\mathbf{k}\cdot(\mathbf{k}-\mathbf{q})}
{\mathbf{k}^2+\chi}
\right]
\Bigg\}
\nonumber\\
&+
\frac{\mathbf{u}\cdot\mathbf{k}}{xE}\,
\frac{
2\mathbf{k}^2
-(\mathbf{k}^2-\chi)
\frac{\mathbf{q}^2}{\mathbf{q}^2+\mu^2}
}
{
(\mathbf{k}^2+\chi)^2
+\left(\frac{4xE}{L}\right)^2
}
+
\frac{
\mathbf{k}^2
\left(4\mathbf{q}^2+\mu^2\right)
}
{
L\,
(\mathbf{k}^2+\chi)\,
\sqrt{\mathbf{q}^2+\mu^2}\,
\left[
(\mathbf{k}^2+\chi)^2
+\left(\frac{4xE}{L}\right)^2
\right]
}
\nonumber\\
&+\frac{\mathbf{u}\cdot\mathbf{q}}{x E}
\frac{
\frac{64x^2E^2}{L^2}\,
\left[
(\mathbf{k}-\mathbf{q})^2+m_g^2-x^2M^2
\right]
}
{
\left[
\left[(\mathbf{k}-\mathbf{q})^2+\chi\right]^2
+\left(\frac{4xE}{L}\right)^2
\right]^{2}
}
\left[
\frac{(\mathbf{k}-\mathbf{q})^2}
{(\mathbf{k}-\mathbf{q})^2+\chi}
-\frac{
\mathbf{k}\cdot(\mathbf{k}-\mathbf{q})
}
{\mathbf{k}^2+\chi}
\right]
\Bigg\} \, .
\label{specExp}
\end{align}
\end{widetext}
Equation~\eqref{specExp} retains the full first-subeikonal
modification of the double-differential radiation spectrum, including
the mass-dependent and phase-induced terms described above.
The flow-dependent contributions are linear in
$\mathbf{u}\cdot\mathbf{k}$, $\mathbf{u}\cdot\mathbf{q}$, or
$\mathbf{u}\cdot(\mathbf{k}-\mathbf{q})$, so their effects are
resolved before averaging over the direction of the emitted gluon.
Their angular properties are considered together with the numerical
results in Sec.~\ref{sec:Num}, rather than being combined with the
azimuthally integrated observables.

The subsequent numerical study therefore addresses, first, the static
first-subeikonal modification of the DGLV spectrum and the integrated
fractional radiative energy loss, using Eqs.~\eqref{spectExpDistr} and
\eqref{ElossExpDstatic}, and, second, the differential radiation in a
transversely flowing medium, evaluated from Eq.~\eqref{specExp}.
This distinction isolates the static correction that survives full
azimuthal averaging from the angular redistribution induced at linear
order by collective flow.

\section{\label{sec:Num}Numerical results and discussion}

\begin{figure*}[t]
\centering
\includegraphics[width=\textwidth]{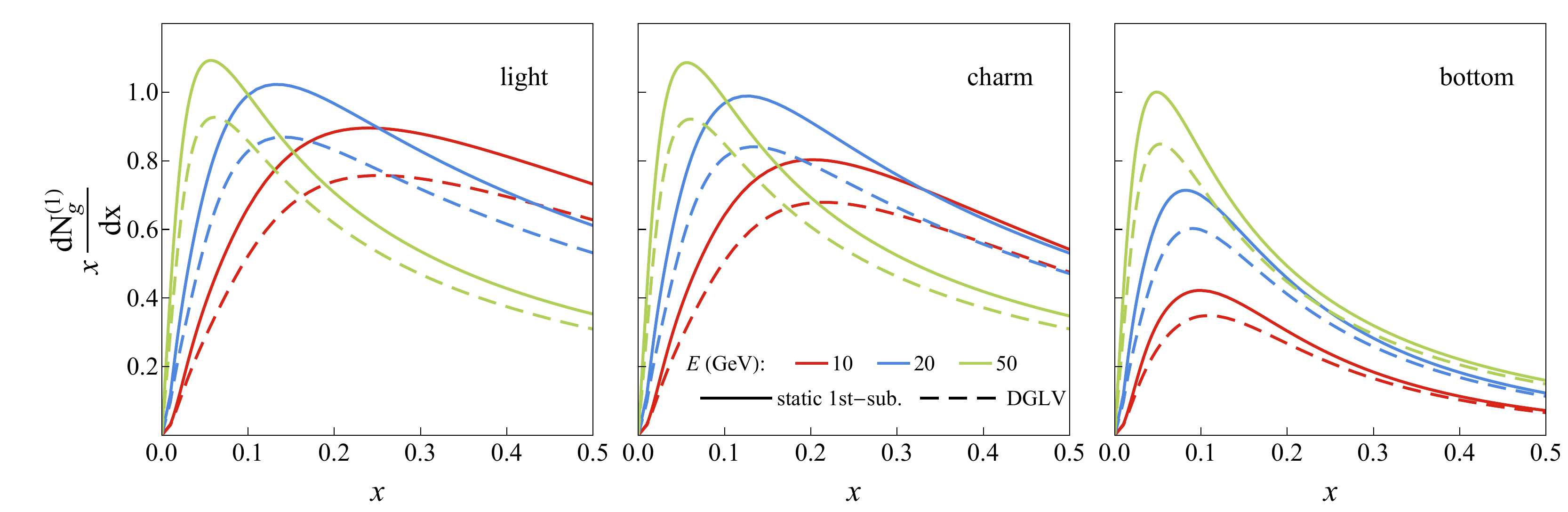}
\caption{\small The effect of the first-subeikonal correction on the fractional differential radiative energy loss at first order in opacity, $x\,dN_g^{(1)}/dx$, as a function of $x$ at $L=5\,\mathrm{fm}$ for light (left panel), charm (middle panel), and bottom (right panel) quarks in a static QCD medium. The solid curves correspond to the full static results through first-subeikonal order (labeled ``static 1st-sub.'' in the panels) given by Eq.~\eqref{spectExpDistr}, while the dashed curves represent the conventional DGLV predictions~\cite{DGLV}. Red, blue, and green curves denote initial quark energies $E=10$, $20$, and $50$ GeV, respectively.}
\label{fig:Fig1}
\end{figure*}

\begin{figure*}[t]
\centering
\includegraphics[width=0.8\textwidth,trim=70bp 0bp 10bp 0bp,clip]{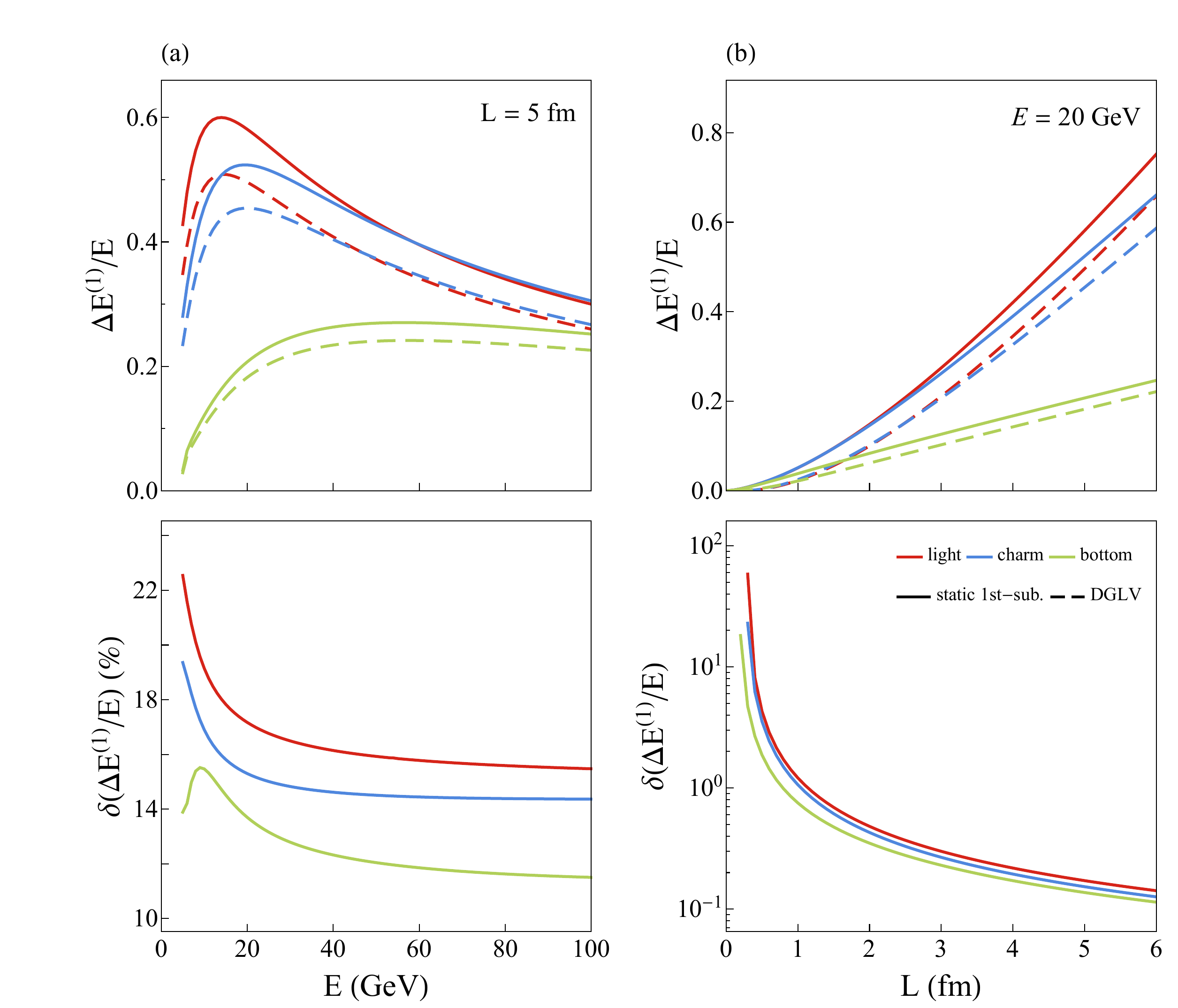}
\caption{\small The effect of the first-subeikonal correction on the integrated fractional radiative energy loss at first order in opacity, $\Delta E^{(1)}/E$, for light, charm, and bottom quarks in a static QCD medium. Panel (a) shows the dependence on the initial energy at $L=5$ fm, while panel (b) presents the path-length dependence at $E=20$ GeV. The solid curves correspond to the full static results through first-subeikonal order (labeled ``static 1st-sub.'' in the panels), while the dashed curves represent the conventional DGLV predictions~\cite{DGLV}. The lower panels show the corresponding relative correction, expressed as a percentage in panel (a) and on a logarithmic scale in panel (b). Red, blue, and green curves denote light, charm, and bottom quarks, respectively.}
\label{fig:Fig21}
\end{figure*}

We first compare the static first-subeikonal results from Eqs.~\eqref{spectExpDistr} and~\eqref{ElossExpDstatic} with the conventional DGLV predictions~\cite{DGLV}, using the same medium parameters and kinematic constraints. For the numerical calculations, we use (see Refs.~\cite{DGLV, bsga}) a constant strong coupling $\alpha_s=0.3$, medium temperature $T=336$ MeV, gluon mean free path $\lambda_g=0.765\,\mathrm{fm}$, and, unless stated otherwise, path length $L=5\,\mathrm{fm}$, with the effective gluon mass $m_g=\mu/\sqrt{2}$, where $\mu=\sqrt{4 \pi \alpha_s (1+n_f/6)}T$ and $n_f=3$. The light, charm, and bottom quark masses are taken as $M_{\rm light}=\mu/\sqrt{6}$, $M_c=1.2$ GeV, and $M_b=4.75$ GeV, respectively.

Figure~\ref{fig:Fig1} shows that the static first-subeikonal correction increases the spectrum throughout the displayed $x$ range, with a relative magnitude that depends on the quark mass and initial energy.

For light quarks (Fig.~\ref{fig:Fig1}, left panel), the energy ordering $E=50>20>10$ GeV of the spectra at small $x$ changes at larger $x$, with the $E=10$ GeV spectrum becoming the largest near the upper end of the displayed interval. Increasing the initial quark energy also shifts the maximum toward smaller $x$. Charm-quark spectra (middle panel) exhibit similar crossings, with finite-mass effects becoming more pronounced at lower energies and intermediate to larger $x$.

Bottom-quark spectra (right panel) are more concentrated at small $x$ and decrease rapidly beyond their maxima, preserving the ordering $E=50>20>10$ GeV throughout the displayed range. These features also occur in the DGLV spectra.

Relative to the DGLV result, the first-subeikonal correction produces its largest relative enhancement at small $x$. At intermediate and larger $x$, the relative enhancement generally decreases with increasing energy and quark mass and is least pronounced for bottom quarks.

The mass and energy dependence reflects finite-mass effects and gluon formation dynamics, as established within the GLV and DGLV frameworks~\cite{GLV1,DGLV}. The gluon mass accounts for the Ter-Mikayelian modification~\cite{DG03} of the in-medium dispersion relation, while the quark mass produces dead-cone suppression~\cite{DokshitzerKharzeev2001}. Both enter through $\chi$ in the radiation amplitudes and formation frequencies. Increasing the quark mass suppresses radiation through the transverse propagator structure but also shortens the formation time, reducing destructive LPM  interference~\cite{DGLV}. The DGLV and first-subeikonal contributions to Eq.~\eqref{spectExpDistr} retain these competing effects, although their formation-dependent denominators involve $(\mathbf{k}-\mathbf{q})^2+\chi$ and $\mathbf{k}^2+\chi$, respectively.
The increasing relevance of $x^2M^2$ at larger $x$ qualitatively explains the stronger suppression and concentration of bottom-quark radiation toward smaller gluon energy fractions. 

The quark and gluon mass contributions become comparable when $xM\sim m_g$, corresponding to $x\sim m_g/M$. For the parameters used here, this occurs at $x\simeq0.1$ for bottom and $x\simeq0.5$ for charm quarks, whereas the corresponding scale for light quarks lies outside the displayed range. Thus, the quark-mass contribution becomes relevant at substantially smaller $x$ for bottom quarks.

The energy dependence reflects the competition between LPM interference and the finite transverse-momentum phase space~\cite{GLV1,DGLV}. After longitudinal averaging, the scale $4xE/L$ enters both contributions to Eq.~\eqref{spectExpDistr}. At fixed $x$ and transverse momenta, increasing $E$ lengthens the formation time and strengthens destructive interference, while enlarging the transverse-momentum integration range. Together with the mass dependence through $\chi$, these effects provide a qualitative interpretation of the shifts of the spectral maxima and the crossings for light and charm quarks. For bottom quarks, stronger suppression at intermediate and larger $x$ restricts the spectrum without reversing its energy ordering. 

The overall mass and energy dependence largely follows the DGLV predictions, whereas the first-subeikonal correction adds an $x$-dependent enhancement whose relative magnitude also depends on the quark mass and initial energy. Nonvanishing subeikonal corrections in static matter, including enhanced radiation at small transverse momenta, were previously established in Ref.~\cite{KML}; our result extends this analysis to finite quark and gluon masses.

The absolute static first-subeikonal correction becomes increasingly concentrated at small $x$ with increasing quark mass, particularly for bottom quarks. Consequently, its contribution to the integrated fractional radiative energy loss depends not only on its magnitude but also on the range of gluon energy fractions over which it remains significant.

We next investigate the effect of the first-subeikonal correction on the integrated fractional radiative energy loss, $\Delta E^{(1)}/E$, in a static QCD medium. In Fig.~\ref{fig:Fig21}, the solid curves represent our static first-subeikonal results obtained from Eq.~\eqref{ElossExpDstatic}, whereas the dashed curves correspond to the DGLV predictions~\cite{DGLV}, recovered by retaining only the first term of that equation. To quantify the relative contribution of the second, subeikonal term, we define $\delta\left(\Delta E^{(1)}/{E}\right)=\left( \left.\Delta E^{(1)}/{E}\right|_{\rm static}
-\left.\Delta E^{(1)}/{E}\right|_{\rm DGLV} \right)/
\left.\Delta E^{(1)}/{E}\right|_{\rm DGLV}$.
The lower panels of Fig.~\ref{fig:Fig21} display this ratio, multiplied by 100 in panel (a) and plotted without this factor on a logarithmic scale in panel (b).

Panel (a) of Fig.~\ref{fig:Fig21} presents the  $\Delta E^{(1)}/E$ energy dependence at $L=5$ fm. The inclusion of the static first-subeikonal correction systematically enhances $\Delta E^{(1)}/E$ for all three quark flavors, as expected from the manifestly nonnegative second term in Eq.~\eqref{ElossExpDstatic}. The light- and charm-quark results exhibit pronounced maxima around $E\simeq15$ and $20$ GeV, reaching approximately $0.6$ and $0.5$, respectively, before gradually decreasing. In contrast, the bottom-quark fractional energy loss remains considerably smaller, with a broader maximum of approximately $0.25$ shifted toward $E\simeq50$--$60$ GeV. The correction slightly shifts these maxima toward lower energies, while its main effect is to increase their magnitude. At higher energies, the light- and charm-quark results approach one another, whereas bottom-quark radiation remains more strongly suppressed.

The lower part of panel (a) shows that the relative enhancement due to the static first-subeikonal correction remains positive throughout the considered energy range and exhibits a distinct mass dependence. For light quarks, it decreases from slightly above $20\%$ at low energies to approximately $15\%$ at $E=100$ GeV. Charm quarks exhibit a similar trend, with a slightly smaller relative contribution. The bottom-quark correction is smaller, reaching a shallow maximum of approximately $15\%$ at low energies before falling to about $10\%$. Thus, the static first-subeikonal correction remains notable even where the absolute fractional radiative energy loss decreases with increasing energy.

The mass and energy dependence of the integrated fractional radiative energy loss is consistent with the behavior of the differential spectra discussed in connection with Fig.~\ref{fig:Fig1}. In particular, the pronounced suppression of bottom-quark radiation at intermediate and larger $x$ contributes to its smaller integrated energy loss, whereas the reduced separation between the light- and charm-quark spectra at higher energies (evident from the left and middle panels of Fig.~\ref{fig:Fig1}) is reflected in their approaching integrated results. The smaller absolute static first-subeikonal contribution for bottom quarks can be understood from the second term of Eq.~\eqref{ElossExpDstatic}, whose positive integrand decreases with increasing $\chi$ at fixed $x$ and transverse momenta. This mass dependence, together with the corresponding DGLV contribution and the available phase space, contributes to the stronger suppression of bottom-quark energy loss observed in Fig.~\ref{fig:Fig21}(a). The nonmonotonic energy dependence further reflects the interplay between LPM interference, finite-mass effects, and the available phase space, as established in previous opacity-expansion studies~\cite{GLV1,DGLV}. At fixed $x$ and transverse momenta, increasing the initial energy strengthens destructive interference through the formation-dependent scale $4xE/L$, while simultaneously enlarging the transverse-momentum integration range. Together with the mass-dependent suppression encoded in $\chi$ and the competing reduction of LPM interference, these effects may provide a qualitative explanation for the broader maximum of the bottom-quark fractional energy loss, shifted toward higher energies compared with light and charm quarks. The relative subeikonal enhancement, however, depends on the ratio of the integrated subeikonal and DGLV contributions. Their different transverse-momentum and formation-dependent structures, together with the changing integration domain, lead to distinct mass and energy dependences of this ratio. These differences provide a qualitative interpretation of the smaller relative correction for heavier quarks and its overall decrease toward higher energies.

Panel (b) of Fig.~\ref{fig:Fig21} shows the path-length dependence of the fractional radiative energy loss at $E=20$ GeV. The static first-subeikonal correction increases $\Delta E^{(1)}/E$ throughout the displayed $L$ range, with the largest absolute contribution occurring for light quarks. Both the corrected and DGLV results grow monotonically with path length, and their separation becomes more pronounced as $L$ increases. In contrast, the relative enhancement decreases rapidly, as illustrated by the logarithmic scale in the lower panel. This behavior is observed for all three quark flavors, with the relative correction remaining largest for light quarks. Particularly large relative contributions occur at short path lengths, although the fractional radiative energy loss remains small in this region.

The contrasting length dependences can be understood from the structure of Eq.~\eqref{ElossExpDstatic}. As in the GLV and DGLV formulations~\cite{GLV1,DGLV}, the medium length enters through the common opacity factor $L/\lambda_g$ and the formation-dependent scale $4xE/L$, whereas the static first-subeikonal correction contains an additional explicit factor $1/L$. In the sufficiently short-$L$ regime, where the formation-dependent terms dominate the denominators, these structures yield an $L^3$ dependence for the DGLV contribution and an $L^2$ dependence for the subeikonal correction, provided the finite integration bounds are kept fixed. A corresponding difference in the path-length dependence of the DGLV and subeikonal contributions was also noted in Ref.~\cite{KML} for the differential radiation spectrum in static matter. Their ratio therefore increases as $L$ decreases, accounting for the pronounced relative enhancement at short path lengths. As $L$ increases, the formation-dependent scale $4xE/L$ decreases, reducing formation suppression in both contributions. However, the DGLV term retains the explicit opacity factor $L/\lambda_g$, which is canceled by the additional $1/L$ factor in the static first-subeikonal correction. The subeikonal contribution to $\Delta E^{(1)}/E$ therefore grows more slowly than the DGLV contribution over the displayed path-length range, explaining the decrease in their ratio despite the increasing absolute difference.

This behavior reflects the competition between formation-dependent interference effects and the mass-dependent kinematic phase space. At small $L$, the scale $4xE/L$ produces strong formation suppression, making the relative contribution particularly sensitive to the balance between the leading and subeikonal terms. Increasing the initial energy also enlarges the accessible phase space and reduces the relative importance of the finite quark mass. These competing effects account for the near coincidence of the light- and charm-quark energy losses and the nontrivial mass ordering of the relative correction. The large relative enhancement at short path lengths does not imply a large absolute subeikonal contribution, because the corresponding leading fractional energy loss is small in this region.

\begin{figure*}[t]
\centering
\includegraphics[width=\textwidth]{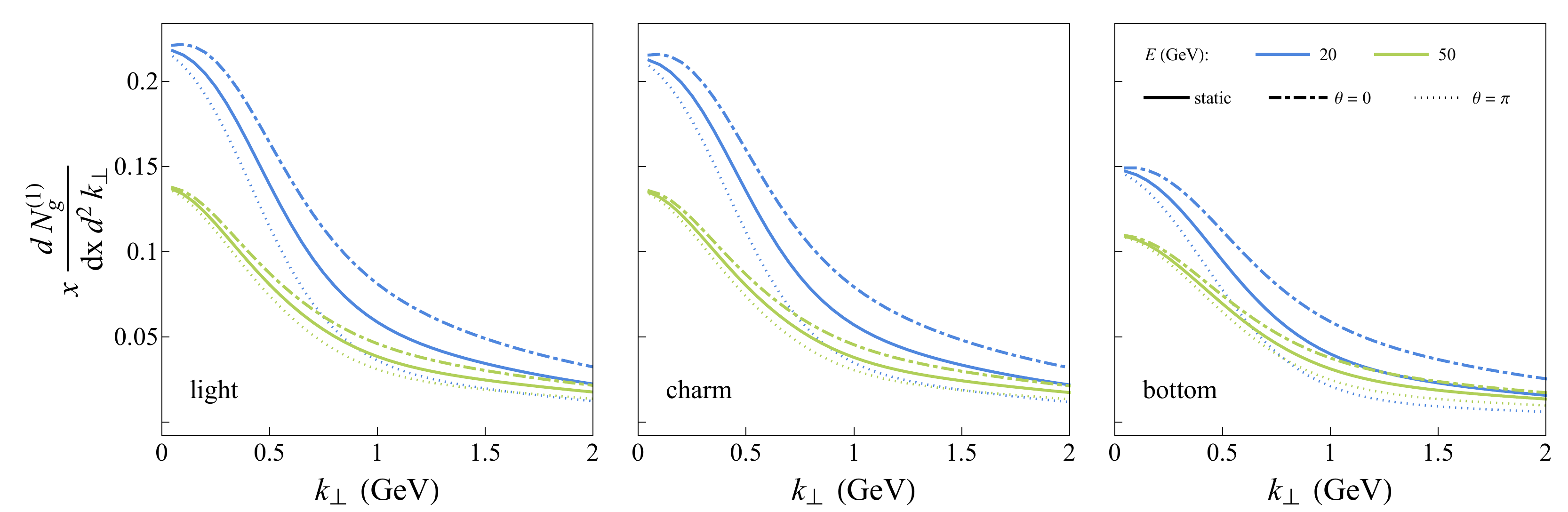}
\caption{\small The effect of transverse collective flow on the double-differential medium-induced gluon radiation spectrum at first order in opacity, $x\,dN_g^{(1)}/(dx\,d^2\mathbf{k})$, in units of $\mathrm{GeV}^{-2}$, for light (left panel), charm (middle panel), and bottom (right panel) quarks. The spectra are shown as functions of $k_\perp$ at fixed $x=0.1$, $L=5$ fm, and $|\mathbf{u}|=0.2$, for initial quark energies $E=20$ and $50$ GeV. The solid curves correspond to the full static first-subeikonal results ($\mathbf{u}=0$), while the dot-dashed and dotted curves represent the flowing-medium results for $\theta=0$ and $\theta=\pi$, respectively, where $\theta$ is the angle between $\mathbf{k}$ and $\mathbf{u}$. Blue and green curves denote $E=20$ and $50$ GeV, respectively.}
\label{fig:Fig3}
\end{figure*}

The azimuthal cancellation established in Sec.~\ref{sec:AzimSymm} applies to quantities integrated over both transverse directions. To examine the angular dependence of the double-differential radiation spectrum, we instead keep the emission angle $\theta$ fixed and integrate only over the azimuthal angle of the exchanged momentum $\mathbf{q}$. Using $\Phi=\theta-\psi$ and the functions $F$ and $G$ introduced in Eq.~\eqref{fullangsym}, the contribution linear in the transverse flow velocity is proportional to
\begin{align}
&\int_0^{2\pi}d\psi\,
\Bigl[
\cos\theta\, F(|\mathbf{k}|,|\mathbf{q}|,\cos\psi)
+\cos(\theta-\psi)\,G(|\mathbf{k}|,|\mathbf{q}|,\cos\psi)
\Bigr]
\nonumber\\
&\quad=
\cos\theta\int_0^{2\pi}d\psi\,
\bigl[F+\cos\psi\,G\bigr]
+\sin\theta\int_0^{2\pi}d\psi\,
\sin\psi\,G
\nonumber\\
&\quad=
\cos\theta\int_0^{2\pi}d\psi\,
\bigl[F+\cos\psi\,G\bigr] \, ,
\label{pi2kperp}
\end{align}
where the last equality follows because $G$ depends on $\psi$ only through $\cos\psi$, so that
\begin{equation}
\int_0^{2\pi}d\psi\,\sin\psi\,
G(|\mathbf{k}|,|\mathbf{q}|,\cos\psi)=0
\end{equation}
under the transformation $\psi\to2\pi-\psi$.

After integration over the azimuthal angle of $\mathbf{q}$, the linear-flow correction to the first-subeikonal radiation spectrum has a characteristic dipolar $\cos\theta$ dependence. For $\theta=\pi/2$, the flow-dependent correction vanishes identically, yielding
\begin{equation}
\left.
x\, \frac{dN_g^{(1)}}{dx\,d^2\mathbf{k}}
\right|_{\theta=\pi/2}
=
\left.
x \, \frac{dN_g^{(1)}}{dx\,d^2\mathbf{k}}
\right|_{\mathbf{u}=0}
+\mathcal{O}(\mathbf{u}^2).
\label{pi2}
\end{equation}
The flow-induced modifications for parallel ($\theta=0$) and antiparallel ($\theta=\pi$) emission are equal in magnitude and opposite in sign. This cancellation holds independently of the finite quark and gluon masses and the retained subleading corrections to the formation phases, which modify the coefficients of the linear-flow terms without changing their dipolar angular structure.

Figure~\ref{fig:Fig3} shows these flow-induced modifications, evaluated from Eq.~\eqref{specExp}.
The spectra generally decrease over the displayed $k_\perp$ range, apart from shallow small-$k_\perp$ maxima in the $E=20$ GeV, $\theta=0$ light- and charm-quark curves. Consistent with the dipolar structure in Eq.~\eqref{pi2kperp}, the numerical results show enhanced radiation for $\theta=0$ and suppressed radiation for $\theta=\pi$ relative to the corresponding static spectrum. The separation between the two angular configurations is small near $k_\perp=0$, becomes most pronounced at intermediate transverse momenta, and decreases toward larger $k_\perp$. This anisotropy is generally stronger at $E=20$ GeV than at $50$ GeV. The light- and charm-quark results remain close, while bottom-quark radiation is more strongly suppressed. Furthermore, the spectra at $E=20$ GeV are larger at small $k_\perp$ for all quark flavors, although their ordering relative to the $E=50$ GeV results can change at larger transverse momenta, depending on the emission angle.

The dipolar structure follows directly from the linear dependence on the transverse flow velocity retained in our calculation. After integration over the azimuthal angle of the exchanged momentum, rotational symmetry requires the flow-dependent contribution to be proportional to $\mathbf{u}\cdot\mathbf{k}=|\mathbf{u}| \, |\mathbf{ k}| \cos\theta$ (see Eqs.~\eqref{pi2kperp} and \eqref{pi2}). Consequently, the modifications for $\theta=0$ and $\theta=\pi$ are equal in magnitude and opposite in sign, whereas for $\theta=\pi/2$ the spectrum coincides with the full static first-subeikonal result at this order. The latter configuration is therefore omitted from the figure. This behavior is consistent with our analytical demonstration that, for the rotationally symmetric transverse-momentum integration domains used here, all contributions linear in the transverse flow velocity vanish separately upon full azimuthal integration (see Eq.~\eqref{fullangsym}). Thus, at linear order in $\mathbf{u}$, transverse flow generates a dipolar angular anisotropy without modifying the azimuthally integrated radiation spectrum.

The factor $|\mathbf{u}|\,|\mathbf{k}|\cos\theta$ suppresses the angular separation near $k_\perp=0$, while the remaining momentum-dependent factors suppress it at large $k_\perp$, producing the maximum at intermediate $k_\perp$ shown in Fig.~\ref{fig:Fig3}.

 The observed mass dependence reflects both the conventional dependence on $\chi$, entering the propagator denominators and leading formation phases, and additional mass-dependent flow corrections. In the mixed quark-gluon contributions, the same subleading mismatch between longitudinal propagator poles generates mass-dependent residue-denominator corrections and shifts of the relative formation phases. The latter involve $(\mathbf{k}-\mathbf{q})^2+m_g^2-x^2M^2$, where the gluon and heavy-quark mass terms enter with opposite signs, unlike in the conventional DGLV combination $\chi$. At $x=0.1$, the quark-mass term is small for light and charm quarks but becomes comparable to $m_g^2$ for bottom quarks. Consequently, finite masses affect both the static radiation spectrum and the flow-induced angular anisotropy, with the latter determined by the combined contributions of all flow-dependent terms in Eq.~\eqref{specExp}.

The energy dependence reflects the combined effects of LPM interference and the explicit inverse-energy factors in the linear-flow contributions of Eq.~\eqref{specExp}. At fixed $x$ and transverse momenta, increasing $E$ raises the formation-dependent scale $4xE/L$, strengthening destructive interference and contributing to the smaller radiation spectra at $E=50$ GeV in the low-$k_\perp$ region. At the same time, the factors $1/(xE)$ in the flow-dependent terms tend to reduce the angular splitting at higher energies, although its detailed behavior also depends on their formation-dependent coefficients. With increasing $k_\perp$, the transverse-momentum scales become more important relative to the formation-dependent scale $4xE/L$, reducing the energy dependence associated with LPM interference. This is consistent with the convergence of the static spectra at $E=20$ and $50$ GeV observed in Fig.~\ref{fig:Fig3}. Since the negative flow-induced contribution for $\theta=\pi$ is more pronounced at the lower quark energy, it further reduces the difference between the corresponding spectra and can contribute to a reversal of their energy ordering at larger $k_\perp$. For $\theta=0$, the positive flow correction acts in the opposite direction, tending to preserve the original ordering.

The flow-induced angular asymmetry is qualitatively consistent  with the preferential gluon emission along the medium flow reported in Ref.~\cite{SSVflow}, where the entire transverse-flow dependence of the radiation spectrum enters through a single term proportional to $\mathbf{u}\cdot\mathbf{k}/(xE)$ in the soft-gluon limit (Eq.~(51) of Ref.~\cite{SSVflow}). A similar anisotropy was obtained in Ref.~\cite{KML}. In the massless limit, our subleading Lorentz-contraction and potential-derivative contributions agree with the corresponding first-subeikonal structures of Ref.~\cite{KML}, including the ordinary sine terms originating from scattering-potential poles. Our calculation further incorporates finite heavy-quark and gluon masses, including the associated mass-dependent residue-denominator and phase-induced $z \sin$ corrections, while preserving the dipolar angular dependence at linear order in $\mathbf{u}$.

\section[Conclusions]{\label{sec:CO} Conclusions}

We calculated medium-induced gluon radiation from a highly energetic heavy quark propagating through a finite-size, finite-temperature QCD medium with transverse collective flow. Within the DGLV framework, we evaluated the radiation spectrum at first order in opacity and to first-subeikonal accuracy, retaining both the finite heavy-quark mass and the effective thermal gluon mass. Moving-medium interactions were incorporated directly at the amplitude level for all single- and two-scattering contributions, with the transverse-flow dependence kept to linear order.

In the mixed quark-gluon contributions, the same subleading pole mismatch generates mass-dependent residue-denominator corrections and shifts of the formation phases. The mass-dependent residue corrections vanish in the massless limit, whereas the phase-induced first-subeikonal $z\sin\phi_0$ terms can remain nonzero. Under the adopted LPM power counting, these phase corrections contribute at first-subeikonal order whenever the corresponding pole shifts survive. Furthermore, the mass dependence of these corrections differs from the conventional DGLV combination $\chi$, with an explicit heavy-quark mass contribution in the residue correction and oppositely signed quark and gluon mass terms in the flow-induced phase shift.

The resulting radiation spectrum separates the static first-subeikonal corrections from the flow-induced contributions. For rotationally symmetric transverse-momentum integration domains, all terms linear in the transverse flow velocity cancel upon full azimuthal integration, while a nonvanishing static correction originating from scattering-potential pole residues survives. This contribution agrees with the corresponding massless result of Ref.~\cite{KML} and is extended here to finite quark and gluon masses.
  
Our numerical results show that the static first-subeikonal correction systematically enhances both the fractional differential and integrated radiative energy loss for light, charm, and bottom quarks. The enhancement of the radiation spectra is nonuniform in $x$ and generally becomes less pronounced for heavier quarks at intermediate and larger gluon energy fractions. Bottom-quark radiation remains concentrated toward smaller $x$, while its integrated fractional energy loss peaks at higher initial-quark energies than for lighter quarks. The integrated enhancement is typically $10$--$20\%$, with larger relative effects at lower energies. The static correction also exhibits weaker suppression at short path lengths than the leading DGLV contribution, resulting in a larger relative enhancement as $L$ decreases, even though the total fractional radiative energy loss remains small. At larger $L$, its relative importance decreases, despite an increasing absolute difference in $\Delta E^{(1)}/E$ over the considered range.

In the presence of transverse flow, the double-differential gluon spectrum exhibits a dipolar angular anisotropy at linear order in the flow velocity. For the parameters considered, radiation is enhanced along the flow direction and suppressed in the opposite direction, with the angular separation most pronounced at intermediate transverse momenta and generally stronger at lower initial-quark energies. For the rotationally symmetric transverse-momentum integration domains used here, the flow-dependent contribution vanishes after full azimuthal integration. Thus, at linear order in the transverse flow velocity, flow modifies the angular distribution of medium-induced radiation, while the azimuthally integrated radiative energy loss is affected only by the static first-subeikonal correction.

{\em Acknowledgments:}
This work is supported by the Ministry of Science,
Technological Development and Innovation of the Republic of Serbia.

\appendix

\section{\label{sec:Assumptions} Assumptions and power counting}

This Appendix summarizes the kinematic conventions, theoretical assumptions, and power-counting scheme employed throughout our calculation, including the treatment of the medium and hard-production source.

Four-momenta are written in ordinary Minkowski components as
\begin{equation}
    P^\mu=(P^0,\mathbf{P},P_z),
\end{equation}
where boldface symbols denote two-dimensional transverse vectors with
respect to the chosen longitudinal $z$ direction,
$\vec{P}=(\mathbf{P},P_z)$. The initial heavy-quark momentum is not
required to be exactly aligned with the $z$ axis.

The frame is chosen such that the flow has no longitudinal component,
\begin{equation}
    u_z=0,
    \qquad
    u^\mu=(1,\mathbf{u},0),
\label{flowu}
\end{equation}
following the frame choice used in Ref.~\cite{KML}. We use the convention in which the common Lorentz factor $\gamma$ is
factored out of $u^\mu$~\cite{SSVflow,KML}. 

\begin{itemize}
%-----------------------------------------------------------------------
\item \textit{Soft-gluon and soft-rescattering approximations.}

The radiation is treated in the soft-gluon limit~\cite{GLV2,DGLV, KML},
\begin{equation}
    x\equiv\frac{\omega}{E}\ll1,
\end{equation}
where $E$ and $\omega$ denote the energies of the energetic parent quark and
the emitted gluon, respectively. We retain the soft-gluon
approximation, as corrections beyond it were found to be numerically
negligible within the DGLV framework~\cite{bsga}.

Following Refs.~\citep{GLV2,DGLV, KML}, we assume that the parton energies and longitudinal momenta are large compared to their transverse momenta, so that both the radiated gluon and the final quark undergo only small transverse deflections with respect to their eikonal trajectories. Thus, the eikonal hierarchy is taken to be
\begin{equation}
\begin{aligned}
   E \gg xE \gg
|\mathbf{p}|,\,
|\mathbf{k}|,\,
|\mathbf{q}|,\,
\mu,\,
m_g,\,
xM .
\end{aligned}
\label{eq:hierarchy}
\end{equation}
 Here $\mathbf{k}$ is the
transverse momentum of the emitted gluon, $\mathbf{q}$ is a transverse
momentum exchanged with the medium, and $\mu$ is the Debye screening
mass.

The quark and gluon masses are nevertheless retained  through first-subeikonal order. In the present calculation we define
the DGLV mass combination
\begin{equation}
\chi \equiv m_g^2+x^2M^2\,  ,
\label{eq:chi}
\end{equation}
which enters the standard DGLV transverse propagator denominators and
leading formation frequencies. Additional mass-dependent terms are
retained separately wherever they arise at first-subeikonal accuracy.

%-----------------------------------------------------------------------
\item \textit{First order in opacity.}

The opacity expansion is truncated at first order, as discussed in~\cite{GLV2,DGLV, KML, MGFprvi,prvi1}. The corresponding
medium-induced contribution consists of the squared single-scattering
amplitude and the interference of the two-scattering amplitude with the
vacuum amplitude~\cite{DGLV},
\begin{equation}
\begin{aligned}
    dN^{(1)}
    \propto {}&
    \left\langle
        |\mathcal{M}_{1}|^2
    \right\rangle
    \\
    &+
    2\,{\rm Re}
    \left\langle
        \mathcal{M}_{0}^{\dagger}
        \mathcal{M}_{2}
    \right\rangle .
\end{aligned}
\label{eq:firstopacity}
\end{equation}
Here $\mathcal{M}_{1}$ denotes the sum of the single-scattering diagrams, 
$\mathcal{M}_{2}$ the sum of the two-scattering contributions, and $\mathcal{M}_0$ is the zeroth-order-in-opacity radiation amplitude, corresponding to gluon emission in the absence of interactions with the medium. Terms
involving a larger number of medium interactions are outside the
present order in opacity.

%-----------------------------------------------------------------------
\item \textit{Flowing-medium background field.}

Medium interactions are represented by an external color field written
as a sum over moving scattering centers~\cite{SSVflow,KML}, 
\begin{align}
A_{\rm ext}^{\mu,a}(q)
&=
\sum_i
e^{-i\left(
       \mathbf{q}\cdot\mathbf{x}_i
      +q_z z_i
     \right)}
a_i^{\mu,a}(q),
\label{Aext}
\end{align}
with
\begin{align}
g_s a_i^{\mu,a}(q)
&=
T_i^a u_i^\mu v_i(q)\,
2\pi\,
\delta\!\left(
    q^0-\vec{u}_i\cdot\vec{q}
\right).
\label{a}
\end{align}
Here $(\mathbf{x}_i,z_i)$ specifies the position of the $i^{th}$
scattering center and $T_i^a$ is the corresponding $SU(N_c)$ generator
in the color space of that medium constituent, whereas the sum encompasses all scattering centers in the medium. The screened
single-center scattering potential is described by
\begin{align}
v_i(q)
&=
\frac{-g_s^2}
{\vec{q}^{\,2}+\mu_i^2-(\vec{u}_i\cdot\vec{q})^2}.
\label{scattpot}
\end{align}

The frame specified above permits the axial-gauge choice
\begin{equation}
    n\cdot A=0,
    \qquad
    n^\mu=(0,\mathbf{0},1),
    \label{eq:axialgauge}
\end{equation}
for which only the two physical gluon polarizations enter the
calculation.

Using the momentum conventions for on-shell heavy quark $(p-k)^\mu$ and massive emitted gluon $k^\mu$ specified in the main text by Eq.~\eqref{vektori}, we impose the following conditions on the emitted-gluon polarization vector
\begin{equation}
\begin{aligned}
&\varepsilon(k)\cdot k = 0,
\qquad
\varepsilon(k)\cdot n = 0,
\\
&\varepsilon(k)^2 = -1.
\end{aligned}
\label{kin0M1}
\end{equation}
The first condition expresses transversality with respect to the gluon momentum, the second implements the axial-gauge condition specified in Eq.~\eqref{eq:axialgauge}, and the third fixes the normalization of the physical polarization state. To first-subeikonal order, the polarization vector can be written as given in Eq.~\eqref{vektori}.

In addition, the flow dependence is expanded only to linear order in the transverse velocity,
\begin{equation}
\mathcal{O}(|\mathbf{u}|)
\quad\text{retained},
\qquad
\mathcal{O}{(|\mathbf{u}|^2})
\quad\text{neglected}.
\label{eq:flowexpansion}
\end{equation}

%-----------------------------------------------------------------------
\item \textit{Subeikonal expansion and finite formation phases.}

At first-subeikonal accuracy, the expansion is organized in powers of
$\perp/(xE)$, where $\perp$ denotes a generic soft transverse or mass
scale, with
$|\mathbf{p}|\sim|\mathbf{k}|\sim|\mathbf{q}|\sim\mu\sim m_g\sim xM
\sim\perp$,
following the power-counting logic of Ref.~\cite{KML}.

The leading formation phase scales as
\begin{equation}
    \phi_0
    \sim
    \frac{\perp^2}{xE}\,z
    \sim
    \mathcal{O}(1).
    \label{eq:phi0counting}
\end{equation}
Thus, the longitudinal extent of the medium is not assumed to be small
compared with the gluon formation length. In particular,
$\perp^2z/(xE)$ is not used as an expansion parameter, under the LPM counting of Refs.~\cite{GLV2,DGLV,KML}.

The longitudinal propagator-pole positions are expanded through $\mathcal{O}\!\left(\perp^3/(xE)^2\right)$ wherever required to achieve first-subeikonal accuracy in residue denominators and formation phases. Schematically,
\begin{equation}
\begin{aligned}
    Q^- &= Q^-_{0}+Q^-_{1} +\delta Q^- ,
    \\
    Q^-_{0}
    &\sim
   \perp,
    \\
    Q^-_{1}
    &\sim
    \frac{\perp^2}{xE},
    \\
    \frac{\delta Q^-}{Q^-_{1}}
    &\sim
    \frac{\perp}{xE}
    \ll 1 .
\end{aligned}
\label{eq:polecounting}
\end{equation}
Equivalently,
\begin{equation}
    \delta Q^-
    \sim
    \frac{\perp^3}{(xE)^2}.
\label{deltaQ}
\end{equation}
In terms of the corresponding dimensionless pole expansion, this means
that terms through
 $  \mathcal{O}\!\left(
        \frac{\perp^2}{(xE)^2}
    \right)$
are retained.

Although $\delta Q^-$ is formally of higher order in the pole expansion, it can contribute to the amplitude at first-subeikonal accuracy. When the leading $\mathcal{O}(\perp)$ terms cancel in a difference of longitudinal propagator poles, the remaining separation is of order $Q^-_1\sim\perp^2/(xE)$, making the relative correction from $\delta Q^-$ of first-subeikonal order, as shown in Eq.~\eqref{eq:polecounting}. By contrast, when the leading pole separation remains of order $\mathcal{O}(\perp)$, the same correction is beyond the working accuracy. Thus, the first-subeikonal relevance of subleading pole corrections must be assessed separately for residue denominators and formation phases.

Writing a generic formation phase as
\begin{equation}
    \phi=\phi_0+\delta\phi ,
    \label{eq:phidecomposition}
\end{equation}
with $\phi_0$ defined by the counting in
Eq.~\eqref{eq:phi0counting}, the subleading contribution satisfies
\begin{equation}
\delta\phi
=
\delta Q^-\,z
\sim
\frac{\perp}{xE}\,\phi_0
\sim
\mathcal{O}\!\left(\frac{\perp}{xE}\right)
\ll 1 .
\label{eq:deltaphicounting}
\end{equation}
Here, following Eq.~\eqref{deltaQ}, $\delta Q^-$ denotes the net subleading correction to the longitudinal pole combination determining the formation phase, after any cancellations between individual pole contributions.

The leading phase $\phi_0$ is therefore left arbitrary and is not
expanded. Only the parametrically small correction $\delta\phi$ is
expanded, giving
\begin{equation}
\begin{aligned}
\cos(\phi_0 \pm\delta\phi)
={}&
\cos\phi_0
\mp
\delta\phi\,\sin\phi_0
\\
&+
\mathcal{O}\!\left(
    \delta\phi^2
\right).
\end{aligned}
\label{eq:cosphaseexpansion}
\end{equation}

For $\phi_0=\mathcal{O}(1)$, the surviving subleading pole shift produces $\delta\phi=\delta Q^-z=\mathcal{O}(\perp/(xE))$. Its contribution to the formation-phase factors therefore enters at the same first-subeikonal order as the retained residue-denominator corrections and must be included. Ref.~\cite{KML} adopts the same LPM power counting  but evaluates the formation phases at leading-pole accuracy, omitting these contributions. Here, the surviving phase shifts are retained according to Eqs.~\eqref{eq:deltaphicounting} and~\eqref{eq:cosphaseexpansion}.

%-----------------------------------------------------------------------
\item \textit{Medium geometry and averaging over scattering centers.}

Only the longitudinal dependence of the medium density is resolved;
transverse variations of the density are neglected. We write the
distribution of scattering centers as
\begin{equation}
    \rho_0(\mathbf{x},z)
    =
    \frac{N}{A_\perp}\,
    \rho(z),
    \qquad
    \int dz\,\rho(z)=1 ,
    \label{eq:density}
\end{equation}
following the standard opacity-expansion ensemble of
Ref.~\cite{DGLV}. The transverse area $A_\perp$ is taken to be large
compared with the characteristic interaction area
$1/\mu^2$~\cite{DGLV}. Transverse gradients of the medium properties
are therefore omitted~\cite{KML}, while the flow dependence is treated
according to Eqs.~\eqref{flowu} and~\eqref{eq:flowexpansion}.

With Eq.~\eqref{eq:density}, the discrete sum over scattering centers
is replaced by
\begin{equation}
\begin{aligned}
    \sum_i
    f(\mathbf{x}_i,z_i)
    \longrightarrow
    \int d^2\mathbf{x}\,dz\,
    \rho_0(\mathbf{x},z)\,
    f(\mathbf{x},z).
\end{aligned}
\label{eq:continuum}
\end{equation}
The transverse position average consequently takes the form 
\begin{equation}
    \mean{\ldots}
    =
    \int
    \frac{d^2\mathbf{x}}{A_\perp}\,
    \ldots ,
    \label{app8}
\end{equation}
and is used below in particular as
\begin{equation}
\mean{
e^{-i(\mathbf{q}_i\pm\mathbf{q}_j)\cdot\mathbf{x}}
}
=
\frac{(2\pi)^2}{A_\perp}\,
\delta^{(2)}
(\mathbf{q}_i\pm\mathbf{q}_j).
\label{app9}
\end{equation}
The upper sign applies to two-scattering contact contributions, whereas the lower sign occurs for the single-scattering terms
after multiplication by the complex-conjugate amplitude.

We summarize here the color-algebra relations used throughout the
calculation. For target (scattering center) partons in a $d_T$-dimensional representation $T$ of
$SU(N_c)$, with quadratic Casimir $C_2(T)$, we use
\begin{equation}
    \Tr\!\left(T_i^aT_j^b\right)
    =
    \delta_{ij}\delta^{ab}
    \frac{C_2(T)d_T}{d_A},
    \label{tr2}
\end{equation}
where $d_A=N_c^2-1$ is the dimension of the adjoint representation
($d_A=8$ for $N_c=3$)~\cite{bsga}. The generators obey
\begin{equation}
    [T^a,T^b]=if^{abc}T^c,
    \label{gen1}
\end{equation}
with $f^{abc}$ denoting the totally antisymmetric structure constants of $SU(N_c)$, a property used frequently in our calculations. We also
use the Hermiticity of the generators, $(T^a)^\dagger=T^a$, together
with the cyclic invariance of the trace.

For the heavy quark in the fundamental representation, we use
\begin{equation}
\Tr\!\left(
T_{jet}^{a}T_{jet}^{b}
\right)
=
\delta^{ab}\,
\frac{C_2(F)d_F}{d_A},
\label{trjet}
\end{equation}

where $d_F=N_c$ and
$C_2(F)=(N_c^2-1)/(2N_c)$.
%-----------------------------------------------------------------------
\item \textit{Hard-production point and separation from the scattering centers.}

The production vertex is placed at the coordinate origin,
\begin{equation}
    x_0^\mu=(0,\mathbf{0},0),
\end{equation}
following Refs.~\cite{SSVflow,KML}. 

The scattering centers contributing to the calculation are taken to
satisfy
\begin{equation}
    z_i \gg \frac{1}{\mu} ,
    \label{eq:separation}
\end{equation}
i.e., we assume that the separation between the source $J$ and the scattering centers is large compared to the characteristic interaction length~\cite{GLV2, DGLV, bsga, KML}.

For contributions containing the associated longitudinal phase, the
scattering-potential pole terms are exponentially damped in this limit.
For some two-scattering contributions, however, the required
$z$-dependent phase factor is absent, and the corresponding potential
poles are therefore retained.

%-----------------------------------------------------------------------
\item \textit{Broad-source approximation.}

In the broad-source approximation, the source current is assumed to vary slowly with momentum, so that
\begin{equation}
    J(E,\mathbf{p}-\mathbf{q}_\ell)
    \simeq
    J(E,\mathbf{p}),
    \label{eq:broadsource}
\end{equation}
as in  Refs.~\cite{KML,GLV2,DGLV, bsga}.

\end{itemize}

\section{\label{sec:FeynmanRules}Propagators and gluonic vertices}

The assumptions and power counting specified in
Sec.~\ref{sec:Assumptions} determine which terms have to be retained in
the diagrammatic calculation. Since the calculation is performed in
finite-temperature pQCD, with a finite heavy-quark mass and a finite
asymptotic thermal gluon mass, we summarize below the propagators and
gluonic vertices employed throughout the calculation. We work in the
axial gauge specified in Eq.~\eqref{eq:axialgauge}.

At first-subeikonal accuracy, the two-scattering amplitude $\mathcal{M}_2$ also receives a contribution involving the four-gluon vertex, which is absent at leading eikonal order in the standard (D)GLV treatment~\cite{GLV2,DGLV,KML}.
\begin{itemize}
    \item heavy-quark propagator with mass $M$:
\begin{equation}
\vcenter{\hbox{%
\begin{tikzpicture}[
    line cap=round,
    line join=round,
    >=latex
]

\coordinate (A) at (-1.10,0);
\coordinate (B) at ( 1.10,0);

\draw[line width=0.85pt]
    (A) -- (B);

\draw[line width=0.75pt,->]
    (-0.55,0.36) -- (0.55,0.36)
    node[midway,above=1pt] {$p$};

\end{tikzpicture}%
}}
\quad = \quad
\frac{i}{p^2-M^2+i\epsilon}.
\label{eq:massive_propagator}
\end{equation}

    \item gluon propagator with asymptotic thermal mass $m_g=\mu/\sqrt{2}$~\cite{DG03}:
\begin{equation}
\vcenter{\hbox{%
\begin{tikzpicture}[
    line cap=round,
    line join=round,
    >=latex
]

\coordinate (A) at (-1.10,0);
\coordinate (B) at ( 1.10,0);

\draw[
    line width=0.85pt,
    decorate,
    decoration={
        coil,
        amplitude=2.1pt,
        segment length=4.8pt
    }
] (A) -- (B);

\node[above left=2pt]  at (A) {$a,\mu$};
\node[above right=2pt] at (B) {$b,\nu$};

\draw[line width=0.75pt,->]
    (-0.55,0.36) -- (0.55,0.36)
    node[midway,above=1pt] {$p$};

\end{tikzpicture}%
}}
\quad = \quad
\frac{-i\,\delta_{ab}P_{\mu\nu}}
     {p^2-m_g^2+i\epsilon},
\label{eq:gluon_propagator}
\end{equation}
where $P_{\mu\nu}$ denotes the transverse gluon projector
of Ref.~\cite{DG03}. In the axial gauge specified in
Eq.~\eqref{eq:axialgauge}, it is given by
\begin{equation}
\label{Proj}
 P_{\mu \nu}= g_{\mu \nu} - \frac{ p_{\mu} p_{\nu}  -n_{\mu} n_{\nu} p^2 +n_{\mu}p_{\nu} (np) +n_{\nu} p_{\mu}(np)}{ p^2+(np)^2} .
\end{equation}

    \item three-gluon vertex:
\begin{widetext}

\noindent
\begin{minipage}[c]{0.24\textwidth}
\centering

\begin{tikzpicture}[
    line cap=round,
    line join=round,
    >=latex
]

\coordinate (V) at (0,0);

\coordinate (A) at (-1.55, 0.00);
\coordinate (B) at ( 1.15, 1.10);
\coordinate (C) at ( 1.15,-1.10);

\draw[
    line width=0.85pt,
    decorate,
    decoration={
        coil,
        amplitude=2.1pt,
        segment length=4.8pt
    }
] (A) -- (V);

\draw[
    line width=0.85pt,
    decorate,
    decoration={
        coil,
        amplitude=2.1pt,
        segment length=4.8pt
    }
] (B) -- (V);

\draw[
    line width=0.85pt,
    decorate,
    decoration={
        coil,
        amplitude=2.1pt,
        segment length=4.8pt
    }
] (C) -- (V);

\fill[black] (V) circle (1.9pt);

\node[left=4pt]        at (A) {$a,\mu$};
\node[above right=2pt] at (B) {$b,\nu$};
\node[below right=2pt] at (C) {$c,\rho$};

\draw[line width=0.75pt,->]
    (-1.25,0.22) -- (-0.60,0.22)
    node[midway,above=1pt] {$p_1$};

\draw[line width=0.75pt,->]
    (0.78,1.07) -- (0.33,0.64)
    node[midway,above=1pt,sloped] {$p_2$};

\draw[line width=0.75pt,->]
    (0.78,-1.07) -- (0.33,-0.64)
    node[midway,below=1pt,sloped] {$p_3$};

\end{tikzpicture}

\end{minipage}
\hfill
\begin{minipage}[c]{0.73\textwidth}

\begin{equation}
\Gamma_{\mu\nu\rho}^{abc}(p_1,p_2,p_3)
=
-g_s f^{abc}
\left[
g_{\mu\rho}(p_1-p_3)_{\nu}
+
g_{\mu\nu}(p_2-p_1)_{\rho}
+
g_{\nu\rho}(p_3-p_2)_{\mu}
\right].
\label{eq:3gluon_vertex}
\end{equation}

\end{minipage}

\end{widetext}
%%%%%%%%%%%%%%%%%%%%%%%%%
 \item four-gluon vertex:

\begin{widetext}

\noindent
\begin{minipage}[c]{0.24\textwidth}
\centering

\begin{tikzpicture}[
    line cap=round,
    line join=round
]

\coordinate (V) at (0,0);

\coordinate (A) at (-1.35,-1.05); % a,mu
\coordinate (B) at ( 1.35,-1.05); % b,nu
\coordinate (C) at ( 1.35, 1.05); % c,rho
\coordinate (D) at (-1.35, 1.05); % d,sigma

\draw[
    line width=0.85pt,
    decorate,
    decoration={
        coil,
        amplitude=2.1pt,
        segment length=4.8pt
    }
] (A) -- (V);

\draw[
    line width=0.85pt,
    decorate,
    decoration={
        coil,
        amplitude=2.1pt,
        segment length=4.8pt
    }
] (B) -- (V);

\draw[
    line width=0.85pt,
    decorate,
    decoration={
        coil,
        amplitude=2.1pt,
        segment length=4.8pt
    }
] (C) -- (V);

\draw[
    line width=0.85pt,
    decorate,
    decoration={
        coil,
        amplitude=2.1pt,
        segment length=4.8pt
    }
] (D) -- (V);

\fill[black] (V) circle (1.9pt);

\node[below left=2pt]  at (A) {$a,\mu$};
\node[below right=2pt] at (B) {$b,\nu$};
\node[above right=2pt] at (C) {$c,\rho$};
\node[above left=2pt]  at (D) {$d,\sigma$};

\end{tikzpicture}

\end{minipage}
\hfill
\begin{minipage}[c]{0.73\textwidth}

\begin{equation}
\begin{aligned}
V_{\mu\nu\rho\sigma}^{abcd}
=
-i g_s^2
\Big[
& f^{eab}f^{ecd}
\left(
g_{\mu\rho}g_{\nu\sigma}
-
g_{\mu\sigma}g_{\nu\rho}
\right)
\\[2pt]
&+
f^{ecb}f^{ead}
\left(
g_{\mu\rho}g_{\nu\sigma}
-
g_{\mu\nu}g_{\rho\sigma}
\right)
\\[2pt]
&+
f^{eac}f^{ebd}
\left(
g_{\mu\nu}g_{\rho\sigma}
-
g_{\mu\sigma}g_{\nu\rho}
\right)
\Big].
\end{aligned}
\label{eq:4gluon_vertex}
\end{equation}

\end{minipage}

\end{widetext}
\end{itemize}

\section{\label{sec:M0} Zeroth-order radiation amplitude}

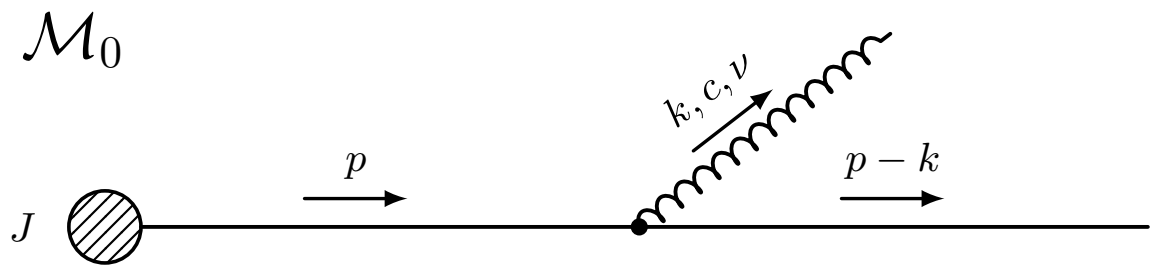
\begin{figure}[t]
\centering

\resizebox{0.43\textwidth}{!}{%
\begin{tikzpicture}[
    line cap=round,
    line join=round,
    >=latex
]

\path[use as bounding box] (-0.8,-0.55) rectangle (8.25,1.65);

\node at (-0.25,1.45) {\Large $\mathcal{M}_0$};

\draw[
    pattern=north east lines,
    pattern color=black,
    line width=0.8pt
] (0,0) circle (0.28);

\node[left=4pt] at (-0.28,0) {$J$};

\coordinate (V) at (4.15,0.00);

\coordinate (QinStart) at (0.28,0.00);
\coordinate (Qout)     at (8.10,0.00);

\coordinate (Gout) at (6.10,1.50);

\draw[line width=0.85pt]
    (QinStart) -- (V);

\draw[line width=0.85pt]
    (V) -- (Qout);

\draw[
    line width=0.85pt,
    decorate,
    decoration={
        coil,
        amplitude=2.3pt,
        segment length=5.2pt
    }
] (V) -- (Gout);

\fill[black] (V) circle (1.9pt);

\draw[line width=0.75pt,->]
    (1.55,0.22) -- (2.35,0.22);

\node[above=1pt] at (1.95,0.22)
    {$p$};

\draw[line width=0.75pt,->]
    (5.72,0.22) -- (6.52,0.22);

\node[above=1pt] at (6.12,0.22)
    {$p-k$};

\draw[line width=0.75pt,->]
    (4.57,0.59) -- (5.19,1.07)
    node[midway, above=1pt, sloped]
    {$k,c,\nu$};

\end{tikzpicture}%
}

\caption{\small Zeroth-order gluon-emission diagram in the absence of interactions
with the QCD medium.
}
\label{fig:M0}

\end{figure}

We first consider the zeroth-order contribution, corresponding to gluon
emission in the absence of interactions with the medium. As shown in
Fig.~\ref{fig:M0}, the hard source produces an off-shell heavy quark with
momentum $p$, which emits a gluon with momentum $k$, polarization
$\varepsilon^\mu(k)$, and color $c$, and emerges as an on-shell heavy
quark with momentum $p-k$.

%%%%%M0

Using the momentum and polarization conventions and definitions introduced in
Sec.~\ref{sec:Assumptions} and Sec.~\ref{sec:FeynmanRules}, the corresponding amplitude is
\begin{align}
\mathcal{M}_0
={}&
\left[
i g_s (2p-k)^\nu
\varepsilon_\nu^{*}(k)
T_{jet}^{c}
\right]
\frac{i}{p^2-M^2+i\epsilon}\,
iJ(p) \, .
\label{M0}
\end{align}
To obtain its reduced form, we employ
 Eq.~\eqref{vektori}.
Retaining terms through
$\mathcal{O}\!\left(\perp^2/(xE)^2\right)$ and applying the
soft-gluon approximation, we obtain
\begin{align}
p^2-M^2
&\simeq
\frac{\mathbf{k}^2+\chi}{x},
\label{pM}
\\
\varepsilon(k)\cdot p
&\simeq
\frac{\boldsymbol{\varepsilon}\cdot\mathbf{k}}{x}.
\label{M0pol}
\end{align}
Equation~\eqref{M0} therefore reduces to
\begin{equation}
\mathcal{M}_0
\simeq
-2i g_s J(p)\,T_{jet}^{c}\,
\frac{
\boldsymbol{\varepsilon}\cdot\mathbf{k}
}{
\mathbf{k}^2+\chi
}.
\label{M0final}
\end{equation}
%%%%%%%%%%%%%%%% 

Squaring Eq.~\eqref{M0final}, summing over the two physical
polarizations and over the color $c$ of the emitted gluon, and using
Eq.~\eqref{trjet}, we obtain
\begin{align}
\Tr
\left\langle
|\mathcal{M}_0|^2
\right\rangle
={}&
4g_s^2
|J(p)|^2
C_2(F)d_F\,
\frac{\mathbf{k}^2}
{\left(\mathbf{k}^2+\chi\right)^2}.
\label{M0sq}
\end{align}
The corresponding zeroth-order contribution is written as~\cite{DGLV}
\begin{equation}
d^3N_g^{(0)}\,d^3N_J
\simeq
\Tr
\left\langle
|\mathcal{M}_0|^2
\right\rangle
\frac{d^3(\vec{p}-\vec{k})}
     {(2\pi)^3\,2(E-xE)}
\frac{d^3\vec{k}}
     {(2\pi)^3\,2xE}.
\label{M0N}
\end{equation}

Although the final heavy quark carries momentum $p-k$, the hard-source
factor can be separated in terms of the parent momentum $p$. Indeed,
the broad-source approximation in Eq.~\eqref{eq:broadsource} reduces
the source current to $J(E,\mathbf{p})$, while
the change of variables
$\vec{p}-\vec{k}\to\vec{p}$ at fixed $k$ has unit Jacobian. We
therefore define the parent-quark distribution as
\begin{equation}
d^3N_J
=
d_F
\left|J(E,\mathbf{p})\right|^2
\frac{d^3\vec{p}}
{(2\pi)^3\,2E} \, .
\label{NJ}
\end{equation}

Using momenta from Eq.~\eqref{vektori}, at fixed
$\mathbf{k}$ we have
$dk_z=E\left[1+(\mathbf{k}^2+m_g^2)/(2x^2E^2)\right]dx$.
The final-state phase space therefore becomes
\begin{align}
&\frac{d^3(\vec{p}-\vec{k})}
{(2\pi)^3\,2(1-x)E}
\frac{d^3\vec{k}}
{(2\pi)^3\,2xE}
\nonumber \\
&\qquad=
\frac{d^3\vec{p}}
{(2\pi)^3\,2E}
\frac{dx\,d^2\mathbf{k}}
{(2\pi)^3\,2x}\,
\left[
1+
\mathcal{O}\left(
\frac{\perp^2}{(xE)^2}
\right)
\right] \nonumber \\
&\qquad\simeq
\frac{d^3\vec{p}}
{(2\pi)^3\,2E}
\frac{dx\,d^2\mathbf{k}}
     {(2\pi)^3\,2x} \, ,
\label{M0phase}
\end{align}
where $1/(1-x)\simeq1$ has been used in the soft-gluon limit (consistently
with Sec.~\ref{sec:Assumptions}), while
the Jacobian correction starts beyond the first-subeikonal accuracy.

Finally, defining $\alpha_s\equiv g_s^2/(4\pi)$, the remaining factors
in Eq.~\eqref{M0N} give the zeroth-order radiation spectrum
\begin{equation}
x\,
\frac{d^3 N_g^{(0)}}
     {dx\,d^2\mathbf{k}}
=
\frac{\alpha_s C_2(F)}{\pi^2}\,
\frac{\mathbf{k}^2}
     {\left(\mathbf{k}^2+\chi\right)^2},
\label{M0spectrum}
\end{equation}
which reproduces  Eq.~(9) from Ref.~\cite{DGLV}.

\section{\label{sec:M1}Single-scattering contributions $\mathcal{M}_{1,0,1}$, $\mathcal{M}_{1,0,0}$, and $\mathcal{M}_{1,1,0}$}

\begin{figure*}[t]
\centering
% ============================================================
% PANEL (a): M_{1,0,1}
% ============================================================
\begin{minipage}[t]{0.32\textwidth}
\centering
\resizebox{\linewidth}{!}{%
\begin{tikzpicture}[
    line cap=round,
    line join=round,
    >=latex
]

\path[use as bounding box] (-0.9,-2.35) rectangle (6.9,1.8);

\node at (-0.55,1.55) {\Large (a)};
\node at (0.75,1.45) {\Large $\mathcal{M}_{1,0,1}$};

\draw[
    pattern=north east lines,
    pattern color=black,
    line width=0.8pt
] (0,0) circle (0.28);
\node[left=4pt] at (-0.28,0) {$J$};

\coordinate (V1) at (2.20,0.00);
\coordinate (V2) at (3.60,-0.95);
\coordinate (QinStart) at (0.28,0.00);
\coordinate (Qout) at (6.55,0.00);
\coordinate (Gout) at (5.70,-0.95);
\coordinate (Qmed) at (3.60,-2.00);

\draw[line width=0.85pt] (QinStart) -- (V1);
\draw[line width=0.85pt] (V1) -- (Qout);

\draw[
    line width=0.85pt,
    decorate,
    decoration={coil, amplitude=2.3pt, segment length=5.2pt}
] (V1) -- (V2);

\draw[
    line width=0.85pt,
    decorate,
    decoration={coil, amplitude=2.3pt, segment length=5.2pt}
] (V2) -- (Gout);

\draw[
    line width=0.85pt,
    decorate,
    decoration={coil, amplitude=2.1pt, segment length=4.8pt}
] (V2) -- (Qmed);

\fill[black] (V1) circle (1.9pt);
\fill[black] (V2) circle (1.9pt);

\node[above=2pt] at (V1) {$b,\rho$};

\node[xshift=9pt,yshift=10pt] at (V2)
    {$b^{\prime},\rho^{\prime}$};

\draw[line width=0.8pt] (Qmed) circle (0.18);
\draw[line width=0.8pt]
    ($(Qmed)+(-0.12,-0.12)$) -- ($(Qmed)+(0.12,0.12)$);
\draw[line width=0.8pt]
    ($(Qmed)+(-0.12,0.12)$) -- ($(Qmed)+(0.12,-0.12)$);

\node[below=5pt] at (Qmed) {$a_1, i, \mu$};

\draw[line width=0.75pt,->]
    (0.94,0.22) -- (1.54,0.22);
\node[above=1pt] at (1.24,0.22) {$p-q$};

\draw[line width=0.75pt,->]
    (4.025,0.22) -- (4.725,0.22);
\node[above=1pt] at (4.375,0.22) {$p-k$};

\draw[line width=0.75pt,->]
    (2.60,-0.461) -- (3.20,-0.869)
    node[midway, below=2pt, sloped] {$k-q$};

\draw[line width=0.75pt,->]
    (4.275,-1.18) -- (5.025,-1.18);
\node[below=1pt] at (4.65,-1.18) {$k,c,\nu$};

\draw[line width=0.75pt,->]
    (3.27,-1.785) -- (3.27,-1.165);
\node[left=2pt] at (3.27,-1.475) {$q$};

\end{tikzpicture}%
}
\end{minipage}\hfill%
% ============================================================
% PANEL (b): M_{1,1,0}
% ============================================================
\begin{minipage}[t]{0.32\textwidth}
\centering
\resizebox{\linewidth}{!}{%
\begin{tikzpicture}[
    line cap=round,
    line join=round,
    >=latex
]

\path[use as bounding box] (-0.9,-2.35) rectangle (6.9,1.8);

\node at (-0.55,1.55) {\Large{(b)}};
\node at (0.75,1.45) {\Large $\mathcal{M}_{1,1,0}$};

\draw[
    pattern=north east lines,
    pattern color=black,
    line width=0.8pt
] (0,0) circle (0.28);
\node[left=4pt] at (-0.28,0) {$J$};

\coordinate (V1) at (2.35,0.00);
\coordinate (V2) at (4.30,0.00);
\coordinate (QinStart) at (0.28,0.00);
\coordinate (Qout) at (6.55,0.00);
\coordinate (Qmed) at (2.35,-2.00);
\coordinate (Gout) at (5.15,1.35);

\draw[line width=0.85pt] (QinStart) -- (V1);
\draw[line width=0.85pt] (V1) -- (V2);
\draw[line width=0.85pt] (V2) -- (Qout);

\draw[
    line width=0.85pt,
    decorate,
    decoration={coil, amplitude=2.1pt, segment length=4.8pt}
] (V1) -- (Qmed);

\draw[
    line width=0.85pt,
    decorate,
    decoration={coil, amplitude=2.3pt, segment length=5.2pt}
] (V2) -- (Gout);

\fill[black] (V1) circle (1.9pt);
\fill[black] (V2) circle (1.9pt);

\draw[line width=0.8pt] (Qmed) circle (0.18);
\draw[line width=0.8pt]
    ($(Qmed)+(-0.12,-0.12)$) -- ($(Qmed)+(0.12,0.12)$);
\draw[line width=0.8pt]
    ($(Qmed)+(-0.12,0.12)$) -- ($(Qmed)+(0.12,-0.12)$);

\node[below=5pt] at (Qmed) {$a_1, i, \mu$};

\draw[line width=0.75pt,->]
    (1.015,0.22) -- (1.615,0.22);
\node[above=1pt] at (1.315,0.22) {$p-q$};

\draw[line width=0.75pt,->]
    (2.975,0.22) -- (3.675,0.22);
\node[above=1pt] at (3.325,0.22) {$p$};

\draw[line width=0.75pt,->]
    (5.075,0.22) -- (5.775,0.22);
\node[above=1pt] at (5.425,0.22) {$p-k$};

\draw[line width=0.75pt,->]
    (2.02,-1.31) -- (2.02,-0.69);
\node[left=2pt] at (2.02,-1.00) {$q$};

\draw[line width=0.75pt,->]
    (4.47,0.685) -- (4.776,1.171)
    node[midway, above=1pt, sloped] {$k,c,\nu$};

\end{tikzpicture}%
}
\end{minipage}\hfill%
% ============================================================
% PANEL (c): M_{1,0,0}
% ============================================================
\begin{minipage}[t]{0.32\textwidth}
\centering
\resizebox{\linewidth}{!}{%
\begin{tikzpicture}[
    line cap=round,
    line join=round,
    >=latex
]

\path[use as bounding box] (-0.9,-2.35) rectangle (6.9,1.8);

\node at (-0.55,1.55) {\Large (c)};
\node at (0.75,1.45) {\Large $\mathcal{M}_{1,0,0}$};

\draw[
    pattern=north east lines,
    pattern color=black,
    line width=0.8pt
] (0,0) circle (0.28);
\node[left=4pt] at (-0.28,0) {$J$};

\coordinate (V1) at (2.20,0.00);
\coordinate (V2) at (4.30,0.00);
\coordinate (QinStart) at (0.28,0.00);
\coordinate (Qout) at (6.55,0.00);
\coordinate (Gout) at (3.00,1.35);
\coordinate (Qmed) at (4.30,-2.00);

\draw[line width=0.85pt] (QinStart) -- (V1);
\draw[line width=0.85pt] (V1) -- (V2);
\draw[line width=0.85pt] (V2) -- (Qout);

\draw[
    line width=0.85pt,
    decorate,
    decoration={coil, amplitude=2.3pt, segment length=5.2pt}
] (V1) -- (Gout);

\draw[
    line width=0.85pt,
    decorate,
    decoration={coil, amplitude=2.1pt, segment length=4.8pt}
] (V2) -- (Qmed);

\fill[black] (V1) circle (1.9pt);
\fill[black] (V2) circle (1.9pt);

\draw[line width=0.8pt] (Qmed) circle (0.18);
\draw[line width=0.8pt]
    ($(Qmed)+(-0.12,-0.12)$) -- ($(Qmed)+(0.12,0.12)$);
\draw[line width=0.8pt]
    ($(Qmed)+(-0.12,0.12)$) -- ($(Qmed)+(0.12,-0.12)$);

\node[below=5pt] at (Qmed) {$a_1, i, \mu$};

\draw[line width=0.75pt,->]
    (0.94,0.22) -- (1.54,0.22);
\node[above=1pt] at (1.24,0.22) {$p-q$};

\draw[line width=0.75pt,->] (2.95,0.22) -- (3.75,0.22);
\node[above=1pt] at (3.35,0.22) {$p-k-q$};

\draw[line width=0.75pt,->]
    (5.075,0.22) -- (5.775,0.22);
\node[above=1pt] at (5.425,0.22) {$p-k$};

\draw[line width=0.75pt,->]
    (2.15,0.56) -- (2.55,1.235)
    node[midway, above=1pt, sloped] {$k,c,\nu$};

\draw[line width=0.75pt,->]
    (3.97,-1.31) -- (3.97,-0.69);
\node[left=2pt] at (3.97,-1.00) {$q$};

\end{tikzpicture}%
}
\end{minipage}
\caption{
Single-scattering diagrams contributing at first order in opacity  $L/{\lambda}$:
(a) interaction of the radiated gluon with the medium after emission,
(b) interaction of the heavy quark with the medium before gluon emission, and
(c) interaction of the outgoing heavy quark with the medium after emission.
In all three diagrams, the final heavy quark and emitted gluon carry momenta $p-k$ and $k$, respectively.
}
\label{fig:M1}

\end{figure*}
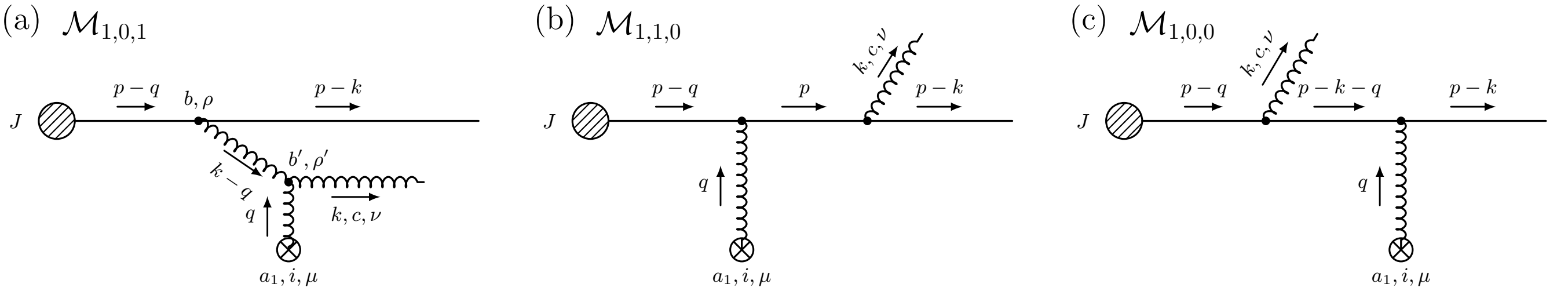

In this section, we evaluate the three single-scattering amplitudes $\mathcal{M}_{1,0,1}$, $\mathcal{M}_{1,1,0}$ and $\mathcal{M}_{1,0,0}$ , corresponding to gluon emission from a heavy quark undergoing one interaction with the medium. The associated diagrams are shown in Fig.~\ref{fig:M1}. These amplitudes enter the first-order-in-opacity contribution through the single-scattering term in Eq.~\eqref{eq:firstopacity}.

Using Eqs.~\eqref{a} and~\eqref{flowu}, the $q^0$ integration fixes
$q^0=\mathbf{u}\cdot\mathbf{q}$. Accordingly, throughout this section we use
\begin{align}~\label{kin4M1}
q=(\boldsymbol{u}\cdot \boldsymbol{q}, \, \boldsymbol{q}, \, q_z).
\end{align}

For later convenience, we also introduce
\begin{align}~\label{vt}   
v_i(\mathbf{q})=\frac{-g^2_s}{\mathbf{q}^2+\mu_i^2} \, .
\end{align}
%%%%%%%%%%%%%%%%%%%%%%%%%%%%%%%%%%%%%%%%%%%%%

\subsection{\label{sec:M101}Single-scattering contribution $\mathcal{M}_{1,0,1}$}

We first consider the amplitude $\mathcal{M}_{1,0,1}$ shown in Fig.~\ref{fig:M1}(a), in which the emitted gluon undergoes an interaction with the medium, receiving momentum transfer $q$, before emerging with momentum $k$. Using the propagators and three-gluon vertex specified in Sec.~\ref{sec:FeynmanRules}, together with the momentum, Lorentz-index, color, and scattering-center assignments of Fig.~\ref{fig:M1} (a), we obtain
\begin{widetext}
\begin{align}
\mathcal{M}_{1,0,1}
{}&=
\int\frac{d^4q}{(2\pi)^4}\,
\varepsilon^{* \nu}(k)\,
\Gamma^{b' c a_1}_{\rho' \nu \mu} (k-q, -k,q)
A_{ext}^{\mu,a_1}(q)
\frac{(-i)P^{\rho\rho'}(k-q)\delta_{bb'}}
{(k-q)^2-m_g^2+i\epsilon}\,
\left[
i g_s(2p-k-q)_{\rho}T_{jet}^{b}
\right]
\nonumber\\
&\times
\frac{i}{(p-q)^2-M^2+i\epsilon}\,
iJ(p-q)
\nonumber\\
{}&=
-ig_s
\sum_i
[c, a_1]\,T_i^{a_1}
\int\frac{d^2\mathbf{q}}{(2\pi)^2}\,
e^{-i\mathbf{q}\cdot\mathbf{x}_i}
\int\frac{dq_z}{2\pi}\,
e^{-iq_z z_i}\,
v_i(q)\,
J(p-q)\,
\frac{I_{1,0,1}}
{\left[(p-q)^2-M^2+i\epsilon\right]
 \left[(k-q)^2-m_g^2+i\epsilon\right]}\,, 
\label{m101p}
\end{align}
\end{widetext}
where $I_{1,0,1}$ denotes the Lorentz-contracted structure obtained from the polarization vector, the three-gluon vertex, the gluon projector, the quark-gluon vertex, and the flow four-vector $u^\mu$ entering through the external field in Eqs.~\eqref{Aext} and~\eqref{a}.

In Eq.~\eqref{m101p}, and henceforth, we use $a$ as a shorthand notation for the color generator $T^a_{jet}$ in the fundamental representation of $SU(3)$.
The second form of Eq.~\eqref{m101p} follows by inserting the external field from Eqs.~\eqref{Aext} and~\eqref{a}, performing the $q^0$ integration, which fixes $q^0=\mathbf{u}\cdot\mathbf{q}$, as displayed in Eq.~\eqref{kin4M1}. Together with Eq.~\eqref{flowu}, this gives
\begin{align}
u\cdot q
={}&
0\,.
\label{qu}
\end{align}
The color-commutator relation in Eq.~\eqref{gen1} is also used. The hard production point is taken to be $x_0=0$, as specified in Sec.~\ref{sec:Assumptions}. 

In obtaining $I_{1,0,1}$
\begin{widetext}
\begin{align}
I_{1,0,1}
={}&
-4E
\left[
\boldsymbol{\varepsilon}\cdot(\mathbf{k}-\mathbf{q})
-
\boldsymbol{\varepsilon}\cdot\mathbf{k}\,
\frac{\mathbf{u}\cdot(\mathbf{k}-\mathbf{q})}{xE}
-
\mathbf{q}\cdot(\mathbf{q}-\mathbf{k})\,
\frac{\boldsymbol{\varepsilon}\cdot\mathbf{u}}{xE}
\right]  \, ,
\label{I101}
\end{align}
\end{widetext}
all scalar products are evaluated directly
from the momentum, polarization, and flow four-vectors specified in
Eq.~\eqref{vektori}, and
\eqref{flowu}, \eqref{kin4M1} together with Eq.~\eqref{qu}. 

The three-gluon vertex $\Gamma_{\rho'\nu\mu}^{b'ca_1}(k-q,-k,q)$, defined in Eq.~\eqref{eq:3gluon_vertex}, is contracted with the polarization vector, the external field, and the gluon projector, all transverse to the axial-gauge vector $n^\mu=(0,\mathbf{0},1)$ defined in Eq.~\eqref{eq:axialgauge}. Specifically, $\varepsilon(k)\cdot n=0$ by Eq.~\eqref{kin0M1}, $n_\rho P^{\rho\rho'}(k-q)=0$, and $n\cdot A_{ext}=0$ by the axial-gauge condition in Eq.~\eqref{eq:axialgauge}  (consistent with $n\cdot u=0$, as follows from Eqs.~\eqref{flowu},  and~\eqref{Aext},~\eqref{a},~\eqref{eq:axialgauge}). Consequently, all explicit $q_z$ dependence cancels in the Lorentz-contracted structure $I_{1,0,1}$ in Eq.~\eqref{I101}, so that no additional longitudinal poles arise.

The remaining longitudinal momentum-transfer integral is
\begin{widetext}
\begin{align}
\mathcal{I}_{1,0,1}
={}&
\int\frac{dq_z}{2\pi}\,
e^{-iq_z z_i}
\frac{
v_i(q)\,J(p-q)
}{
\left[(p-q)^2-M^2+i\epsilon\right]
\left[(k-q)^2-m_g^2+i\epsilon\right]
}\,.
\label{IM101p}
\end{align}
\end{widetext}
Since $z_i>0$, the contour is closed in the lower half-plane; its clockwise orientation supplies the corresponding overall minus sign. We also employ the broad-source approximation of Eq.~\eqref{eq:broadsource},
$J(p-q)\simeq J(p)$.

There are three lower-half-plane singularities: one from the scattering potential, one from the heavy-quark propagator, and one from the gluon propagator. Using Eq.~\eqref{vt}, the scattering potential can be written as
\begin{align}
v_i(q)
={}&
\frac{-g_s^2}
{(q_z+iR)(q_z-iR)}\,,
\label{scattpot1}
\end{align}
where
\begin{equation}
R^2
=
\mathbf{q}^2+\mu^2
-
(\mathbf{u}\cdot\mathbf{q})^2
>0\,.
\label{R}
\end{equation}
Since $R\sim\mu\sim\perp$ according to the hierarchy in Eq.~\eqref{eq:hierarchy}, the separation assumption in Eq.~\eqref{eq:separation} implies
$Rz_i\sim\mathcal{O}(\perp z_i)\gg1$.
The residue of the scattering-potential pole $q_z=-iR$ is therefore exponentially suppressed and can be neglected.

We denote the relevant lower-half-plane propagator roots by a superscript $-$. The heavy-quark propagator gives
\begin{align}
q^-_{z,p-q}
={}&
\mathbf{u}\cdot\mathbf{q}
-
\frac{\mathbf{k}^2+\chi}{2xE}
+
\frac{\mathbf{u}\cdot\mathbf{q}}{xE}\,
\frac{x^2M^2}{2xE}
-i\epsilon\, .
\label{qzp_q}
\end{align}
The $\mathcal{O}\!\left(\perp^3/(xE)^2\right)$ correction is retained according to the pole-counting prescription of Sec.~\ref{sec:Assumptions}, Eqs.~\eqref{eq:polecounting} and~\eqref{deltaQ}. Its contribution to the formation phases is governed by Eq.~\eqref{eq:deltaphicounting}.

Within the same accuracy, the gluon propagator pole is
\begin{align}
q^-_{z,k-q}
={}&
\mathbf{u}\cdot\mathbf{q}
+
\frac{(\mathbf{k}-\mathbf{q})^2-\mathbf{k}^2}{2xE}
+
\frac{\mathbf{u}\cdot\mathbf{q}}{xE}\,
\frac{(\mathbf{k}-\mathbf{q})^2+m_g^2}{2xE}
-i\epsilon\,.
\label{qzk-q}
\end{align}
For later use, we note the identity
\begin{equation}
(\mathbf{k}-\mathbf{q})^2+m_g^2-x^2M^2
=
(\mathbf{k}-\mathbf{q})^2+\chi-2x^2M^2\,,
\label{2m2x2}
\end{equation}
which will be used repeatedly when factorizing differences of propagator poles entering the residue denominators.

The common $\mathcal{O}(\perp)$ terms in Eqs.~\eqref{qzp_q} and~\eqref{qzk-q} cancel, leaving a pole difference of order $\perp^2/(xE)$:
\begin{widetext}
\begin{align}
q^-_{z,k-q}
-
q^-_{z,p-q}
={}&
\frac{(\mathbf{k}-\mathbf{q})^2+\chi}{2xE}
\left[
1
+
\frac{\mathbf{u}\cdot\mathbf{q}}{xE}
-
\frac{\mathbf{u}\cdot\mathbf{q}}{xE}\,
\frac{2x^2M^2}
{(\mathbf{k}-\mathbf{q})^2+\chi}
\right].
\label{razlikapolova}
\end{align}
\end{widetext}
The higher-order terms in Eq.~\eqref{razlikapolova} therefore contribute at relative first-subeikonal order to the residue denominator and the corresponding formation phases. 

Substituting the Lorentz-contracted structure from Eq.~\eqref{I101} and the residues evaluated from Eq.~\eqref{IM101p} at the heavy-quark and gluon propagator poles given in Eqs.~\eqref{qzp_q} and~\eqref{qzk-q}, respectively, into Eq.~\eqref{m101p}, we obtain
\begin{widetext}
\begin{align}
\mathcal{M}_{1,0,1}
\approx{}&
2g_s
\sum_i
[c,a_1]\,
T_i^{a_1}
\int\frac{d^2\mathbf{q}}{(2\pi)^2}\,
e^{-i\mathbf{q}\cdot\mathbf{x}_i}\,
e^{-i\mathbf{u}\cdot\mathbf{q}\,z_i}\,
v_i(\mathbf{q})\,
J(E,\mathbf{p})\,
\frac{1}{(\mathbf{k}-\mathbf{q})^2+\chi}
\nonumber\\
&\times
\left[
\boldsymbol{\varepsilon}\cdot(\mathbf{k}-\mathbf{q})
-
\frac{
\boldsymbol{\varepsilon}\cdot\mathbf{k}\,
\mathbf{u}\cdot(\mathbf{k}-\mathbf{q})
}{xE}
-
\frac{
\boldsymbol{\varepsilon}\cdot\mathbf{u}\,
\mathbf{q}\cdot(\mathbf{q}-\mathbf{k})
}{xE}
\right]
\nonumber\\
&\times
\Bigg[
e^{i\frac{\mathbf{k}^2-(\mathbf{k}-\mathbf{q})^2}{2xE}z_i}
e^{-i\frac{\mathbf{u}\cdot\mathbf{q}}{xE}
\frac{(\mathbf{k}-\mathbf{q})^2+m_g^2}{2xE}z_i}
\left(
1
-
\frac{\mathbf{u}\cdot\mathbf{q}}{2xE}\,
\frac{\mathbf{k}^2-(\mathbf{k}-\mathbf{q})^2}
{v_i^2(\mathbf{q})}\,
\frac{\partial v_i^2(\mathbf{q})}{\partial\mathbf{q}^2}
+
\frac{\mathbf{u}\cdot\mathbf{q}}{xE}\,
\frac{2x^2M^2}{(\mathbf{k}-\mathbf{q})^2+\chi}
\right)
\nonumber\\
&\qquad
-
e^{i\frac{\mathbf{k}^2+\chi}{2xE}z_i}
e^{-i\frac{\mathbf{u}\cdot\mathbf{q}}{xE}
\frac{x^2M^2}{2xE}z_i}
\left(
1
-
\frac{\mathbf{u}\cdot\mathbf{q}}{2xE}\,
\frac{\mathbf{k}^2+\chi}
{v_i^2(\mathbf{q})}\,
\frac{\partial v_i^2(\mathbf{q})}{\partial\mathbf{q}^2}
+
\frac{\mathbf{u}\cdot\mathbf{q}}{xE}\,
\frac{2x^2M^2}{(\mathbf{k}-\mathbf{q})^2+\chi}
\right)
\Bigg]\,.
\label{m101}
\end{align}
\end{widetext}
Here, for the potential introduced in Eq.~\eqref{vt},
\begin{equation}
\frac{1}{v_i^2(\mathbf{q})}\,
\frac{\partial v_i^2(\mathbf{q})}{\partial\mathbf{q}^2}
=
-\frac{2}{\mathbf{q}^2+\mu_i^2}\,.
\label{vt2}
\end{equation}
The identity in Eq.~\eqref{vt2} will be used repeatedly in the subsequent calculation.

The three terms in the first square bracket of Eq.~\eqref{m101} originate from the Lorentz contraction $I_{1,0,1}$ in Eq.~\eqref{I101}. The first is the leading-eikonal contribution, whereas the remaining two are generated at first-subeikonal order by the transverse medium flow. The terms proportional to $\partial v_i^2/\partial\mathbf{q}^2$ arise from expanding the scattering potential about the gluon- and heavy-quark-propagator poles when evaluating the corresponding residues.

The mass-dependent  residue-denominator correction $\frac{\mathbf{u}\cdot\mathbf{q}}{xE}\,\frac{2x^2M^2}{(\mathbf{k}-\mathbf{q})^2+\chi}$ in Eq.~\eqref{m101} (see also Eq.~\eqref{eq:flow_massresidue} and classification in Sec.~\ref{FullSpectrumClassification}) originates from the $\mathcal{O}\!\left(\perp^3/(xE)^2\right)$ mismatch between quark and gluon propagator poles in Eq.~\eqref{razlikapolova}. Since their common $\mathcal{O}(\perp)$
contribution cancels, this mismatch enters the residue denominators already at first-subeikonal accuracy, consistently with the power counting of Sec.~\ref{sec:Assumptions}. The interpretation of this term as a modification of the dead-cone denominator, induced by the flow-dependent shift of the intermediate-state energy, is discussed in the main text, see Sec.~\ref{KinInter}.

The subleading pole corrections also enter the second exponential associated with each residue in  the second square bracket of Eq.~\eqref{m101}. As established in Eqs.~\eqref{eq:phidecomposition} and~\eqref{eq:deltaphicounting} (see also Eq.~\eqref{eq:flow_deltaphi}), the corresponding flow-induced formation phase shift $\delta\phi=\delta Q^-\,z=\mathcal{O}\!\left(\perp/(xE)\right)$ contributes at first-subeikonal order, as discussed in Sec.~\ref{KinInter}. Upon interference (including those between the propagator-pole residues within $|\mathcal{M}_{1,0,1}|^2$), these corrections give rise to the phase-induced $z\sin$ terms in the radiation spectrum, as pointed out in Eq.~\eqref{eq:cosphaseexpansion} and classified in Sec.~\ref{FullSpectrumClassification}.

Which combinations of these subleading phases survive is determined only after the individual first-order amplitudes are squared and their pairwise interference terms are evaluated. The exponential form in Eq.~\eqref{m101} is therefore kept for compactness and consistency, while the resulting squared-amplitude and interference contributions are expanded only to the working first-subeikonal accuracy when constructing the physical quantity.

In the static-medium limit, $\mathbf{u}=0$, Eq.~\eqref{m101} reproduces the structure of the $\mathcal{M}_{1,0,1}$ result in Eq.~(41) of Ref.~\cite{DGLV}, up to the sign convention adopted there for the scattering potential. 

In the massless limit, $M=m_g=0$, Eq.~\eqref{m101} reduces, to linear order in $\mathbf{u}$, to the single-Born contribution $R_C$ in Eq.~(19) of Ref.~\cite{KML}, provided the subleading propagator-pole corrections are omitted from the formation phases. Ref.~\cite{KML} retains the $\mathcal{O}\left(\perp^3/(xE)^2\right)$ pole corrections in the denominators of the propagator-pole residues, but not in the formation-phase arguments. In our calculation, these corrections are retained in both, generating an additional first-subeikonal phase contribution already at the amplitude level.

\subsection{\label{sec:M110}Single-scattering contribution $\mathcal{M}_{1,1,0}$}

We next consider the amplitude $\mathcal{M}_{1,1,0}$, for which the heavy quark interacts with the medium before emitting the gluon. In contrast to $\mathcal{M}_{1,0,1}$, the remaining longitudinal integration contains only one propagator singularity. Using the conventions introduced above, from Fig.~\ref{fig:M1} (b) we obtain
\begin{widetext}
\begin{align}
\mathcal{M}_{1,1,0}
{}&=
\int\frac{d^4q}{(2\pi)^4}\,
\left[
i g_s(2p-k)^{\nu}
\varepsilon^{*}_{\nu}(k)
T_{ jet}^{c}
\right]
\frac{i}{p^2-M^2+i\epsilon}\,
\left[
i g_s(2p-q)_{\mu}
A_{ext}^{\mu,a_1}(q)
T_{jet}^{a_1}
\right]
\frac{i}{(p-q)^2-M^2+i\epsilon}\,
iJ(p-q)
\nonumber \\
&=
4 i g_s E
\sum_i
(ca_1)\,
T_i^{a_1}
\frac{\boldsymbol{\varepsilon}\cdot\mathbf{k}}
{\mathbf{k}^2+\chi}\,
\int\frac{d^2\mathbf{q}}{(2\pi)^2}\,
e^{-i\mathbf{q}\cdot\mathbf{x}_i}
\int\frac{dq_z}{2\pi}\,
e^{-iq_z z_i}
\frac{v_i(q)\,J(p-q)}
{(p-q)^2-M^2+i\epsilon}\,.
\label{m110p}
\end{align}
\end{widetext}
The second form of Eq.~\eqref{m110p} follows from the same $q^0$ reduction used for $\mathcal{M}_{1,0,1}$: inserting Eqs.~\eqref{Aext} and~\eqref{a} fixes $q^0=\mathbf{u}\cdot\mathbf{q}$ through Eq.~\eqref{kin4M1}, while Eq.~\eqref{qu} eliminates $u\cdot q$. The emission vertex and adjacent off-shell propagator are reduced as in the zeroth-order amplitude, Eq.~\eqref{M0final}. 

The remaining longitudinal integral is
\begin{align}
\mathcal{I}_{1,1,0}
{}&=
\int\frac{dq_z}{2\pi}\,
e^{-iq_z z_i}
\frac{v_i(q)}
{(p-q)^2-M^2+i\epsilon}\,.
\label{I110p}
\end{align}
For $z_i>0$, the contour is closed in the lower half-plane. Note that the scattering-potential pole is exponentially suppressed by the separation condition in Eq.~\eqref{eq:separation}, exactly as in the evaluation of Eq.~\eqref{IM101p}. The integral is therefore determined solely by the heavy-quark propagator pole already given in Eq.~\eqref{qzp_q}. Retaining that pole to the accuracy specified in Eqs.~\eqref{eq:polecounting} and~\eqref{deltaQ}, we find
\begin{widetext}
\begin{align}
\mathcal{M}_{1,1,0}
{}&\approx
2 g_s
\sum_i
(ca_1)\,
T_i^{a_1}
\int\frac{d^2\mathbf{q}}{(2\pi)^2}\,
e^{-i\mathbf{q}\cdot\mathbf{x}_i}\,
e^{-i\mathbf{u}\cdot\mathbf{q}\,z_i}\,
v_i(\mathbf{q})\,
J(E,\mathbf{p})\,
\frac{\boldsymbol{\varepsilon}\cdot\mathbf{k}}
{\mathbf{k}^2+\chi}\,
e^{i\frac{\mathbf{k}^2+\chi}{2xE}z_i}
\nonumber \\
&\times
e^{-i\frac{\mathbf{u}\cdot\mathbf{q}}{xE}
\frac{x^2M^2}{2xE}z_i}
\left[
1-
\frac{\mathbf{u}\cdot\mathbf{q}}{2xE}\,
\frac{\mathbf{k}^2+\chi}
{v_i^2(\mathbf{q})}\,
\frac{\partial v_i^2(\mathbf{q})}
{\partial\mathbf{q}^2}
\right] \, .
\label{m110}
\end{align}
\end{widetext}
The first-subeikonal term in the square bracket of Eq.~\eqref{m110} arises from expanding the scattering potential about the heavy-quark propagator pole when evaluating the residue. 
Unlike $\mathcal{M}_{1,0,1}$, no corresponding difference of propagator poles enters the residue denominator of this amplitude. Consequently, $\mathcal{M}_{1,1,0}$ does not generate a mass-dependent  residue-denominator correction of the form displayed in Eq.~\eqref{razlikapolova}.

The additional phase factor in Eq.~\eqref{m110} instead originates from the $\mathcal{O}\left(\perp^3/(xE)^2\right)$ correction $\delta Q^-$ retained in the heavy-quark-propagator pole of Eq.~\eqref{qzp_q}. According to the counting in Eq.~\eqref{eq:deltaphicounting}, this correction must also be kept in the formation phase at the amplitude level and contributes to the phase-induced first-subeikonal structure.  The subleading pole correction enters $\mathcal{M}_{1,1,0}$ as an overall phase and therefore cancels from the diagonal contribution $|\mathcal{M}_{1,1,0}|^2$; its effect can arise only through interference with the other single-scattering amplitudes.

In the static-medium limit, $\mathbf{u}=0$, Eq.~\eqref{m110} reproduces the $\mathcal{M}_{1,1,0}$ result in Eq.~(51) of Ref.~\cite{DGLV}, after accounting for the opposite sign convention used there for the scattering potential.

To linear order in $\mathbf{u}$, Eq.~\eqref{m110} reduces in the massless limit, $M=m_g=0$, to the single-Born amplitude $R_A$ in Eq.~(13) of Ref.~\cite{KML}. For finite $M$, however, retaining the subleading heavy-quark propagator-pole correction in the formation phase generates an additional exponential factor in Eq.~\eqref{m110}, absent from the massless result of Ref.~\cite{KML}. As for $\mathcal{M}_{1,0,1}$ in Eq.~\eqref{m101}, this correction is retained according to the first-subeikonal phase counting established in Eq.~\eqref{eq:deltaphicounting}.

\subsection{\label{sec:M100}Single-scattering contribution $\mathcal{M}_{1,0,0}$}

The remaining quark-rescattering topology is $\mathcal{M}_{1,0,0}$, shown in Fig.~\ref{fig:M1}(c), where the gluon is emitted before the heavy quark exchanges momentum $q$ with the medium. With the assignments of Fig.~\ref{fig:M1}, the corresponding amplitude is
\begin{widetext}
\begin{align}
\mathcal{M}_{1,0,0}
{}&=
\int\frac{d^4q}{(2\pi)^4}\,
\left[
i g_s(2p-2k-q)_{\mu}
A_{ext}^{\mu,a_1}(q)
T^{a_1}_{jet}
\right]
\frac{i}{(p-k-q)^2-M^2+i\epsilon}\,
\left[
i g_s(2p-k-2q)^{\nu}
\varepsilon^{*}_{\nu}(k)
T^c_{jet}
\right]
\nonumber \\
&\times
\frac{i}{(p-q)^2-M^2+i\epsilon}\,
iJ(p-q)\,.
\label{m100p}
\end{align}
\end{widetext}
Proceeding as in the previous single-scattering contributions, the Lorentz products at the scattering and emission vertices are evaluated using the momentum, polarization, and flow four-vectors defined by Eq.~\eqref{vektori} and in Sec.~\ref{sec:Assumptions}, together with $u\cdot q=0$ from Eq.~\eqref{qu}. At the working soft-gluon accuracy, the resulting vertex factors are reduced according to the same hierarchy used above.

After the $q^0$ integration, the remaining longitudinal structure contains two heavy-quark propagators,
\begin{widetext}
\begin{align}
\mathcal{I}_{1,0,0}
{}&=
\int\frac{dq_z}{2\pi}\,
e^{-iq_z z_i}
\frac{v_i(q)}
{\left[(p-k-q)^2-M^2+i\epsilon\right]
\left[(p-q)^2-M^2+i\epsilon\right]}\,.
\label{IM100p}
\end{align}
\end{widetext}
The broad-source approximation of Eq.~\eqref{eq:broadsource} is understood. As in Eqs.~\eqref{IM101p} and~\eqref{I110p}, the scattering-potential pole is exponentially suppressed by Eq.~\eqref{eq:separation}, so only the heavy-quark-propagator poles need to be retained. One of them is the heavy-quark pole $q^-_{z,p-q}$ already given in Eq.~\eqref{qzp_q}. The second originates from the  propagator with momentum $p-k-q$ and, to the same pole accuracy, is
\begin{align}
q^-_{z,p-k-q}
{}&=
\mathbf{u}\cdot\mathbf{q}
+
\frac{\mathbf{u}\cdot\mathbf{q}}{xE}\,
\frac{x^2M^2}{2xE}
-i\epsilon\,.
\label{qzp_k-q}
\end{align}

Evaluating the residues at the two heavy-quark propagator poles gives
\begin{widetext}
\begin{align}
\mathcal{M}_{1,0,0}
\approx{}&
2 g_s
\sum_i
(a_1c)\,
T_i^{a_1}
\int\frac{d^2\mathbf{q}}{(2\pi)^2}\,
e^{-i\mathbf{q}\cdot\mathbf{x}_i}\,
e^{-i\mathbf{u}\cdot\mathbf{q}\,z_i}\,
v_i(\mathbf{q})\,
J(E,\mathbf{p})\,
\frac{\boldsymbol{\varepsilon}\cdot\mathbf{k}}
{\mathbf{k}^2+\chi}\,
e^{-i\frac{\mathbf{u}\cdot\mathbf{q}}{xE}
\frac{x^2M^2}{2xE}z_i}
\nonumber \\
&\times
\left[
1-
e^{i\frac{\mathbf{k}^2+\chi}{2xE}z_i}
\left(
1-
\frac{\mathbf{u}\cdot\mathbf{q}}{2xE}\,
\frac{\mathbf{k}^2+\chi}
{v_i^2(\mathbf{q})}\,
\frac{\partial v_i^2(\mathbf{q})}
{\partial\mathbf{q}^2}
\right)
\right]\,.
\label{m100}
\end{align}
\end{widetext}
The derivative term in Eq.~\eqref{m100} arises from expanding the scattering potential at the $q^-_{z,p-q}$ pole when evaluating its residue. Unlike $\mathcal{M}_{1,0,1}$, the two heavy-quark propagator poles contain identical $\mathcal{O}\!\left(\perp^3/(xE)^2\right)$ mass-dependent  corrections (on top of the same leading $\mathbf{u}\cdot\mathbf{q}$ contribution), which cancel in their difference. Consequently, no additional mass-dependent  residue-denominator correction analogous to that in Eqs.~\eqref{razlikapolova} and~\eqref{eq:flow_massresidue} arises. The common subleading pole  correction nevertheless survives as an overall phase factor in Eq.~\eqref{m100}, retained at first-subeikonal order according to Eq.~\eqref{eq:deltaphicounting}. This phase cancels in $|\mathcal{M}_{1,0,0}|^2$ but can contribute to interference terms involving amplitudes with different phase shifts.

For $\mathbf{u}=0$, Eq.~\eqref{m100} reproduces the static-medium $\mathcal{M}_{1,0,0}$ result of Eq.~(45) in Ref.~\cite{DGLV}, after accounting for the opposite sign convention used there for the scattering potential. 

To linear order in $\mathbf{u}$, Eq.~\eqref{m100} reduces in the massless limit, $M=m_g=0$, to the single-Born amplitude $R_B$ in Eq.~(15) of Ref.~\cite{KML}. For finite $M$, however, retaining the subleading mass-dependent correction to the heavy-quark propagator poles introduces an additional overall phase factor at first-subeikonal order in Eq.~\eqref{m100}, absent from the massless result of Ref.~\cite{KML}.

\subsection{\label{sec:M1sq}Sum of the single-scattering contributions}

Having derived the three single-scattering amplitudes, we sum them to obtain
\begin{equation}
\label{M1}
\mathcal{M}_1= \mathcal{M}_{1,1,0}+ \mathcal{M}_{1,0,0} + \mathcal{M}_{1,0,1} \, .
\end{equation}

Its contribution to the first-order-in-opacity gluon radiativion spectum in Eq.~\eqref{spectum}
contains three diagonal terms and three independent interference
combinations,
\begin{widetext}
\begin{align}
\frac{1}{d_T}\Tr\mean{|\mathcal{M}_1|^2}
={}&
\frac{1}{d_T}\Tr\mean{|\mathcal{M}_{1,0,1}|^2}
+
\frac{1}{d_T}\Tr\mean{|\mathcal{M}_{1,1,0}|^2}
+
\frac{1}{d_T}\Tr\mean{|\mathcal{M}_{1,0,0}|^2}
\nonumber \\
&+
\frac{1}{d_T}\Tr\mean{
\mathcal{M}_{1,0,1}\mathcal{M}_{1,1,0}^{\dagger}
+\mathrm{c.c.}
}
+
\frac{1}{d_T}\Tr\mean{
\mathcal{M}_{1,0,1}\mathcal{M}_{1,0,0}^{\dagger}
+\mathrm{c.c.}
}
\nonumber \\
&+
\frac{1}{d_T}\Tr\mean{
\mathcal{M}_{1,1,0}\mathcal{M}_{1,0,0}^{\dagger}
+\mathrm{c.c.}
}.
\label{M1p}
\end{align}
\end{widetext}

We evaluate these contributions in the order displayed above, beginning
with the three diagonal terms.

We first present a detailed derivation of the contribution from $\mathcal{M}_{1,0,1}$ given by Eq.~\eqref{m101}
\begin{widetext}
\begin{align}
\frac{1}{d_T}
\operatorname{Tr}
\left\langle
\mathcal{M}_{1,0,1}^{\dagger}
\mathcal{M}_{1,0,1}
\right\rangle 
{}&=
\frac{1}{d_T}\operatorname{Tr}
\Bigg\langle
\sum_{i,j}
4g_s^2\,
[c,a_1][a_2,c]\,
T_i^{a_1}T_j^{a_2}
\int\frac{d^2\mathbf{q}}{(2\pi)^2}
\int\frac{d^2\mathbf{q}'}{(2\pi)^2}\,
e^{-i\mathbf{q}\cdot\mathbf{x}_i}
e^{i\mathbf{q}'\cdot\mathbf{x}_j}
v_i(\mathbf{q})v^{*}_j(\mathbf{q}')\,
\left|J(E,\mathbf{p})\right|^2
\nonumber\\
&\times
\frac{
\left[
\boldsymbol{\varepsilon}\cdot(\mathbf{k}-\mathbf{q})
-\boldsymbol{\varepsilon}\cdot\mathbf{k} \, \frac{
\mathbf{u}\cdot(\mathbf{k}-\mathbf{q})
}{xE}
+\boldsymbol{\varepsilon}\cdot\mathbf{u} \, \frac{
\mathbf{q}\cdot(\mathbf{k}-\mathbf{q})
}{xE}
\right]
}{
(\mathbf{k}-\mathbf{q})^2+\chi
}
\frac{
\left[
\boldsymbol{\varepsilon}\cdot(\mathbf{k}-\mathbf{q}')
-\boldsymbol{\varepsilon}\cdot\mathbf{k}\frac{
\mathbf{u}\cdot(\mathbf{k}-\mathbf{q}')
}{xE}
+\boldsymbol{\varepsilon}\cdot\mathbf{u} \, \frac{
\mathbf{q}'\cdot(\mathbf{k}-\mathbf{q}')
}{xE}
\right]
}{
(\mathbf{k}-\mathbf{q}')^2+\chi
}
\nonumber\\
&\times
e^{-i \mathbf{u} \cdot \mathbf{q} \, z_i}
e^{i \mathbf{u} \cdot \mathbf{q}' \, z_j}
\Bigg\{
\Bigg[
1
-\frac{\mathbf{u}\cdot\mathbf{q}}{2xE}\,
\frac{\mathbf{k}^2+\chi}{v_i^2(\mathbf{q})}\,
\frac{\partial v_i^2(\mathbf{q})}{\partial\mathbf{q}^2}
+\frac{\mathbf{u}\cdot\mathbf{q}}{xE}\,
\frac{2x^2M^2}{(\mathbf{k}-\mathbf{q})^2+\chi}
\Bigg]
e^{\,i\frac{\mathbf{k}^2+\chi}{2xE}z_i}
e^{-i\frac{\mathbf{u}\cdot\mathbf{q}}{xE}
\frac{x^2M^2}{2xE}z_i}
\nonumber\\
&%\hspace{1.4cm}
-
\Bigg[
1
+\frac{\mathbf{u}\cdot\mathbf{q}}{2xE}\,
\frac{(\mathbf{k}-\mathbf{q})^2-\mathbf{k}^2}{v_i^2(\mathbf{q})}\,
\frac{\partial v_i^2(\mathbf{q})}{\partial\mathbf{q}^2}
+\frac{\mathbf{u}\cdot\mathbf{q}}{xE}\,
\frac{2x^2M^2}{(\mathbf{k}-\mathbf{q})^2+\chi}
\Bigg]
e^{i\frac{\mathbf{k}^2 - (\mathbf{k}-\mathbf{q})^2}{2xE}z_i}
e^{-i\frac{\mathbf{u}\cdot\mathbf{q}}{xE}
\frac{(\mathbf{k}-\mathbf{q})^2+m_g^2}{2xE}z_i}
\Bigg\}
\nonumber\\
&\times
\Bigg\{
\Bigg[
1
-\frac{\mathbf{u}\cdot\mathbf{q}'}{2xE}\,
\frac{\mathbf{k}^2+\chi}{v_j^2(\mathbf{q}')}\,
\frac{\partial v_j^2(\mathbf{q}')}{\partial\mathbf{q}'^{\,2}}
+\frac{\mathbf{u}\cdot\mathbf{q}'}{xE}\,
\frac{2x^2M^2}{(\mathbf{k}-\mathbf{q}')^2+\chi}
\Bigg]
e^{-i\frac{\mathbf{k}^2+\chi}{2xE}z_j}
e^{\,i\frac{\mathbf{u}\cdot\mathbf{q}'}{xE}
\frac{x^2M^2}{2xE}z_j}
\nonumber\\
&%\hspace{1.4cm}
-
\Bigg[
1
+\frac{\mathbf{u}\cdot\mathbf{q}'}{2xE}\,
\frac{(\mathbf{k}-\mathbf{q}')^2-\mathbf{k}^2}{v_j^2(\mathbf{q}')}\,
\frac{\partial v_j^2(\mathbf{q}')}{\partial\mathbf{q}'^{\,2}}
+\frac{\mathbf{u}\cdot\mathbf{q}'}{xE}\,
\frac{2x^2M^2}{(\mathbf{k}-\mathbf{q}')^2+\chi}
\Bigg]
e^{\,i\frac{(\mathbf{k}-\mathbf{q}')^2-\mathbf{k}^2}{2xE}z_j}
e^{\,i\frac{\mathbf{u}\cdot\mathbf{q}'}{xE}
\frac{(\mathbf{k}-\mathbf{q}')^2+m_g^2}{2xE}z_j}
\Bigg\}
\Bigg\rangle \, .
\label{EM101p}
\end{align}
\end{widetext}

Starting from Eq.~\eqref{EM101p}, we first perform the target-color
trace. Equation~\eqref{tr2} identifies the scattering center and color
indices in the amplitude and its conjugate, $i=j$ and $a_1=a_2$,
leaving a single sum over scattering centers. The remaining \text{blue}{quark}-color
structure can be written as
\[
\operatorname{Tr}\!\left([c,a_1][a_1,c]\right)=2\alpha,
\qquad
\alpha\equiv
\operatorname{Tr}\!\left(c^2a_1^2-ca_1ca_1\right),
\]
where the commutator algebra and the cyclic invariance of the trace are
those specified below Eqs.~\eqref{gen1} and~\eqref{trjet}. The identity $\sum_{a_1} {a_1} c {a_1}=[C_2(F)-C_A/2] c$ gives
\[
\alpha=\frac{1}{2}C_A C_2(F)d_F,
\qquad
\frac{\alpha}{d_F}=\frac{1}{2}C_A C_2(F)\, .
\]

The remaining sum over scattering centers is converted to the
continuum form using Eqs.~\eqref{eq:density} and~\eqref{eq:continuum}.
Since the product of the single-scattering amplitude and its conjugate
contains the transverse phase
$e^{-i(\mathbf{q}-\mathbf{q}')\cdot\mathbf{x}}$, the lower-sign case
of Eq.~\eqref{app9} gives
\begin{equation}
\mathbf{q}'=\mathbf{q} \,
\label{qq'M1}
\end{equation}
which then holds throughout this subsection.

Consequently, the two momentum-transfer integrations reduce to a
single transverse integral. The source factor
is treated according to the broad-source approximation in
Eq.~\eqref{eq:broadsource}. Summing over the two physical transverse
gluon polarizations specified by Eqs.~\eqref{eq:axialgauge}
and~\eqref{vektori}, and retaining only terms linear in $\mathbf{u}$ as
prescribed by Eq.~\eqref{eq:flowexpansion}, leaves the phase difference
between the two pole contributions in the form
\[
\phi_0+\delta\phi
=
\frac{(\mathbf{k}-\mathbf{q})^2+\chi}{2xE}\,z
+
\frac{\mathbf{u}\cdot\mathbf{q}}{xE}\,
\frac{(\mathbf{k}-\mathbf{q})^2+m_g^2-x^2M^2}{2xE}\,z .
\]

Here the first term is the leading formation phase, whereas the second
is the first-subeikonal correction discussed in
Eqs.~\eqref{eq:phidecomposition} and~\eqref{eq:deltaphicounting}.
Accordingly, the factor $1-\cos(\phi_0+\delta\phi)$ is expanded using
Eq.~\eqref{eq:cosphaseexpansion}, while $\phi_0$ itself is kept
unexpanded. Keeping all terms through first-subeikonal accuracy, we obtain
\begin{widetext}
\begin{align}~\label{E_M101}
\frac{1}{d_T}\Tr & \mean{|\mathcal{M}_{1,0,1}|^2} = {}  \frac{16g^2_s C_2(T) N}{A_{\perp} d_A} \alpha \int_0^L{dz \rho{(z)}} \int{ \frac{d^2 \mathbf{q}}{(2 \pi)^2}}  |v({\mathbf{q}})|^2 |J(E, \mathbf{p})|^2 \frac{(\mathbf{k}- \mathbf{q})^2}{((\mathbf{k} - \mathbf{q})^2 +\chi)^2} \nonumber \\
\times& \left\{ \left[ 1-  2 \frac{\mathbf{u} \cdot (\mathbf{k} -\mathbf{q})}{x E}- \frac{\mathbf{u} \cdot \mathbf{q}}{2 x E}\, \frac{2 \mathbf{k}^2- (\mathbf{k} -\mathbf{q})^2 +\chi}{v^2(\mathbf{q})} \, \frac{\partial v^2( \mathbf{q})}{\partial  \mathbf{q}^2}  + \frac{\mathbf{u} \cdot \mathbf{q}}{x E} \,  \frac{4 x^2 M^2}{(\mathbf{k} - \mathbf{q})^2 +\chi} \right]   \left[ 1-\cos \left( \frac{(\mathbf{k} - \mathbf{q})^2 +\chi}{2 x E} \, z \right) \right] \right.
   \nonumber \\
&\left. + \frac{\mathbf{u} \cdot \mathbf{q}}{x E} \frac{(\mathbf{k}-\mathbf{q})^2 +m^2_g - x^2 M^2}{2 x E}\, z \sin{\left( \frac{(\mathbf{k} - \mathbf{q})^2 +\chi}{2 x E} \, z\right)}  \right\} \, .
\end{align}
\end{widetext}

The correction proportional to $-2\,\mathbf{u}\cdot(\mathbf{k}-\mathbf{q})/(xE)$ in Eq.~\eqref{E_M101} arises from the subleading Lorentz contraction, while the term involving $\partial v^2(\mathbf{q})/\partial\mathbf{q}^2$ is the flow-induced potential-derivative correction. The mass-dependent  residue-denominator correction, proportional to $4x^2M^2/[(\mathbf{k}-\mathbf{q})^2+\chi]$, and the phase-induced $z\sin$ term both originate from the higher-order ($\mathcal{O}\!\left(\perp^3/(xE)^2\right)$ ) quark-gluon propagator-pole mismatch in Eq.~\eqref{razlikapolova}. These contributions are identified according to the classification in Sec.~\ref{FullSpectrumClassification}.

To linear order in $\mathbf{u}$, Eq.~\eqref{E_M101} reduces in the massless limit, $M=m_g=0$, to the $|R_C|^2$ contribution in Eq.~(25) of Ref.~\cite{KML}, provided the subleading propagator-pole corrections are omitted from the formation phases. Retaining these corrections in our calculation generates an additional phase-induced $z\sin$  first-subeikonal contribution.

For the diagonal contribution associated with $\mathcal{M}_{1,1,0}$, we follow the same procedure using Eq.~\eqref{m110}. The target-color trace and medium averaging, together with the transverse-position integration in Eq.~\eqref{app9}, identify the scattering-center indices and impose Eq.~\eqref{qq'M1}, reducing the result to a single transverse-momentum integral. Applying the continuum prescription of Eqs.~\eqref{eq:density} and~\eqref{eq:continuum}, the corresponding quark-color trace, the broad-source approximation of Eq.~\eqref{eq:broadsource}, and the physical-polarization sum, we obtain, through first-subeikonal order and linear order in $\mathbf{u}$,
\begin{widetext}
\begin{align}~\label{E_M110}
\frac{1}{d_T}\Tr  \mean{|\mathcal{M}_{1,1,0}|^2} =& {}  \frac{4g^2_s C_2(T) N}{A_{\perp} d_A} \Tr({c^2a^2_1)} \int_0^L{dz \rho{(z)}} \int{ \frac{d^2 \mathbf{q}}{(2 \pi)^2}}  |v({\mathbf{q}})|^2 |J(E, \mathbf{p})|^2 \frac{\mathbf{k}^2}{(\mathbf{k}^2 +\chi)^2} \nonumber \\
\times& \left[ 1-   \frac{\mathbf{u} \cdot \mathbf{q}}{x E}\, \frac{\mathbf{k}^2 +\chi}{v^2(\mathbf{q})} \, \frac{\partial v^2( \mathbf{q})}{\partial  \mathbf{q}^2}   \right] \nonumber \\
\approx& \frac{4g^2_s C_2(T) N}{A_{\perp} d_A} \Tr({c^2a^2_1)} \int_0^L{dz \rho{(z)}} \int{ \frac{d^2 \mathbf{q}}{(2 \pi)^2}}  |v({\mathbf{q}})|^2 |J(E, \mathbf{p})|^2 \frac{\mathbf{k}^2}{(\mathbf{k}^2 +\chi)^2} \, .
\end{align}
\end{widetext}
In the last step, we have omitted the term that is odd under
$\mathbf{q}\to-\mathbf{q}$ and therefore vanishes upon angular
integration. We further note that, although $\mathcal{M}_{1,1,0}$
contains a longitudinal phase, it cancels against its complex conjugate
after the medium average sets Eq.~\eqref{qq'M1}. Hence, no
formation-phase dependence survives in this diagonal contribution, and
there is no corresponding phase-induced z sin first-subeikonal correction. Moreover, no first-subeikonal correction survives in this diagonal contribution at all.

In the massless limit, this agrees with the corresponding $|R_A|^2$
contribution in Eq.~(23) of Ref.~\cite{KML}, where the flow-dependent
term is likewise odd in $\mathbf{q}$ and vanishes after angular
integration, leaving no surviving first-subeikonal correction.
  
Applying the same procedure to $M_{1,0,0}$ in Eq.~\eqref{m100}, the corresponding diagonal contribution at the first-subeikonal order is

\begin{widetext}
\begin{align}~\label{E_M100}
\frac{1}{d_T}\operatorname{Tr}  \mean{|\mathcal{M}_{1,0,0}|^2} =& {}  \frac{8g^2_s C_2(T) N}{A_{\perp} d_A} \Tr({c^2a^2_1)} \int_0^L{dz \rho{(z)}} \int{ \frac{d^2 \mathbf{q}}{(2 \pi)^2}}  |v({\mathbf{q}})|^2 |J(E, \mathbf{p})|^2 \frac{\mathbf{k}^2}{(\mathbf{k}^2 +\chi)^2} \nonumber \\
\times& \left[ 1-   \frac{\mathbf{u} \cdot \mathbf{q}}{2 x E}\, \frac{\mathbf{k}^2 +\chi}{v^2(\mathbf{q})} \, \frac{\partial v^2( \mathbf{q})}{\partial  \mathbf{q}^2} \right]   \left[ 1-\cos \left( \frac{\mathbf{k}^2 +\chi}{2 x E} \, z \right) \right]   \nonumber \\
\approx&  \frac{8g^2_s C_2(T) N}{A_{\perp} d_A} \Tr({c^2a^2_1)} \int_0^L{dz \rho{(z)}} \int{ \frac{d^2 \mathbf{q}}{(2 \pi)^2}}  |v({\mathbf{q}})|^2 |J(E, \mathbf{p})|^2 \frac{\mathbf{k}^2}{(\mathbf{k}^2 +\chi)^2} \nonumber \\
\times& \left[ 1-\cos \left( \frac{\mathbf{k}^2 +\chi}{2 x E} \, z \right) \right] \, .
\end{align}
\end{widetext}
In the last step, the term odd under
$\mathbf{q}\to-\mathbf{q}$ has been omitted, since it vanishes upon
angular integration. The common first-subeikonal correction to the
longitudinal phase cancels between $\mathcal{M}_{1,0,0}$ and its complex
conjugate, leaving only the leading formation-phase dependence.
Consequently, no first-subeikonal correction survives in this diagonal
contribution.

 In the massless limit, this reduces to the corresponding
$|R_B|^2$ contribution in Eq.~(24) of Ref.~\cite{KML}.

Having completed the three diagonal contributions, we now turn to the
interference terms entering $|\mathcal{M}_1|^2$ in Eq.~\eqref{M1p}. Applying the same color and medium averaging, transverse-position integration, polarization sum, and broad-source approximation, the interference between $\mathcal{M}_{1,0,1}$ and  $\mathcal{M}_{1,1,0}$ yields, through first-subeikonal order and to linear order in $\mathbf{u}$,
\begin{widetext}
\begin{align}
\label{E_M110M101}
\frac{1}{d_T}\Tr \mean{\mathcal{M}_{1,0,1} \mathcal{M}^\dagger_{1,1,0}}
+\mathrm{c.c.}
&{}=
-\frac{8g_s^2 C_2(T) N}{A_{\perp} d_A}\,
\alpha
\int_0^L dz\,\rho(z)
\int \frac{d^2\mathbf{q}}{(2\pi)^2}\,
\frac{|v(\mathbf{q})|^2
|J(E,\mathbf{p})|^2}{(\mathbf{k}^2+\chi)
[(\mathbf{k}-\mathbf{q})^2+\chi]}
\nonumber\\
&\quad\times
\Biggl\{
\Biggl[
\mathbf{k}\cdot(\mathbf{k}-\mathbf{q})
-\mathbf{k}^2
\frac{\mathbf{u}\cdot(\mathbf{k}-\mathbf{q})}{xE}
-\mathbf{q}\cdot(\mathbf{q}-\mathbf{k})
\frac{\mathbf{u}\cdot\mathbf{k}}{xE}
-\mathbf{k}\cdot(\mathbf{k}-\mathbf{q})
\frac{\mathbf{u}\cdot\mathbf{q}}{xE}
\frac{\mathbf{k}^2+\chi}{v^2(\mathbf{q})}
\frac{\partial v^2(\mathbf{q})}{\partial\mathbf{q}^2}
\nonumber\\
&\qquad
+\mathbf{k}\cdot(\mathbf{k}-\mathbf{q})
\frac{\mathbf{u}\cdot\mathbf{q}}{xE}
\frac{2x^2M^2}{
(\mathbf{k}-\mathbf{q})^2+\chi
}
\Biggr]
\Biggl\{
1-
\cos\left(
\frac{(\mathbf{k}-\mathbf{q})^2+\chi}{2xE}\,z
\right)
\nonumber\\
&\qquad \times
\Biggl[
1+
\frac{\mathbf{u}\cdot\mathbf{q}}{2xE}
\frac{(\mathbf{k}-\mathbf{q})^2+\chi}{v^2(\mathbf{q})}
\frac{\partial v^2(\mathbf{q})}{\partial\mathbf{q}^2}
\Biggr]
\Biggr\}
\nonumber\\
&\qquad
+
\mathbf{k} \cdot (\mathbf{k}-\mathbf{q}) \, \frac{\mathbf{u}\cdot\mathbf{q}}{xE}
\frac{(\mathbf{k}-\mathbf{q})^2+m_g^2-x^2M^2}{2xE} \, z
\sin\left(
\frac{(\mathbf{k}-\mathbf{q})^2+\chi}{2xE}\,z
\right)
\Biggr\}.
\end{align}
\end{widetext}
Following the classification in Sec.~\ref{FullSpectrumClassification}, the first-subeikonal terms in Eq.~\eqref{E_M110M101} include subleading Lorentz-contraction and flow-induced potential-derivative corrections. The mass-dependent  residue-denominator correction is inherited from $\mathcal{M}_{1,0,1}$, whereas the final $z\sin$ term results from the subleading shift of the relative formation phase between the two amplitudes. This relative phase combines the longitudinal-pole contributions of $\mathcal{M}_{1,0,1}$ and $\mathcal{M}_{1,1,0}^{\dagger}$, including the gluon-heavy-quark pole difference $q^-_{z,k-q}-q^-_{z,p-q}$ discussed in Sec.~\ref{sec:M101}, with its subleading correction expanded according to Eq.~\eqref{eq:cosphaseexpansion}. The same mechanism generates phase-induced first-subeikonal corrections between different pole residues within $|\mathcal{M}_{1,0,1}|^2$ and in its interference with the other single-scattering amplitudes.

Note that all products of flow-dependent and subeikonal corrections in Eq.~\eqref{E_M110M101} are understood to be expanded consistently and retained only through linear order in $\mathbf{u}$ and first-subeikonal accuracy; terms beyond either order are discarded in accordance with Eq.~\eqref{eq:flowexpansion} and the power counting of Sec.~\ref{sec:Assumptions}.

In the massless limit, $M=m_g=0$, and upon restricting the formation
phase to leading-pole accuracy, this result reduces to the
$R_A R_C^{*}+\mathrm{c.c.}$ interference contribution in Eq.~(27) of
Ref.~\cite{KML}. The additional z sin term reflects the subleading pole
correction retained also in the formation phase in the present
calculation.

The next interference contribution involves $\mathcal{M}_{1,0,1}, \mathcal{M}_{1,0,0}^{\dagger}$ and their complex conjugate. Applying the same procedure as for the preceding terms, through first-subeikonal order and to linear order in $\mathbf{u}$, we obtain
\begin{widetext}
\begin{align}
\label{E_M100M101}
\frac{1}{d_T}\operatorname{Tr} \mean{\mathcal{M}_{1,0,1} \mathcal{M}^\dagger_{1,0,0}}
+\mathrm{c.c.}
&{}=
-\frac{8g_s^2 C_2(T) N}{A_{\perp} d_A}\,
\alpha
\int_0^L dz\,\rho(z)
\int \frac{d^2\mathbf{q}}{(2\pi)^2}\,
\frac{|v(\mathbf{q})|^2
|J(E,\mathbf{p})|^2}{(\mathbf{k}^2+\chi)
[(\mathbf{k}-\mathbf{q})^2+\chi]}
\nonumber\\
&\quad\times
\Biggl\{
\Biggl[
\mathbf{k}\cdot(\mathbf{k}-\mathbf{q})
-\mathbf{k}^2
\frac{\mathbf{u}\cdot(\mathbf{k}-\mathbf{q})}{xE}
-\mathbf{q}\cdot(\mathbf{q}-\mathbf{k})
\frac{\mathbf{u}\cdot\mathbf{k}}{xE}
\nonumber\\
&\qquad
-\mathbf{k}\cdot(\mathbf{k}-\mathbf{q})
\frac{\mathbf{u}\cdot\mathbf{q}}{xE}
\frac{\mathbf{k}^2+\chi}{v^2(\mathbf{q})}
\frac{\partial v^2(\mathbf{q})}{\partial\mathbf{q}^2}
+\mathbf{k}\cdot(\mathbf{k}-\mathbf{q})
\frac{\mathbf{u}\cdot\mathbf{q}}{xE}
\frac{2x^2M^2}{(\mathbf{k}-\mathbf{q})^2+\chi}
\Biggr]
\nonumber\\
&\qquad\times
\Biggl\{
1
-\cos\left(
\frac{\mathbf{k}^2+\chi}{2xE}\,z
\right)
\Biggl[
1+
\frac{\mathbf{u}\cdot\mathbf{q}}{2xE}
\frac{\mathbf{k}^2+\chi}{v^2(\mathbf{q})}
\frac{\partial v^2(\mathbf{q})}{\partial\mathbf{q}^2}
\Biggr]
\nonumber\\
&\qquad\quad
-\cos\left(
\frac{(\mathbf{k}-\mathbf{q})^2+\chi}{2xE}\,z
\right)
\Biggl[
1+
\frac{\mathbf{u}\cdot\mathbf{q}}{2xE}
\frac{(\mathbf{k}-\mathbf{q})^2+\chi}{v^2(\mathbf{q})}
\frac{\partial v^2(\mathbf{q})}{\partial\mathbf{q}^2}
\Biggr]
\nonumber\\
&\qquad\quad
+\cos\left(
\frac{\mathbf{k}^2-(\mathbf{k}-\mathbf{q})^2}{2xE}\,z
\right)
\Biggl[
1+
\frac{\mathbf{u}\cdot\mathbf{q}}{2xE}
\frac{\mathbf{k}^2+(\mathbf{k}-\mathbf{q})^2+2\chi}
     {v^2(\mathbf{q})}
\frac{\partial v^2(\mathbf{q})}{\partial\mathbf{q}^2}
\Biggr]
\Biggr\}
\nonumber\\
&\qquad
+
\mathbf{k}\cdot(\mathbf{k}-\mathbf{q})\,
\frac{\mathbf{u}\cdot\mathbf{q}}{xE}
\frac{(\mathbf{k}-\mathbf{q})^2+m_g^2-x^2M^2}{2xE}\,
z\,
\sin\left(
\frac{(\mathbf{k}-\mathbf{q})^2+\chi}{2xE}\,z
\right)
\nonumber\\
&\qquad
+
\mathbf{k}\cdot(\mathbf{k}-\mathbf{q})\,
\frac{\mathbf{u}\cdot\mathbf{q}}{xE}
\frac{(\mathbf{k}-\mathbf{q})^2+m_g^2-x^2M^2}{2xE}\,
z\,
\sin\left(
\frac{\mathbf{k}^2-(\mathbf{k}-\mathbf{q})^2}{2xE}\,z
\right)
\Biggr\}.
\end{align}
\end{widetext}
Equation~\eqref{E_M100M101} likewise contains subleading Lorentz-contraction and flow-induced potential-derivative. The mass-dependent pole-difference residue term proportional to $2x^2M^2/[(\mathbf{k}-\mathbf{q})^2+\chi]$ originates from the residue-denominator correction in $\mathcal{M}_{1,0,1}$. The two $z\sin$ terms arise instead from subleading corrections to the relative formation phases in the interference with $\mathcal{M}_{1,0,0}^{\dagger}$. These corrections cancel for the phase proportional to $\mathbf{k}^2+\chi$ but survive in the other two phase combinations, with the common coefficient $[(\mathbf{k}-\mathbf{q})^2+m_g^2-x^2M^2]/(2xE)$, following the expansion in Eq.~\eqref{eq:cosphaseexpansion}.
 Again, all products in Eq.~\eqref{E_M100M101} are understood to be consistently truncated at linear order in $\mathbf{u}$ and first-subeikonal accuracy, with higher-order contributions implicitly discarded.

In the massless limit, at linear order in $\mathbf{u}$, and upon evaluating the formation phases at leading-pole accuracy, Eq.~\eqref{E_M100M101} reduces to the $R_B R_C^{*}+\mathrm{c.c.}$ contribution in Eq.~(28) of Ref.~\cite{KML}. The two additional $z\sin$ terms arise from  retaining the subleading pole corrections in the corresponding relative phase arguments.

The remaining single-scattering interference contribution involves $\mathcal{M}_{1,1,0}$, $\mathcal{M}_{1,0,0}^{\dagger}$, and their complex conjugate. Applying the same procedure as above, through first-subeikonal order and to linear order in $\mathbf{u}$, gives
\begin{widetext}
\begin{align}~\label{E_M110M100}
\frac{1}{d_T}\operatorname{Tr}  \mean{\mathcal{M}_{1,1,0} \mathcal{M}^\dagger_{1,0,0}} +\mathrm{c.c.}=& {}  \frac{-8g^2_s C_2(T) N}{A_{\perp} d_A} \Tr({c a_1 c a_1)} \int_0^L{dz \rho{(z)}} \int{ \frac{d^2 \mathbf{q}}{(2 \pi)^2}}  |v({\mathbf{q}})|^2 |J(E, \mathbf{p})|^2 \frac{\mathbf{k}^2}{(\mathbf{k}^2 +\chi)^2} \nonumber \\
\times& \left\{ 1-   \frac{\mathbf{u} \cdot \mathbf{q}}{ x E}\, \frac{\mathbf{k}^2 +\chi}{v^2(\mathbf{q})} \, \frac{\partial v^2( \mathbf{q})}{\partial  \mathbf{q}^2} - \cos \left(  \frac{\mathbf{k}^2 +\chi}{2 x E} \, z   \right)  \left[1-  \frac{\mathbf{u} \cdot \mathbf{q}}{ 2 x E}\, \frac{\mathbf{k}^2 +\chi}{v^2(\mathbf{q})} \, \frac{\partial v^2( \mathbf{q})}{\partial  \mathbf{q}^2}  \right] \right\}  \nonumber \\
\approx&  \frac{-8g^2_s C_2(T) N}{A_{\perp} d_A} \Tr({c a_1 c a_1)} \int_0^L{dz \rho{(z)}} \int{ \frac{d^2 \mathbf{q}}{(2 \pi)^2}}  |v({\mathbf{q}})|^2 |J(E, \mathbf{p})|^2 \frac{\mathbf{k}^2}{(\mathbf{k}^2 +\chi)^2} \nonumber \\
\times& \left[ 1-\cos \left( \frac{\mathbf{k}^2 +\chi}{2 x E} \, z \right) \right] \, .
\end{align}
\end{widetext}
In the last step, the flow-dependent term has been omitted because it
is odd under $\mathbf{q}\to-\mathbf{q}$ and therefore vanishes upon
angular integration. Moreover, the subleading corrections to the
longitudinal phases of $\mathcal{M}_{1,1,0}$ and
$\mathcal{M}_{1,0,0}^{\dagger}$ cancel in their relative phase, so that
no $z\sin$ term is generated. Consequently, no first-subeikonal flow
correction survives in this interference contribution.

In the massless limit and at linear order in $\mathbf{u}$,
Eq.~\eqref{E_M110M100} reduces to the
$R_A R_B^{*}+\mathrm{c.c.}$ contribution in Eq.~(26) of
Ref.~\cite{KML}, which likewise contains no surviving
first-subeikonal flow correction after angular integration.

Combining the three diagonal contributions (Eqs.~\eqref{E_M101}, \eqref{E_M110}, and~\eqref{E_M100}) with the three interference terms (Eqs.~\eqref{E_M110M101}, \eqref{E_M100M101}, and~\eqref{E_M110M100}), we obtain the complete single-scattering squared-amplitude contribution to the first-order-in-opacity spectrum in Eq.~\eqref{spectum}, retaining terms through first-subeikonal order and linear order in $\mathbf{u}$
\begin{widetext}
\begin{align}
\frac{1}{d_T}\Tr\mean{|\mathcal{M}_1|^2}
{}&=
\frac{4g_s^2 C_2(T)N}{A_\perp d_A}
\int_0^L dz\,\rho(z)
\int \frac{d^2\mathbf{q}}{(2\pi)^2}\,
|v(\mathbf{q})|^2
|J(E,\mathbf{p})|^2
\Bigg\{
\Tr(c^2a_1^2)\,
\frac{\mathbf{k}^2}{(\mathbf{k}^2+\chi)^2}
\nonumber \\
\quad&
+2\alpha \, 
\frac{\mathbf{k}^2}{(\mathbf{k}^2+\chi)^2}
\left[
1-
\cos\left(
\frac{\mathbf{k}^2+\chi}{2xE}\,z
\right)
\right]
+
4 \alpha \, \frac{(\mathbf{k}-\mathbf{q})^2}
{\left[(\mathbf{k}-\mathbf{q})^2+\chi\right]^2}
\left[
1-
\cos\left(
\frac{(\mathbf{k}-\mathbf{q})^2+\chi}{2xE}\,z
\right)
\right]
\nonumber \\
\qquad&
-2\alpha \, \frac{\mathbf{k}\cdot(\mathbf{k}-\mathbf{q})}
{(\mathbf{k}^2+\chi)
\left[(\mathbf{k}-\mathbf{q})^2+\chi\right]}
\Bigg[
2
-\cos\left(
\frac{\mathbf{k}^2+\chi}{2xE}\,z
\right)
-2\cos\left(
\frac{(\mathbf{k}-\mathbf{q})^2+\chi}{2xE}\,z
\right)
+
\cos\left(
\frac{\mathbf{k}^2-(\mathbf{k}-\mathbf{q})^2}{2xE}\,z
\right)
\Bigg]
\nonumber \\
\qquad&
-8\alpha \, \frac{\mathbf{u}\cdot(\mathbf{k}-\mathbf{q})}{xE}\,
\frac{(\mathbf{k}-\mathbf{q})^2}
{\left[(\mathbf{k}-\mathbf{q})^2+\chi\right]^2}
\left[
1-
\cos\left(
\frac{(\mathbf{k}-\mathbf{q})^2+\chi}{2xE}\,z
\right)
\right]
\nonumber \\
\qquad&
+
\frac{2\alpha}{(\mathbf{k}-\mathbf{q})^2+\chi} \, \left[ \frac{\mathbf{u}\cdot(\mathbf{k}-\mathbf{q})}{xE} \,
\frac{\mathbf{k}^2}{\mathbf{k}^2+\chi}
+\frac{\mathbf{u}\cdot\mathbf{k}}{xE}\, \frac{\mathbf{q}\cdot(\mathbf{q}-\mathbf{k})}{\mathbf{k}^2+\chi}
\right]
\Bigg[
2
-\cos\left(
\frac{\mathbf{k}^2+\chi}{2xE}\,z
\right)
-2\cos\left(
\frac{(\mathbf{k}-\mathbf{q})^2+\chi}{2xE}\,z
\right)
\nonumber \\
&\qquad \hspace{12mm}
+
\cos\left(
\frac{\mathbf{k}^2-(\mathbf{k}-\mathbf{q})^2}{2xE}\,z
\right)
\Bigg]
\nonumber \\
\qquad&
-2\alpha \, \frac{\mathbf{u}\cdot\mathbf{q}}{xE}\,
\frac{
2\mathbf{k}^2-(\mathbf{k}-\mathbf{q})^2+\chi
}
{\left[(\mathbf{k}-\mathbf{q})^2+\chi\right]^2}
\frac{(\mathbf{k}-\mathbf{q})^2}{v^2(\mathbf{q})}\,
\frac{\partial v^2(\mathbf{q})}{\partial\mathbf{q}^2}
\left[
1-
\cos\left(
\frac{(\mathbf{k}-\mathbf{q})^2+\chi}{2xE}\,z
\right)
\right]
\nonumber \\
\qquad&
+
\frac{\alpha}{(\mathbf{k}-\mathbf{q})^2+\chi} \, \frac{\mathbf{u}\cdot\mathbf{q}}{x E}
\frac{\mathbf{k}\cdot(\mathbf{k}-\mathbf{q})}{v^2(\mathbf{q})}\,
\frac{\partial v^2(\mathbf{q})}{\partial\mathbf{q}^2}
\Bigg\{
4
-
\cos\left(
\frac{\mathbf{k}^2+\chi}{2xE}\,z
\right)
\nonumber \\
&\hspace{12mm}
-
2\frac{
2\mathbf{k}^2-(\mathbf{k}-\mathbf{q})^2
+\chi}{\mathbf{k}^2+\chi}
\cos\left(
\frac{(\mathbf{k}-\mathbf{q})^2+\chi}{2xE}\,z
\right)
+
\frac{
\mathbf{k}^2-(\mathbf{k}-\mathbf{q})^2
}{\mathbf{k}^2 +\chi}
\cos\left(
\frac{\mathbf{k}^2-(\mathbf{k}-\mathbf{q})^2}{2xE}\,z
\right)
\Bigg\}
\nonumber \\
\qquad&
+
\frac{2\alpha}{(\mathbf{k}-\mathbf{q})^2+\chi} \, \frac{\mathbf{u}\cdot\mathbf{q}}{xE} \, \frac{2x^2M^2}{(\mathbf{k}-\mathbf{q})^2+\chi}\,
\Bigg\{
4\frac{(\mathbf{k}-\mathbf{q})^2}
{(\mathbf{k}-\mathbf{q})^2+\chi}
\left[
1-
\cos\left(
\frac{(\mathbf{k}-\mathbf{q})^2+\chi}{2xE}\,z
\right)
\right]
\nonumber \\
&\hspace{12mm}
-\frac{\mathbf{k}\cdot(\mathbf{k}-\mathbf{q})}
{\mathbf{k}^2+\chi
}
\Bigg[
2 -\cos\left(
\frac{\mathbf{k}^2+\chi}{2xE}\,z
\right)
-2\cos\left(
\frac{(\mathbf{k}-\mathbf{q})^2+\chi}{2xE}\,z
\right)
+
\cos\left(
\frac{\mathbf{k}^2-(\mathbf{k}-\mathbf{q})^2}{2xE}\,z
\right)
\Bigg]
\Bigg\}
\nonumber \\
\qquad&
+
2\alpha \, \frac{\mathbf{u}\cdot\mathbf{q}}{xE}\,
\frac{
(\mathbf{k}-\mathbf{q})^2+m_g^2-x^2M^2
}{2xE}\,
z
\Bigg\{
2\frac{(\mathbf{k}-\mathbf{q})^2}
{\left[(\mathbf{k}-\mathbf{q})^2+\chi\right]^2}
\sin\left(
\frac{(\mathbf{k}-\mathbf{q})^2+\chi}{2xE}\,z
\right)
\nonumber \\
&\hspace{12mm}
-\frac{\mathbf{k}\cdot(\mathbf{k}-\mathbf{q})}
{(\mathbf{k}^2+\chi)
\left[(\mathbf{k}-\mathbf{q})^2+\chi\right]}
\Bigg[
2\sin\left(
\frac{(\mathbf{k}-\mathbf{q})^2+\chi}{2xE}\,z
\right)
+
\sin\left(
\frac{\mathbf{k}^2-(\mathbf{k}-\mathbf{q})^2}{2xE}\,z
\right)
\Bigg]
\Bigg\}
\Bigg\} \, .
\end{align}
\end{widetext}

\section{\label{sec:M220}Two-scattering contribution $\mathcal{M}_{2,2,0}$}

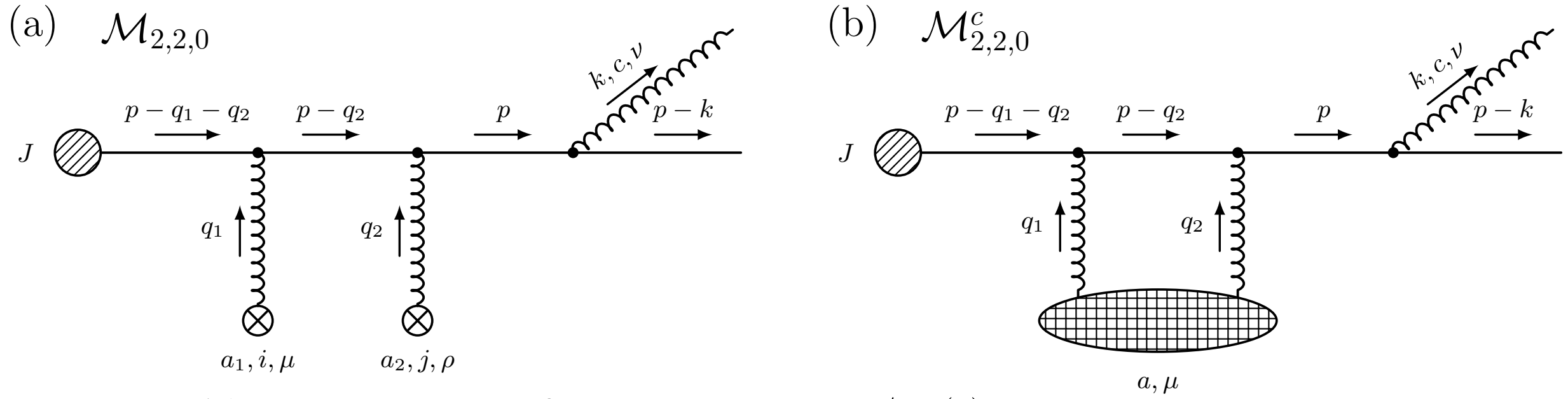
\begin{figure*}[t]
\centering
% ============================================================
% PANEL (a): non-CL
% ============================================================
\begin{minipage}[t]{0.43\textwidth}
\centering
\resizebox{\linewidth}{!}{%
\begin{tikzpicture}[
    line cap=round,
    line join=round,
    >=latex
]
\path[use as bounding box] (-0.8,-2.55) rectangle (8.25,1.65);
\node at (-0.55,1.55) {\Large (a)};
\node at (0.95,1.45) {\Large $\mathcal{M}_{2,2,0}$};
\draw[
    pattern=north east lines,
    pattern color=black,
    line width=0.8pt
] (0,0) circle (0.28);
\node[left=4pt] at (-0.28,0) {$J$};
\coordinate (V1) at (2.20,0.00);   
\coordinate (V2) at (4.15,0.00);   
\coordinate (V3) at (6.05,0.00);   
\coordinate (QinStart) at (0.28,0.00);
\coordinate (Qout)     at (8.10,0.00);
\coordinate (Gout)     at (8.00,1.50);
\coordinate (C1) at (2.20,-2.05);
\coordinate (C2) at (4.15,-2.05);
\coordinate (B1) at (2.20,-1.87);
\coordinate (B2) at (4.15,-1.87);
\draw[line width=0.85pt] (QinStart) -- (V1);
\draw[line width=0.85pt] (V1) -- (V2);
\draw[line width=0.85pt] (V2) -- (V3);
\draw[line width=0.85pt] (V3) -- (Qout);
\draw[
    line width=0.85pt,
    decorate,
    decoration={coil, amplitude=2.1pt, segment length=4.8pt}
] (V1) -- (B1);
\draw[
    line width=0.85pt,
    decorate,
    decoration={coil, amplitude=2.1pt, segment length=4.8pt}
] (V2) -- (B2);
\draw[
    line width=0.85pt,
    decorate,
    decoration={coil, amplitude=2.3pt, segment length=5.2pt}
] (V3) -- (Gout);
\fill[black] (V1) circle (1.9pt);
\fill[black] (V2) circle (1.9pt);
\fill[black] (V3) circle (1.9pt);
\draw[line width=0.8pt] (C1) circle (0.18);
\draw[line width=0.8pt] ($(C1)+(-0.12,-0.12)$) -- ($(C1)+(0.12,0.12)$);
\draw[line width=0.8pt] ($(C1)+(-0.12,0.12)$) -- ($(C1)+(0.12,-0.12)$);
\draw[line width=0.8pt] (C2) circle (0.18);
\draw[line width=0.8pt] ($(C2)+(-0.12,-0.12)$) -- ($(C2)+(0.12,0.12)$);
\draw[line width=0.8pt] ($(C2)+(-0.12,0.12)$) -- ($(C2)+(0.12,-0.12)$);
\node[below=7pt] at (C1) {$a_1,i, \mu$};
\node[below=7pt] at (C2) {$a_2,j, \rho$};
\draw[line width=0.75pt,->] (0.95,0.22) -- (1.75,0.22);
\node[above=1pt] at (1.35,0.22) {$p-q_1-q_2$};
\draw[line width=0.75pt,->] (2.75,0.22) -- (3.45,0.22);
\node[above=1pt] at (3.10,0.22) {$p-q_2$};
\draw[line width=0.75pt,->] (4.85,0.22) -- (5.55,0.22);
\node[above=1pt] at (5.20,0.22) {$p$};
\draw[line width=0.75pt,->] (7.05,0.22) -- (7.75,0.22);
\node[above=1pt] at (7.40,0.22) {$p-k$};
\draw[line width=0.75pt,->] (1.98,-1.26) -- (1.98,-0.64);
\node[left=2pt] at (1.98,-0.95) {$q_1$};
\draw[line width=0.75pt,->] (3.93,-1.26) -- (3.93,-0.64);
\node[left=2pt] at (3.93,-0.95) {$q_2$};

\draw[line width=0.75pt,->]
    (6.47,0.59) -- (7.09,1.07)
    node[midway, above=1pt, sloped] {$k,c,\nu$};
\end{tikzpicture}%
}
\end{minipage}\hspace{0.04\textwidth}
\begin{minipage}[t]{0.43\textwidth}
\centering
\resizebox{\linewidth}{!}{%
\begin{tikzpicture}[
    line cap=round,
    line join=round,
    >=latex
]
\path[use as bounding box] (-0.8,-2.55) rectangle (8.25,1.65);
\node at (-0.55,1.55) {\Large (b)};
\node at (0.95,1.45) {\Large $\mathcal{M}^{c}_{2,2,0}$};
\draw[
    pattern=north east lines,
    pattern color=black,
    line width=0.8pt
] (0,0) circle (0.28);
\node[left=4pt] at (-0.28,0) {$J$};
\coordinate (V1) at (2.20,0.00);
\coordinate (V2) at (4.15,0.00);
\coordinate (V3) at (6.05,0.00);
\coordinate (QinStart) at (0.28,0.00);
\coordinate (Qout)     at (8.10,0.00);
\coordinate (Gout)     at (8.00,1.50);
\coordinate (B1) at (2.20,-1.77);
\coordinate (B2) at (4.15,-1.77);
\draw[line width=0.85pt] (QinStart) -- (V1);
\draw[line width=0.85pt] (V1) -- (V2);
\draw[line width=0.85pt] (V2) -- (V3);
\draw[line width=0.85pt] (V3) -- (Qout);
\draw[
    line width=0.85pt,
    decorate,
    decoration={coil, amplitude=2.1pt, segment length=4.8pt}
] (V1) -- (B1);
\draw[
    line width=0.85pt,
    decorate,
    decoration={coil, amplitude=2.1pt, segment length=4.8pt}
] (V2) -- (B2);
\draw[
    line width=0.85pt,
    decorate,
    decoration={coil, amplitude=2.3pt, segment length=5.2pt}
] (V3) -- (Gout);
\fill[black] (V1) circle (1.9pt);
\fill[black] (V2) circle (1.9pt);
\fill[black] (V3) circle (1.9pt);
\draw[
    line width=0.8pt,
    pattern=grid,
    pattern color=black
] (3.175,-2.05) ellipse (1.45 and 0.38);
\node[below=7pt] at (3.175,-2.38) {$a,\mu$};
\draw[line width=0.75pt,->] (0.95,0.22) -- (1.75,0.22);
\node[above=1pt] at (1.35,0.22) {$p-q_1-q_2$};
\draw[line width=0.75pt,->] (2.75,0.22) -- (3.45,0.22);
\node[above=1pt] at (3.10,0.22) {$p-q_2$};
\draw[line width=0.75pt,->] (4.85,0.22) -- (5.55,0.22);
\node[above=1pt] at (5.20,0.22) {$p$};
\draw[line width=0.75pt,->] (7.05,0.22) -- (7.75,0.22);
\node[above=1pt] at (7.40,0.22) {$p-k$};
\draw[line width=0.75pt,->] (1.98,-1.20) -- (1.98,-0.58);
\node[left=2pt] at (1.98,-0.89) {$q_1$};
\draw[line width=0.75pt,->] (3.93,-1.20) -- (3.93,-0.58);
\node[left=2pt] at (3.93,-0.89) {$q_2$};
\draw[line width=0.75pt,->]
    (6.47,0.59) -- (7.09,1.07)
    node[midway, above=1pt, sloped] {$k,c,\nu$};
\end{tikzpicture}%
}
\end{minipage}
\caption{
Diagram $\mathcal{M}_{2,2,0}$ contributing at first order in opacity  $L/{\lambda}$:
(a) two successive interactions with the medium before gluon emission and
(b) the corresponding contact-limit configuration.
}
\label{fig:M220}
\end{figure*}

We next consider the eight two-scattering diagrams contributing at first order in opacity through their interference with the zeroth-order amplitude $\mathcal{M}_0$, as specified in Eq.~\eqref{spectum}. We begin with $\mathcal{M}_{2,2,0}$, shown in Fig.~\ref{fig:M220}, in which the heavy quark undergoes two successive interactions with the medium, exchanging momenta $q_1$ and $q_2$, before emitting a gluon with momentum $k$ and emerging with momentum $p-k$.

Using the four-momentum conventions in Eq.~\eqref{vektori}, the propagators and interaction vertices specified in Appendix~\ref{sec:FeynmanRules}, the conventions of Sec.~\ref{sec:Assumptions}, and the momentum assignments in Fig.~\ref{fig:M220}, we obtain

\begin{widetext}
\begin{align}~\label{m220_p}
\mathcal{M}_{2, 2, 0}={}& \int{\frac{d^4q_1}{(2\pi)^4}\int \frac{d^4q_2}{(2\pi)^4}}[i g_s (2p - k)^{\nu} \varepsilon^{*}_{\nu}(k) T^c_{jet}] \, \frac{i}{p^2 -M^2 +i \epsilon} \,  [i g_s (2p-q_2)_{\rho} A^{\rho, a_2}_{ext}(q_2) T^{a_2}_{jet}] \, \frac{i}{(p-q_2)^2 -M^2 +i \epsilon} \nonumber \\
\times&  [i g_s (2p-q_1 -2 q_2)_{\mu} A^{\mu, a_1}_{ext}(q_1) T^{a_1}_{jet}]\, \frac{i}{(p-q_1-q_2)^2 -M^2 +i \epsilon} \, i J(p-q_1-q_2) \nonumber \\
\approx& \int \frac{d^2 \mathbf{q}_{1}}{(2\pi)^2}
  \int   \frac{d^2 \mathbf{q}_{2}}{(2\pi)^2}   \int \frac{d {q}_{1z}}{2\pi}  \int \frac{d {q}_{2z}}{2\pi}
\sum_{i,j}
e^{-i\mathbf{q}_{1}\cdot\mathbf{x}_{i}}
e^{-i\mathbf{q}_{2}\cdot\mathbf{x}_{j}}
e^{-iq_{1z}z_i}
e^{-iq_{2z}z_j}
(-i) g_s J(p)
c a_2 a_1
%T^c_{jet} T_{jet}^{a_2} T_{jet}^{a_1} 
%\nonumber\\
%&\times
T_{j}^{a_2} T_{i}^{a_1}
v_i(q_1) \, v_j(q_2)
\nonumber\\
&\times
\left(2\varepsilon^{*}(k)\!\cdot p\right)
\left(2u_j\!\cdot p\right)
\left[2u_i\!\cdot\left(p-q_2\right)\right]
\frac{1}{p^2-M^2+i\epsilon}
\frac{1}{(p-q_2)^2-M^2+i\epsilon}
\frac{1}{(p-q_1-q_2)^2-M^2+i\epsilon} \, .
\end{align}
\end{widetext}
The reduced expression follows from inserting the external field in Eqs.~\eqref{Aext} and~\eqref{a} and integrating over $q_1^0$ and $q_2^0$, which fixes $q_\ell^0=\mathbf{u}_\ell\cdot\mathbf{q}_\ell$ ($\ell=1,2$), according to Eq.~\eqref{flowu}. The vertex contractions are simplified using the polarization conditions in Eq.~\eqref{kin0M1} and $u_\ell\cdot q_\ell=0$, together with the broad-source approximation of Eq.~\eqref{eq:broadsource}.

The corresponding interference with $\mathcal{M}_0^\dagger$ in Eq.~\eqref{M0final} takes the form
\begin{widetext}
\begin{align}
\label{m220_interference}
\frac{2}{d_T}\,
\mathrm{Re}\,
\mathrm{Tr}
\left\langle
\mathcal{M}_0^{\dagger} \mathcal{M}_{2,2,0}
\right\rangle
{}&=
\frac{2}{d_T}\,
\mathrm{Re}\,
\mathrm{Tr}\,
\Bigl\langle
\Bigl\{
\sum_{i,j}
\int
\frac{d^2\mathbf{q}_1}{(2\pi)^2}
\int \frac{d^2\mathbf{q}_2}{(2\pi)^2}
\, g_s^2 |J(p)|^2
\int\frac{dq_{1z}}{2\pi}
\int\frac{dq_{2z}}{2\pi}
c^2 a_2 a_1
e^{-i\mathbf{q}_1\cdot\mathbf{x}_i}
e^{-i\mathbf{q}_2\cdot\mathbf{x}_j}
\nonumber\\
&\times
e^{-iq_{1z}z_i}
e^{-iq_{2z}z_j}
T_j^{a_2}T_i^{a_1}\,
\,v_i(q_1)\,v_j(q_2)
\left(2u_j\!\cdot p\right)
\left[2u_i\!\cdot\left(p-q_2\right)\right]
\left(2\varepsilon(k)\!\cdot p\right) \left(2\varepsilon^{*}(k)\!\cdot p\right)
\nonumber\\
&\times
\frac{1}{(p^2-M^2-i\epsilon)}
\frac{1}{(p^2-M^2+i\epsilon)}
\frac{1}{(p-q_2)^2-M^2+i\epsilon}
\frac{1}{(p-q_1-q_2)^2-M^2+i\epsilon}
\Bigr\}
\Bigr\rangle \, .
\end{align}
\end{widetext}
The target-color trace in Eq.~\eqref{tr2} enforces $i=j$ and $a_1=a_2$, restricting both interactions to the same scattering center. Consequently, only the contact-limit (CL) configurations contribute at first order in opacity. This restriction applies to all eight two-scattering diagrams considered in the following sections.

Applying Eqs.~\eqref{eq:density} and~\eqref{eq:continuum}, followed by Eq.~\eqref{app9} with the upper sign, gives 
\begin{equation}\label{q1q2}
\mathbf{q}\equiv \mathbf{q}_1=-\mathbf{q}_2 \, ,
\end{equation}
where we introduced $\mathbf{q}$, a notation adopted throughout the subsequent $\mathcal{M}_2$ sections. Thus we obtain
\begin{widetext}
\begin{align}
\frac{2}{d_T}
\operatorname{Re}\operatorname{Tr}
\left\langle
\mathcal{M}^\dagger_0{\mathcal{M}}^{c}_{2,2,0}
\right\rangle
={}&
\frac{2 g_s^2 N C_2(T)}{A_\perp d_A}\operatorname{Tr}(c^2 a^2_1)\operatorname{Re}
\Bigg\{
\int \frac{d^2\mathbf{q}}{(2\pi)^2}
\int  dz \rho(z)\,\,\lvert J(E, \mathbf{p})\rvert^2
\int \frac{dq_{1z}}{2\pi}
\int \frac{dq_{2z}}{2\pi}\,
e^{-iq_{1z}z}
e^{-iq_{2z}z}
 \nonumber\\
&\times
v(q_1) \, v(q_2)
\frac{
\left(2\varepsilon(k)\cdot p\right)
\left(2\varepsilon^{*}(k)\cdot p\right)
\left(2u\!\cdot p\right)^2}
{(p^2-M^2-i\epsilon)(p^2-M^2+i\epsilon)
\left[(p-q_2)^2-M^2+i\epsilon\right]
\left[(p-q_1-q_2)^2-M^2+i\epsilon\right]}
\Bigg\},
\label{eq:M0M220}
\end{align}
\end{widetext}
where the arguments of the two scattering potentials are understood to be
\begin{align}~\label{kin1M2}
q_1=(\boldsymbol{u}\cdot \boldsymbol{q}, \, \boldsymbol{q}, \, q_{1z}),
\end{align}
\begin{align}~\label{kin2M2}
q_2=(-\boldsymbol{u}\cdot \boldsymbol{q}, \,- \boldsymbol{q}, \, q_{2z}).
\end{align}

We first perform the longitudinal integration over $q_{1z}$,
\begin{align}~\label{IM220p}
\mathcal{I}_{2,2,0}= \int{\frac{dq_{1z}}{2\pi}}\frac{e^{-iq_{1z} z}\, v(q_{1z},{\mathbf{q}})}{(p-q_1-q_2)^2 -M^2+i\epsilon}.
\end{align}
Closing the $q_{1z}$ contour in the lower half-plane gives an additional minus sign from the clockwise orientation. Since the scattering-potential poles ($q^-_{1z, v} = -i R$, Eq.~\eqref{R}) are exponentially suppressed by Eq.~\eqref{eq:separation}, only the heavy-quark propagator pole contributes,
\begin{align}~\label{m220_pol}
q^-_{1z,p-q_1-q_2}= {} & -\frac{{\mathbf{k}}^2 +\chi}{2 x E} -q_{2z}-i\epsilon.
\end{align}
At the pole accuracy specified in Eq.~\eqref{deltaQ}, no $\mathcal{O}\!\left(\perp^3/(xE)^2\right)$ correction arises in Eq.~\eqref{m220_pol}. The corresponding residue gives
\begin{align}~\label{IM220}
\mathcal{I}_{2,2,0}\approx \frac{i g^2_s e^{i \, \frac{\mathbf{k}^2 +\chi}{2 x E} z}  e^{i q_{2z} z}}{2E \left[ \left(\frac{\mathbf{k}^2 +\chi}{2 x E} + q_{2z} \right)^2 + R^2 \right]}.
\end{align}

We next evaluate the remaining longitudinal momentum-transfer integral over $q_{2z}$:
\begin{align}\label{JM220}
\mathcal{J}_{2,2,0} =& {} \int{\frac{dq_{2z}}{2\pi}}
\frac{e^{-iq_{2z} z} \, v(q_{2z},{-\mathbf{q}})}
{(p-q_2)^2 - M^2 +i\epsilon}
\, \mathcal{I}_{2,2,0} \nonumber \\
=& \frac{i g^2_s e^{i \, \frac{\mathbf{k}^2 +\chi}{2 x E} z}}{2E } \int{\frac{dq_{2z}}{2\pi}}
\frac{v(q_{2z},{-\mathbf{q}})}
{(p-q_2)^2 - M^2 +i\epsilon} \nonumber \\
\times&
\frac{1}{\left[ \left(\frac{\mathbf{k}^2 +\chi}{2 x E} + q_{2z} \right)^2 + R^2 \right]}.
\end{align}
The factor $e^{iq_{2z}z}$ generated by the $q_{1z}$ residue in Eq.~\eqref{IM220} cancels the remaining $e^{-iq_{2z}z}$ factor. Consequently, the $q_{2z}$ integration in Eq.~\eqref{JM220} carries no $q_{2z}$-dependent phase, while the scattering-potential poles must still be retained, as discussed following Eq.~\eqref{eq:separation}.  Thus, the phase-induced $z\sin$ mechanism associated with Eq.~\eqref{eq:cosphaseexpansion} is absent for $\mathcal{M}_{2,2,0}$. 

The poles contributing to $\mathcal{J}_{2,2,0}$ are
\begin{align}\label{q2pq2}
q^-_{2z,p-q_2}= -\mathbf{u} \cdot \mathbf{q} - \frac{\mathbf{k}^2 +\chi}{2 x E} -\frac{\mathbf{u} \cdot \mathbf{q}}{x E} \frac{x^2 M^2}{2 x E} - i \epsilon \, ,
\end{align}
\begin{align}\label{q2v1}
q^-_{2z,v_1}= - \frac{\mathbf{k}^2 +\chi}{2 x E}  - i R \, ,
\end{align}
\begin{align}\label{q2v2}
q^-_{2z,v_2}= - i R \, ,
\end{align}
where the pole in Eq.~\eqref{q2pq2} originates from the corresponding heavy-quark propagator, while the poles in Eqs.~\eqref{q2v1} and~\eqref{q2v2} arise from the corresponding scattering potentials.

Unlike the propagator-pole difference in Eq.~\eqref{razlikapolova}, the separations between the heavy-quark and scattering-potential poles remain of order $\perp$. Consequently, the $\mathcal{O}\!\left(\perp^3/(xE)^2\right)$ correction (specified in Eq.~\eqref{deltaQ}) in Eq.~\eqref{q2pq2} is not promoted to first-subeikonal order in the residue denominators. More generally, throughout our calculation, corrections of this order to the propagator-pole positions do not contribute to the regular prefactors at the working accuracy, but can become relevant through small pole differences in residue denominators or through formation-phase shifts. Among the remaining two-scattering amplitudes, this mechanism contributes only to $\mathcal{M}_{2,0,1}$ and $\mathcal{M}_{2,0,2}$, as discussed in Sec.~\ref{FullSpectrumClassification}.

Closing the contour in the lower half-plane, as in the preceding longitudinal integrations, we sum the corresponding residues. The same prescription is applied to the subsequent diagrams. The Lorentz contractions in Eq.~\eqref{eq:M0M220} are evaluated using Eq.~\eqref{pM} and the momentum, polarization, and flow conventions in Eqs.~\eqref{vektori} and~\eqref{flowu}. Retaining the subleading propagator-pole corrections prescribed in Sec.~\ref{sec:Assumptions}, we obtain, through first-subeikonal order and linear order in $\mathbf{u}$,
\begin{widetext}
\begin{align}
\frac{2}{d_T}
\operatorname{Re}\operatorname{Tr}
\left\langle
{\mathcal{M}}^\dagger_0
\mathcal{M}^{c}_{2,2,0}
\right\rangle
={}&
-\frac{4 g^2_s  C_2(T) N}{ A_\perp d_A}\operatorname{Tr}(c^2 a_1^2)\,
\int_0^L dz\,\rho(z)
\int \frac{d^2\mathbf{q}}{(2\pi)^2}\,
\left|v(\mathbf{q})\right|^2
\left|J(E, \mathbf{p})\right|^2
\frac{\mathbf{k}^2}
{\left(\mathbf{k}^2+\chi\right)^2}
\nonumber\\
&\hspace{-1.5cm}\times
\Bigg\{ \left[ 1+  \frac{\mathbf{u} \cdot \mathbf{q}}{2 x E}\, \frac{\mathbf{k}^2 +\chi}{v^2(\mathbf{q})} \, \frac{\partial v^2( \mathbf{q})}{\partial  \mathbf{q}^2}   \right]  
\cos\left(
\frac{\mathbf{k}^2+\chi}{2xE}\,z
\right)
+\frac{3}{4}\, \left[ \mathbf{u} \cdot \mathbf{q} + \frac{\mathbf{k}^2+\chi}{4xE} \right]
\frac{R_0}{v^2(\mathbf{q})}\,
\frac{\partial v^2(\mathbf{q})}{\partial \mathbf{q}^2}\,
\sin\left(
\frac{\mathbf{k}^2+\chi}{2xE}\,z
\right)
\Bigg\} \nonumber \\
\approx&
-\frac{4 g^2_s  C_2(T) N}{ A_\perp d_A}\operatorname{Tr}(c^2 a_1^2)\,
\int_0^L dz\,\rho(z)
\int \frac{d^2\mathbf{q}}{(2\pi)^2}\,
\left|v(\mathbf{q})\right|^2
\left|J(E, \mathbf{p})\right|^2
\frac{\mathbf{k}^2}
{\left(\mathbf{k}^2+\chi\right)^2}
\nonumber\\
\times&
\Bigg\{
\cos\left(
\frac{\mathbf{k}^2+\chi}{2xE}\,z
\right)
+\frac{3}{16}\,
\frac{R_0}{xE}\,
\frac{\left(\mathbf{k}^2+\chi \right)}{v^2(\mathbf{q})}\,
\frac{\partial v^2(\mathbf{q})}{\partial \mathbf{q}^2}\,
\sin\left(
\frac{\mathbf{k}^2+\chi}{2xE}\,z
\right)
\Bigg\} 
\label{eq:M0M220final} \, .
\end{align}
\end{widetext}
In the second equality of Eq.~\eqref{eq:M0M220final}, terms odd under $\mathbf{q}\to-\mathbf{q}$ have been omitted, since they vanish upon integration over $\mathbf{q}$. The same simplification is applied to the remaining two-scattering contributions. To linear order in $\mathbf{u}$, consistently with Eq.~\eqref{eq:flowexpansion}, the screening scale reduces to
\begin{equation}
 R^2_0=\mathbf{q}^2 +\mu^2 \, . 
\label{R0}
\end{equation}

The scattering-potential-pole sine contribution in Eq.~\eqref{eq:M0M220final} originates from the residues of the scattering-potential poles.

As a consistency check, to linear order in $\mathbf{u}$, Eq.~\eqref{eq:M0M220final} reduces in the massless limit, $M=m_g=0$, to Eq.~(33) of Ref.~\cite{KML}.

\section{\label{sec:M203}Two-scattering contribution $\mathcal{M}_{2,0,3}$}

\begin{figure*}[!t]
\centering

% ============================================================
% PANEL (a): M_{2,0,3}, non-CL
% ============================================================
\begin{minipage}[t]{0.40\textwidth}
\centering
\resizebox{\linewidth}{!}{%
\begin{tikzpicture}[
    line cap=round,
    line join=round,
    >=latex
]

\path[use as bounding box] (-0.8,-2.65) rectangle (8.25,1.65);

\node at (-0.55,1.55) {\Large (a)};
\node at (0.95,1.45) {\Large $\mathcal{M}_{2,0,3}$};

\draw[
    pattern=north east lines,
    pattern color=black,
    line width=0.8pt
] (0,0) circle (0.28);

\node[left=4pt] at (-0.28,0) {$J$};

\coordinate (V1) at (2.20,0.00);

\coordinate (QinStart) at (0.28,0.00);
\coordinate (Qout)     at (8.10,0.00);

\coordinate (G1)   at (4.15,-0.95);
\coordinate (G2)   at (6.05,-0.95);
\coordinate (Gout) at (8.10,-0.95);

\coordinate (C1) at (4.15,-2.20);
\coordinate (C2) at (6.05,-2.20);

\coordinate (B1) at (4.15,-2.02);
\coordinate (B2) at (6.05,-2.02);

\draw[line width=0.85pt] (QinStart) -- (V1);
\draw[line width=0.85pt] (V1) -- (Qout);

\draw[
    line width=0.85pt,
    decorate,
    decoration={
        coil,
        amplitude=2.3pt,
        segment length=5.2pt
    }
] (V1) -- (G1);

\draw[
    line width=0.85pt,
    decorate,
    decoration={
        coil,
        amplitude=2.3pt,
        segment length=5.2pt
    }
] (G1) -- (G2);

\draw[
    line width=0.85pt,
    decorate,
    decoration={
        coil,
        amplitude=2.3pt,
        segment length=5.2pt
    }
] (G2) -- (Gout);

\draw[
    line width=0.85pt,
    decorate,
    decoration={
        coil,
        amplitude=2.1pt,
        segment length=4.8pt
    }
] (G1) -- (B1);

\draw[
    line width=0.85pt,
    decorate,
    decoration={
        coil,
        amplitude=2.1pt,
        segment length=4.8pt
    }
] (G2) -- (B2);

\fill[black] (V1) circle (1.9pt);
\fill[black] (G1) circle (1.9pt);
\fill[black] (G2) circle (1.9pt);

\draw[line width=0.8pt] (C1) circle (0.18);
\draw[line width=0.8pt]
    ($(C1)+(-0.12,-0.12)$) --
    ($(C1)+(0.12,0.12)$);
\draw[line width=0.8pt]
    ($(C1)+(-0.12,0.12)$) --
    ($(C1)+(0.12,-0.12)$);

\draw[line width=0.8pt] (C2) circle (0.18);
\draw[line width=0.8pt]
    ($(C2)+(-0.12,-0.12)$) --
    ($(C2)+(0.12,0.12)$);
\draw[line width=0.8pt]
    ($(C2)+(-0.12,0.12)$) --
    ($(C2)+(0.12,-0.12)$);

\node[below=7pt] at (C1) {$a_1,i,\mu$};
\node[below=7pt] at (C2) {$a_2,j,\rho$};

\draw[line width=0.75pt,->] (0.95,0.22) -- (1.75,0.22);
\node[above=1pt] at (1.35,0.22) {$p-q_1-q_2$};

\draw[line width=0.75pt,->] (5.95,0.22) -- (6.75,0.22);
\node[above=1pt] at (6.35,0.22) {$p-k$};

\draw[line width=0.75pt,->]
    (2.53,-0.38) -- (3.20,-0.72)
    node[midway, below=0.9pt, sloped] {$k-q_1-q_2$};

\draw[line width=0.75pt,->] (4.72,-0.70) -- (5.47,-0.70);
\node[above=1pt] at (5.095,-0.70) {$k-q_2$};

\draw[line width=0.75pt,->] (6.72,-1.17) -- (7.47,-1.17);
\node[below=1pt] at (7.095,-1.17) {$k,c,\nu$};

\draw[line width=0.75pt,->] (3.93,-1.82) -- (3.93,-1.20);
\node[left=2pt] at (3.93,-1.60) {$q_1$};

\draw[line width=0.75pt,->] (5.83,-1.82) -- (5.83,-1.20);
\node[left=2pt] at (5.83,-1.60) {$q_2$};

\end{tikzpicture}%
}
\end{minipage}%
\hspace{0.035\textwidth}%
%
% ============================================================
% PANEL (b): M^c_{2,0,3}, CL
% ============================================================
\begin{minipage}[t]{0.40\textwidth}
\centering
\resizebox{\linewidth}{!}{%
\begin{tikzpicture}[
    line cap=round,
    line join=round,
    >=latex
]

\path[use as bounding box] (-0.8,-2.65) rectangle (8.25,1.65);

\node at (-0.55,1.55) {\Large (b)};
\node at (0.95,1.45) {\Large $\mathcal{M}^{c}_{2,0,3}$};

\draw[
    pattern=north east lines,
    pattern color=black,
    line width=0.8pt
] (0,0) circle (0.28);

\node[left=4pt] at (-0.28,0) {$J$};

\coordinate (V1) at (2.20,0.00);

\coordinate (QinStart) at (0.28,0.00);
\coordinate (Qout)     at (8.10,0.00);

\coordinate (G1)   at (4.15,-0.95);
\coordinate (G2)   at (6.05,-0.95);
\coordinate (Gout) at (8.10,-0.95);

\coordinate (B1) at (4.15,-1.991);
\coordinate (B2) at (6.05,-1.991);

%...

\draw[
    line width=0.8pt,
    pattern=grid,
    pattern color=black
] (5.10,-2.18) ellipse (1.45 and 0.25);

\node at (5.10,-2.75) {$a,\mu$};

\draw[line width=0.85pt] (QinStart) -- (V1);
\draw[line width=0.85pt] (V1) -- (Qout);

\draw[
    line width=0.85pt,
    decorate,
    decoration={
        coil,
        amplitude=2.3pt,
        segment length=5.2pt
    }
] (V1) -- (G1);

\draw[
    line width=0.85pt,
    decorate,
    decoration={
        coil,
        amplitude=2.3pt,
        segment length=5.2pt
    }
] (G1) -- (G2);

\draw[
    line width=0.85pt,
    decorate,
    decoration={
        coil,
        amplitude=2.3pt,
        segment length=5.2pt
    }
] (G2) -- (Gout);

\draw[
    line width=0.85pt,
    decorate,
    decoration={
        coil,
        amplitude=2.1pt,
        segment length=4.8pt
    }
] (G1) -- (B1);

\draw[
    line width=0.85pt,
    decorate,
    decoration={
        coil,
        amplitude=2.1pt,
        segment length=4.8pt
    }
] (G2) -- (B2);

\fill[black] (V1) circle (1.9pt);
\fill[black] (G1) circle (1.9pt);
\fill[black] (G2) circle (1.9pt);

\draw[
    line width=0.8pt,
    pattern=grid,
    pattern color=black
] (5.10,-2.18) ellipse (1.45 and 0.25);

\draw[line width=0.75pt,->] (0.95,0.22) -- (1.75,0.22);
\node[above=1pt] at (1.35,0.22) {$p-q_1-q_2$};

\draw[line width=0.75pt,->] (5.95,0.22) -- (6.75,0.22);
\node[above=1pt] at (6.35,0.22) {$p-k$};

\draw[line width=0.75pt,->]
    (2.53,-0.38) -- (3.20,-0.72)
    node[midway, below=0.9pt, sloped] {$k-q_1-q_2$};

\draw[line width=0.75pt,->] (4.72,-0.70) -- (5.47,-0.70);
\node[above=1pt] at (5.095,-0.70) {$k-q_2$};

\draw[line width=0.75pt,->] (6.72,-1.17) -- (7.47,-1.17);
\node[below=1pt] at (7.095,-1.17) {$k,c,\nu$};

\draw[line width=0.75pt,->] (3.93,-1.82) -- (3.93,-1.20);
\node[left=2pt] at (3.93,-1.60) {$q_1$};

\draw[line width=0.75pt,->] (5.83,-1.82) -- (5.83,-1.20);
\node[left=2pt] at (5.83,-1.60) {$q_2$};

\end{tikzpicture}%
}
\end{minipage}
\caption{
Diagram $\mathcal{M}_{2,0,3}$ contributing at first order in opacity  $L/{\lambda}$:
(a) two successive interactions of the emitted gluon with the medium and
(b) the corresponding contact-limit configuration.
}

\label{fig:M203}

\end{figure*}
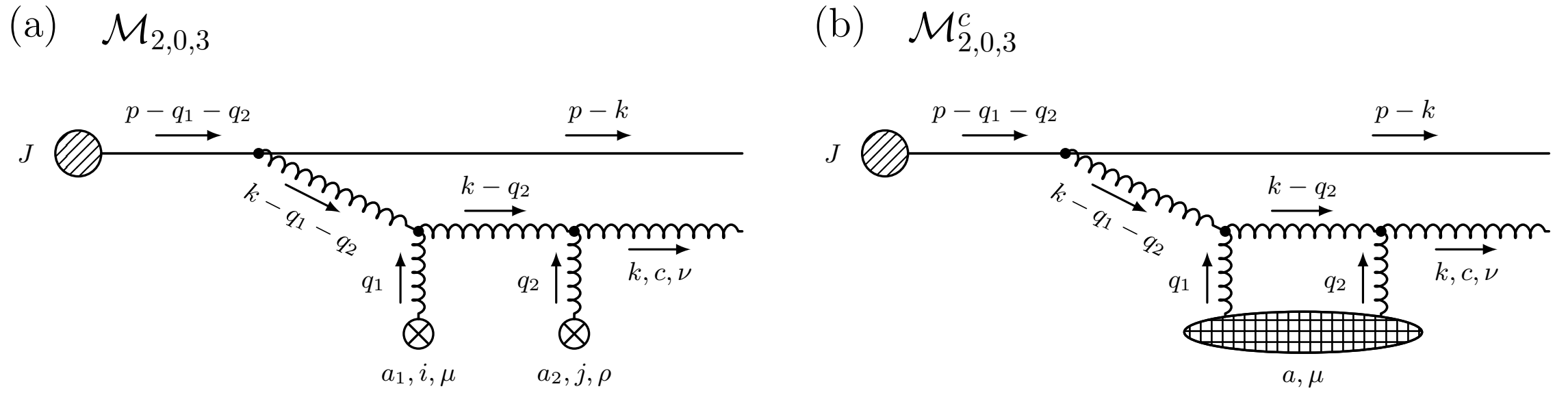

We next consider the two-scattering amplitude $\mathcal{M}_{2,0,3}$, shown in Fig.~\ref{fig:M203}. In this diagram, the heavy quark first emits a gluon, which subsequently undergoes two successive interactions with the medium, with momentum transfers $q_1$ and $q_2$, before emerging with momentum $k$, while the outgoing quark carries momentum $p-k$. 
Using the momentum labeling in Fig.~\ref{fig:M203} and the conventions introduced above, we obtain
\begin{widetext}
\begin{align}
\mathcal{M}_{2,0,3}
={}&
\int \frac{d^4q_1}{(2\pi)^4}
\int \frac{d^4q_2}{(2\pi)^4}
\varepsilon^{*\nu}(k) \,
\Gamma_{\xi'\nu\rho}^{d' c a_2}(k-q_2, -k, q_2) \,
A_{ext}^{\rho, a_2}(q_2) \,
\frac{(-i)P^{\xi\xi'}(k-q_2)\delta_{dd'}}
{(k-q_2)^2-m_g^2+i\epsilon} \,
\Gamma_{\delta'\xi\mu}^{a'da_1} (k-q_1-q_2, -k+q_2, q_1) \,
\nonumber \\
&\times
A_{ext}^{\mu, a_1}(q_1) \,
\frac{(-i)P^{\delta\delta'}(k-q_1-q_2)\delta_{aa'}}
{(k-q_1-q_2)^2-m_g^2+i\epsilon}
\left[
i g_s(2p-k-q_1-q_2)_\delta T_{jet}^{a}
\right]
\frac{i}{(p-q_1-q_2)^2-M^2+i\epsilon} \,
iJ(p-q_1-q_2)
\nonumber \\
\approx{}&
i g_s J(p)
\int \frac{d^2\mathbf{q}_1}{(2\pi)^2}
\int \frac{d^2\mathbf{q}_2}{(2\pi)^2}
\int \frac{dq{_{1z}}}{2\pi}
\int \frac{dq_{2z}}{2\pi}
\sum_{i,j}
e^{-i\mathbf{q}_1\cdot\mathbf{x}_i}
e^{-i\mathbf{q}_2\cdot\mathbf{x}_j}
e^{-iq_{1z}z_i}
e^{-iq_{2z}z_j}
T_j^{a_2}T_i^{a_1} a\, 
v_j(q_2) v_i(q_1)
\nonumber \\
&\times
f^{d c a_2} \, f^{a d a_1} \, I_{2,0,3}\, 
\frac{1}{(p-q_1-q_2)^2-M^2+i\epsilon} \, 
\frac{1}{(k-q_1-q_2)^2-m_g^2+i\epsilon} \, 
\frac{1}{(k-q_2)^2-m_g^2+i\epsilon} \, ,
\label{m203_p}
\end{align}
\end{widetext}
where 
\begin{widetext}
\begin{align}\label{I203}
I_{2,0,3}& {}= \varepsilon^{*\nu}(k)
\left[
g_{\xi'\rho}(k-2q_2)_\nu
+g_{\xi'\nu}(-2k+q_2)_\rho
+g_{\nu\rho}(k+q_2)_{\xi'}
\right]
u_j^\rho \,
P^{\xi\xi'}(k-q_2)
\nonumber \\
&\times
\left[
g_{\delta'\mu}(k-2q_1-q_2)_\xi
+g_{\delta'\xi}(-2k+2q_2+q_1)_\mu
+g_{\xi\mu}(k+q_1-q_2)_{\delta'}
\right]
u_i^\mu 
P^{\delta\delta'}(k-q_1-q_2)
(2p-k-q_1-q_2)_\delta \, .
\end{align}
\end{widetext}
The color factor $f^{dca_2}f^{ada_1}T_j^{a_2}T_i^{a_1}$ is evaluated separately using the target-color trace in Eq.~\eqref{tr2}, the antisymmetry of the structure constants, and the commutator and  quark-color trace identities in Eqs.~\eqref{gen1} and~\eqref{trjet}. Upon interference with $\mathcal{M}_0^\dagger$, the resulting quark-color factor is $-2\alpha$.

Interfering $\mathcal{M}_{2,0,3}$ with $\mathcal{M}_0^\dagger$ from Eq.~\eqref{M0final} and applying the same CL color and medium averaging procedure as in Sec.~\ref{sec:M220}, with the upper-sign case of Eq.~\eqref{app9} reproducing Eq.~\eqref{q1q2}, we find
\begin{widetext}
\begin{align}
\frac{2}{d_T}
\operatorname{Re}\operatorname{Tr}
\left\langle
\mathcal{M}_0^\dagger \mathcal{M}^c_{2,0,3}
\right\rangle
={}&
\frac{8 g_s^2 C_2(T) N}{A_\perp d_A} \, \alpha
\operatorname{Re}
\Bigg\{
\int \frac{d^2\mathbf{q}}{(2\pi)^2}
\int_0^L dz \rho(z) \left|J(E, \mathbf{p})\right|^2
\int \frac{dq_{1z}}{2\pi}
\int \frac{dq_{2z}}{2\pi}
e^{-iq_{1z}z}
e^{-iq_{2z}z}
v(q_1) v(q_2) 
\nonumber \\
&\qquad\times
I_{2,0,3} \frac{\boldsymbol{\varepsilon} \cdot \mathbf{k}}{\mathbf{k}^2 + \chi}
\frac{1}
{(p-q_1-q_2)^2-M^2+i\epsilon}
\frac{1}
{(k-q_1-q_2)^2-m_g^2+i\epsilon}
\frac{1}
{(k-q_2)^2-m_g^2+i\epsilon}
\Bigg\} \, ,
\label{eq:M0M203}
\end{align}
\end{widetext}
where $q_1$ and $q_2$ acqiure the form given by Eqs.~\eqref{kin1M2} and~\eqref{kin2M2}, respectively.

As in the single-scattering case discussed in Sec.~\ref{sec:M101}, the Lorentz contraction in Eq.~\eqref{I203} introduces no additional longitudinal poles. Although the three-gluon vertices themselves contain the full four-momenta, all their Lorentz indices are contracted with the polarization vector, the medium four-velocities, and the gluon-propagator projectors, which are transverse to the axial vector \(n^\mu\). Upon performing these contractions, all explicit dependence on $q_{1z}$ and $q_{2z}$ vanishes from $I_{2,0,3}$, leaving the longitudinal pole structure determined solely by the propagator denominators and scattering potentials. After a lengthy algebraic reduction involving both three-gluon vertices and the corresponding projector structures, Eq.~\eqref{I203} takes the remarkably simple form
\begin{equation}
I_{2,0,3}=8 x E^2 \boldsymbol{\varepsilon} \cdot \mathbf{k} \left( 1- 2 \frac{\mathbf{u} \cdot \mathbf{k}}{x E} \right) \, .
\label{Izr203}
\end{equation}

Following the procedure of Sec.~\ref{sec:M220}, we first evaluate
\begin{widetext}
\begin{align}~\label{IM203p}
\mathcal{I}_{2,0,3}= \int{\frac{dq_{1z}}{2\pi}}\frac{e^{-iq_{1z} z}\, v(q_{1z},{\mathbf{q}})}{(p-q_1-q_2)^2 -M^2+i\epsilon} \frac{1}{(k-q_1-q_2)^2 -m^2_g+i\epsilon} \, .
\end{align}
\end{widetext}
As in Sec.~\ref{sec:M220}, the scattering-potential poles are exponentially suppressed by Eq.~\eqref{eq:separation}, so only the heavy-quark and gluon propagator poles contribute. The former is given by Eq.~\eqref{m220_pol}, while the latter is
\begin{equation}\label{q1kq1q2}
q^-_{1z,k-q_1-q_2}= -q_{2z} -i \epsilon.
\end{equation}
Neither pole receives a correction at $\mathcal{O}\left(\perp^3/(xE)^2\right)$; the next nonvanishing term lies beyond the pole accuracy specified in Eqs.~\eqref{eq:polecounting} and~\eqref{deltaQ}. Furthermore, we denote by $\mathcal{I}^{\ell}_{2,0,3}$, with $\ell=1,2$, the contributions from the residues of the integrand in Eq.~\eqref{IM203p} at the poles $q^-_{1z,p-q_1-q_2}$ and $q^-_{1z,k-q_1-q_2}$, respectively. 

The remaining longitudinal $q_{2z}$ integrals are then
\begin{widetext}
\begin{align}\label{JM203p}
\mathcal{J}^{\ell}_{2,0,3}= \int{\frac{dq_{2z}}{2\pi}}\frac{e^{-iq_{2z} z}\, v(q_{2z},{-\mathbf{q}})}{(k-q_2)^2 -m^2_g+i\epsilon} \, \mathcal{I}^{\ell}_{2,0,3} \, .
\end{align}
\end{widetext}

As in Eq.~\eqref{JM220}, the $-q_{2z}$ term in each contributing $q_{1z}$ pole generates a factor $e^{iq_{2z}z}$, canceling the explicit $e^{-iq_{2z}z}$ in Eq.~\eqref{JM203p}. Consequently, $\mathcal{J}^{\ell}_{2,0,3}$ contains no $q_{2z}$-dependent phase, and the remaining longitudinal integration generates no phase-induced first-subeikonal correction of the type classified in Sec.~\ref{FullSpectrumClassification}.

The $q_{2z}$ poles contributing to $\mathcal{J}^{1}_{2,0,3}$ include the two scattering-potential poles $q^-_{2z,v_1}$ and $q^-_{2z,v_2}$, given by Eqs.~\eqref{q2v1} and~\eqref{q2v2}, respectively, together with the gluon-propagator pole
\begin{equation}
\label{q2kq2}
q^-_{2z,k-q_2}
=
-\mathbf{u}\cdot\mathbf{q}
+\frac{(\mathbf{k}+\mathbf{q})^2-\mathbf{k}^2}{2xE}
-\frac{\mathbf{u}\cdot\mathbf{q}}{xE}
\frac{(\mathbf{k}+\mathbf{q})^2+m_g^2}{2xE}
-i\epsilon \, .
\end{equation}
For $\mathcal{J}^{2}_{2,0,3}$, the two scattering-potential factors share the pole $q^-_{2z,v_2}$, which therefore becomes a double pole, while the gluon-propagator pole remains that of Eq.~\eqref{q2kq2}.

As for $\mathcal{M}_{2,2,0}$ in Sec.~\ref{sec:M220}, the gluon-propagator and scattering-potential poles remain separated by $\mathcal{O}(\perp)$. Consequently, the $\mathcal{O}\!\left(\perp^3/(xE)^2\right)$ correction in Eq.~\eqref{q2kq2} does not contribute to the residue denominators or regular prefactors at first-subeikonal accuracy.

Summing the corresponding residues gives the complete longitudinal contribution $\mathcal{J}^{1}_{2,0,3}+\mathcal{J}^{2}_{2,0,3}$. Substituting this result, together with Eq.~\eqref{Izr203} into Eq.~\eqref{eq:M0M203}, we obtain
\begin{widetext}
\begin{align}
\frac{2}{d_T}
\operatorname{Re}\operatorname{Tr}
\left\langle
\mathcal{M}_0^{\dagger}
\mathcal{M}^c_{2,0,3}
\right\rangle
{}&=
-\frac{8  g_s^2 C_2(T) N}{ A_\perp d_A}\, \alpha 
\int_0^L dz\,\rho(z)
\int \frac{d^2\mathbf{q}}{(2\pi)^2}\,
\left|v(\mathbf{q})\right|^2
\left|J(E,\mathbf{p})\right|^2
\frac{\mathbf{k}^2}
{\left(\mathbf{k}^2+\chi \right)^2}
\nonumber\\
&\times
\Bigg\{
\left[
1
-2\,\frac{\mathbf{u}\cdot\mathbf{k}}{xE}
-\frac{\mathbf{u}\cdot\mathbf{k}}{xE}\,
\frac{\mathbf{q}^2}{v^2(\mathbf{q})}\,
\frac{\partial v^2(\mathbf{q})}{\partial\mathbf{q}^2}
\right]
\left[
1-\cos\left(
\frac{\mathbf{k}^2+\chi}{2xE}\,z
\right)
\right]
\nonumber\\
&\hspace{0.5cm}
-\frac{1}{16}\,
\frac{R_0}{xE}\,
\frac{1}{v^2(\mathbf{q})}\,
\frac{\partial v^2(\mathbf{q})}{\partial\mathbf{q}^2}
\left[
2\mu^2
-4\mathbf{q}^2
-3\left(
\mathbf{k}^2+\chi
\right)
\right]
\sin\left(
\frac{\mathbf{k}^2+\chi}{2xE}\,z
\right)
\Bigg\}
\label{eq:M0M203final} \, ,
\end{align}
\end{widetext}
where terms odd in $\mathbf{q}$ have been omitted, as they vanish upon the subsequent integration over $\mathbf{q}$.

The first-subeikonal terms have distinct origins. The correction proportional to $-2\,\mathbf{u}\cdot\mathbf{k}/(xE)$ arises from subleading Lorentz contraction in Eq.~\eqref{Izr203}. The term proportional to $\partial v^2(\mathbf{q})/\partial\mathbf{q}^2$ multiplying the $1-\cos$ structure is the flow-induced potential-derivative correction generated by expanding the scattering potentials evaluated at the gluon-propagator pole $q^-_{2z,k-q_2}$. The sine term, in turn, is the scattering-potential-pole sine contribution originating from the residues of the scattering-potential poles. No additional mass-dependent  residue-denominator correction of the type identified in Eq.~\eqref{eq:flow_massresidue} arises in this contribution. All these contributions are identified according to the classification established in Sec.~\ref{FullSpectrumClassification}.

It is worth noting that, to linear order in $\mathbf{u}$, Eq.~\eqref{eq:M0M203final} reduces in the massless limit, $M=m_g=0$, to Eq.~(43) of Ref.~\cite{KML}, apart from a typographical error in that equation, where $\mu$ should read $\mu^2$.

%%%%%%%%%%%%%%%%%%%%%%%%
\section{\label{sec:M200}Two-scattering contribution $\mathcal{M}_{2,0,0}$}

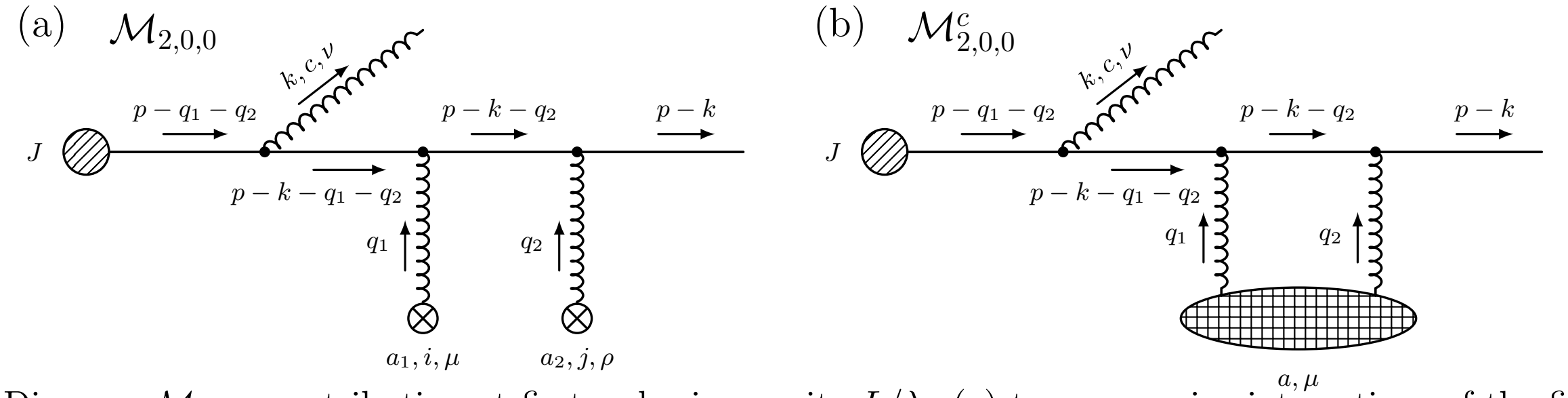
\begin{figure*}[!t]
\centering

% ============================================================
% PANEL (c): non-CL
% ============================================================
\begin{minipage}[t]{0.40\textwidth}
\centering
\resizebox{\linewidth}{!}{%
\begin{tikzpicture}[
    line cap=round,
    line join=round,
    >=latex
]

\path[use as bounding box] (-0.8,-2.45) rectangle (8.25,1.65);

\node at (-0.55,1.55) {\Large (a)};
\node at (0.95,1.45) {\Large $\mathcal{M}_{2,0,0}$};

\draw[
    pattern=north east lines,
    pattern color=black,
    line width=0.8pt
] (0,0) circle (0.28);

\node[left=4pt] at (-0.28,0) {$J$};

\coordinate (V1) at (2.20,0.00);  
\coordinate (V2) at (4.15,0.00);   
\coordinate (V3) at (6.05,0.00);   

\coordinate (QinStart) at (0.28,0.00);
\coordinate (Qout)     at (8.10,0.00);
\coordinate (Gout)     at (4.15,1.50);

\coordinate (C1) at (4.15,-2.05);
\coordinate (C2) at (6.05,-2.05);

\coordinate (B1) at (4.15,-1.87);
\coordinate (B2) at (6.05,-1.87);

\draw[line width=0.85pt] (QinStart) -- (V1);
\draw[line width=0.85pt] (V1) -- (V2);
\draw[line width=0.85pt] (V2) -- (V3);
\draw[line width=0.85pt] (V3) -- (Qout);

\draw[
    line width=0.85pt,
    decorate,
    decoration={
        coil,
        amplitude=2.1pt,
        segment length=4.8pt
    }
] (V2) -- (B1);

\draw[
    line width=0.85pt,
    decorate,
    decoration={
        coil,
        amplitude=2.1pt,
        segment length=4.8pt
    }
] (V3) -- (B2);

\draw[
    line width=0.85pt,
    decorate,
    decoration={
        coil,
        amplitude=2.3pt,
        segment length=5.2pt
    }
] (V1) -- (Gout);

\fill[black] (V1) circle (1.9pt);
\fill[black] (V2) circle (1.9pt);
\fill[black] (V3) circle (1.9pt);

\draw[line width=0.8pt] (C1) circle (0.18);

\draw[line width=0.8pt]
    ($(C1)+(-0.12,-0.12)$) --
    ($(C1)+(0.12,0.12)$);

\draw[line width=0.8pt]
    ($(C1)+(-0.12,0.12)$) --
    ($(C1)+(0.12,-0.12)$);

\draw[line width=0.8pt] (C2) circle (0.18);

\draw[line width=0.8pt]
    ($(C2)+(-0.12,-0.12)$) --
    ($(C2)+(0.12,0.12)$);

\draw[line width=0.8pt]
    ($(C2)+(-0.12,0.12)$) --
    ($(C2)+(0.12,-0.12)$);

\node[below=7pt] at (C1) {$a_1,i,\mu$};
\node[below=7pt] at (C2) {$a_2,j,\rho$};

\draw[line width=0.75pt,->]
    (0.95,0.22) -- (1.75,0.22);

\node[above=1pt] at (1.35,0.22)
    {$p-q_1-q_2$};

\draw[line width=0.75pt,->]
    (2.80,-0.22) -- (3.70,-0.22);

\node[below=1pt] at (2.85,-0.22)
    {$p-k-q_1-q_2$};

\draw[line width=0.75pt,->]
    (4.75,0.22) -- (5.45,0.22);

\node[above=1pt] at (5.10,0.22)
    {$p-k-q_2$};

\draw[line width=0.75pt,->]
    (7.05,0.22) -- (7.75,0.22);

\node[above=1pt] at (7.40,0.22)
    {$p-k$};

\draw[line width=0.75pt,->]
    (3.93,-1.46) -- (3.93,-0.84);

\node[left=2pt] at (3.93,-1.15)
    {$q_1$};

\draw[line width=0.75pt,->]
    (5.83,-1.46) -- (5.83,-0.84);

\node[left=2pt] at (5.83,-1.15)
    {$q_2$};

\draw[line width=0.75pt,->]
    (2.62,0.59) -- (3.24,1.07)
    node[midway, above=1pt, sloped] {$k,c,\nu$};

\end{tikzpicture}%
}
\end{minipage}%
\hspace{0.035\textwidth}%
%
% ============================================================
% PANEL (d): CL
% ============================================================
\begin{minipage}[t]{0.40\textwidth}
\centering
\resizebox{\linewidth}{!}{%
\begin{tikzpicture}[
    line cap=round,
    line join=round,
    >=latex
]

\path[use as bounding box] (-0.8,-2.45) rectangle (8.25,1.65);

\node at (-0.55,1.55) {\Large (b)};
\node at (0.95,1.45) {\Large $\mathcal{M}^{c}_{2,0,0}$};

\draw[
    pattern=north east lines,
    pattern color=black,
    line width=0.8pt
] (0,0) circle (0.28);

\node[left=4pt] at (-0.28,0) {$J$};

\coordinate (V1) at (2.20,0.00);   
\coordinate (V2) at (4.15,0.00);   
\coordinate (V3) at (6.05,0.00);  

\coordinate (QinStart) at (0.28,0.00);
\coordinate (Qout)     at (8.10,0.00);
\coordinate (Gout)     at (4.15,1.50);

\coordinate (B1) at (4.15,-1.77);
\coordinate (B2) at (6.05,-1.77);

\draw[line width=0.85pt] (QinStart) -- (V1);
\draw[line width=0.85pt] (V1) -- (V2);
\draw[line width=0.85pt] (V2) -- (V3);
\draw[line width=0.85pt] (V3) -- (Qout);

\draw[
    line width=0.85pt,
    decorate,
    decoration={
        coil,
        amplitude=2.1pt,
        segment length=4.8pt
    }
] (V2) -- (B1);

\draw[
    line width=0.85pt,
    decorate,
    decoration={
        coil,
        amplitude=2.1pt,
        segment length=4.8pt
    }
] (V3) -- (B2);

\draw[
    line width=0.85pt,
    decorate,
    decoration={
        coil,
        amplitude=2.3pt,
        segment length=5.2pt
    }
] (V1) -- (Gout);

\fill[black] (V1) circle (1.9pt);
\fill[black] (V2) circle (1.9pt);
\fill[black] (V3) circle (1.9pt);

\draw[
    line width=0.8pt,
    pattern=grid,
    pattern color=black
] (5.10,-2.05) ellipse (1.45 and 0.38);

\node[below=7pt] at (5.10,-2.38)
    {$a,\mu$};

\draw[line width=0.75pt,->]
    (0.95,0.22) -- (1.75,0.22);

\node[above=1pt] at (1.35,0.22)
    {$p-q_1-q_2$};

\draw[line width=0.75pt,->]
    (2.80,-0.22) -- (3.70,-0.22);

\node[below=1pt] at (2.85,-0.22)
    {$p-k-q_1-q_2$};

\draw[line width=0.75pt,->]
    (4.75,0.22) -- (5.45,0.22);

\node[above=1pt] at (5.10,0.22)
    {$p-k-q_2$};

\draw[line width=0.75pt,->]
    (7.05,0.22) -- (7.75,0.22);

\node[above=1pt] at (7.40,0.22)
    {$p-k$};

\draw[line width=0.75pt,->]
    (3.93,-1.36) -- (3.93,-0.74);

\node[left=2pt] at (3.93,-1.05)
    {$q_1$};

\draw[line width=0.75pt,->]
    (5.83,-1.36) -- (5.83,-0.74);

\node[left=2pt] at (5.83,-1.05)
    {$q_2$};

\draw[line width=0.75pt,->]
    (2.62,0.59) -- (3.24,1.07)
    node[midway, above=1pt, sloped] {$k,c,\nu$};

\end{tikzpicture}%
}
\end{minipage}
\caption{Diagram $\mathcal{M}_{2,0,0}$ contributing at first order in opacity  $L/{\lambda}$: (a) two successive interactions of the final-state heavy quark with the medium after gluon emission and (b) the corresponding contact-limit configuration.}
\label{fig:M200}
\end{figure*}

We proceed to the $\mathcal{M}_{2,0,0}$ contribution shown in Fig.~\ref{fig:M200}, where the gluon $k$ is emitted before both medium interactions. The final-state heavy quark then undergoes two successive scatterings with momentum transfers $q_1$ and $q_2$, emerging with momentum $p-k$.

%%%%M200

Using the momentum labeling in Fig.~\ref{fig:M200} and the conventions introduced above, the corresponding amplitude is
\begin{widetext}
\begin{align}
\mathcal{M}_{2,0,0}
{}&=
\int\frac{d^4q_1}{(2\pi)^4}
\int\frac{d^4q_2}{(2\pi)^4}
\left[
i g_s
(2p-2k-q_2)_{\rho}
A_{ext}^{\rho, a_2}(q_2)
T_{ jet}^{a_2}
\right]
\frac{i}{(p-k-q_2)^2-M^2+i\epsilon}
\nonumber \\
&\times
\left[
i g_s
(2p-2k-q_1-2q_2)_{\mu}
A_{ext}^{\mu, a_1}(q_1)
T_{ jet}^{a_1}
\right]
\frac{i}{(p-k-q_1-q_2)^2-M^2+i\epsilon}
\nonumber \\
&\times
\left[
i g_s
(2p-k-2q_1-2q_2)^{\nu}
\varepsilon_{\nu}^{*}(k)
T_{ jet}^{c}
\right]
\frac{i}{(p-q_1-q_2)^2-M^2+i\epsilon} \,
iJ(p-q_1-q_2) \, .
\label{m200_p}
\end{align}
\end{widetext}
Following the same procedure as for the preceding two-scattering contributions, we insert Eqs.~\eqref{Aext} and Eq.~\eqref{a} to perform the $q_1^0$ and $q_2^0$ integrations.

The medium average again selects the CL configuration. Using Eq.~\eqref{q1q2}, the interference with $\mathcal{M}_0^\dagger$ reduces to
\begin{widetext}
\begin{align}
\frac{2}{d_T}
\operatorname{Re}\operatorname{Tr}
\left\langle
\mathcal{M}_0^\dagger
\mathcal{M}_{2,0,0}^{c}
\right\rangle
{}&=
\frac{32 g_s^2 C_2(T)  N}{A_\perp d_A} \operatorname{Tr}(c^2 a_1^2)\,
\operatorname{Re}
\Bigg\{
\int_0^L dz\,\rho(z)
\int\frac{d^2\mathbf{q}}{(2\pi)^2}\, \left|J(E, \mathbf{p})\right|^2
\frac{
E^2 \,  \mathbf{k}^2
}{
x\left(
\mathbf{k}^2+\chi
\right)
} \int\frac{dq_{1z}}{2\pi}
\int\frac{dq_{2z}}{2\pi}\,
\nonumber\\
&\times
e^{-iq_{1z}z}
e^{-iq_{2z}z}\,
\frac{
 v(q_1) v(q_2)
}{
\left[
(p-q_1-q_2)^2-M^2+i\epsilon
\right]
\left[
(p-k-q_1-q_2)^2-M^2+i\epsilon
\right]
}
\nonumber\\
&\times
\frac{1}{
(p-k-q_2)^2-M^2+i\epsilon
}
\Bigg\}.
\label{eq:M0M200}
\end{align}
\end{widetext}

We first perform the $q_{1z}$ integration,
\begin{widetext}
\begin{align}
\mathcal{I}_{2,0,0}
{}&=
\int\frac{dq_{1z}}{2\pi}
\frac{
e^{-iq_{1z}z}
v(q_1)
}{
\left[
(p-q_1-q_2)^2-M^2+i\epsilon
\right]
\left[
(p-k-q_1-q_2)^2-M^2+i\epsilon
\right]
}.
\label{IM200p}
\end{align}
\end{widetext}
Closing the contour in the lower half-plane, the scattering-potential pole is exponentially suppressed according to Eq.~\eqref{eq:separation}, whereas both heavy-quark propagators contribute. The pole originating from $(p-q_1-q_2)^2-M^2+i\epsilon$ is the one already obtained in Eq.~\eqref{m220_pol}. As discussed there, it has no correction at $\mathcal{O}\left(\perp^3/(xE)^2\right)$. The second propagator gives
\begin{align}
q^-_{1z,p-k-q_1-q_2}
={}&
-q_{2z}-i\epsilon \, .
\label{q1pkq1q2}
\end{align}
In the CL configuration, $q_1^0+q_2^0=0$ and $\mathbf{q}_1+\mathbf{q}_2=0$, so the corresponding heavy-quark propagator depends on the longitudinal momentum transfers only through $q_{1z}+q_{2z}$. Together with the on-shell condition $(p-k)^2=M^2$, this fixes the real part of the lower-half-plane pole at $q_{1z}+q_{2z}=0$, with no subleading shift of the order specified in Eq.~\eqref{deltaQ}. We denote the residues at the poles in Eqs.~\eqref{m220_pol} and~\eqref{q1pkq1q2} by $\mathcal{I}^1_{2,0,0}$ and $\mathcal{I}^2_{2,0,0}$, respectively.

The subsequent longitudinal integrations can then be written as
\begin{align}
\mathcal{J}^{1,2}_{2,0,0}
={}&
\int\frac{dq_{2z}}{2\pi}\,
\frac{
e^{-iq_{2z}z}
v(q_2)
}{
(p-k-q_2)^2-M^2+i\epsilon
} \,
\mathcal{I}^{1,2}_{2,0,0} \, .
\label{JM200}
\end{align}
The phase cancellation encountered for $\mathcal{M}_{2,2,0}$ in Sec.~\ref{sec:M220} and for $\mathcal{M}_{2,0,3}$ in Sec.~\ref{sec:M203} also occurs here. Each $q_{1z}$ residue supplies a factor $e^{iq_{2z}z}$, which removes the explicit $e^{-iq_{2z}z}$ in Eq.~\eqref{JM200}. Thus, both $\mathcal{J}^1_{2,0,0}$ and $\mathcal{J}^2_{2,0,0}$ are free of $q_{2z}$-dependent phases. Moreover, the formation phase receives no $\delta Q^-$ correction of the type considered in Eq.~\eqref{eq:cosphaseexpansion}.

For $\mathcal{J}^1_{2,0,0}$, the poles in the lower half-plane are the heavy-quark propagator pole
\begin{align}
q^-_{2z,p-k-q_2}
={}&
-\mathbf{u} \cdot \mathbf{q}
-\frac{\mathbf{u} \cdot \mathbf{q}}{xE}
\frac{x^2M^2}{2xE}
-i\epsilon \, ,
\label{q2pkq2M200}
\end{align}
and scattering potential poles  $q^-_{2z,v_1}$ (from the first scattering potential, whose longitudinal argument after taking the $\mathcal{I}^1_{2,0,0}$ residue is $q_{1z}=-(\mathbf{k}^2+\chi)/(2xE)-q_{2z}$) and $q^-_{2z,v_2}$, given by Eqs.~\eqref{q2v1} and~\eqref{q2v2}, respectively.

For $\mathcal{J}^2_{2,0,0}$, the heavy-quark pole remains Eq.~\eqref{q2pkq2M200}, whereas the two scattering-potential poles coincide,
\begin{align}
q^-_{2z,v_1}=
q^-_{2z,v_2}
{}&=
-iR \, ,
\label{q2vM200J2}
\end{align}
and therefore form a double pole.

For both $\mathcal{J}^1_{2,0,0}$ and $\mathcal{J}^2_{2,0,0}$, the subleading pole correction $\delta Q^-$ in Eq.~\eqref{deltaQ} appears only in the heavy-quark propagator pole,
 Eq.~\eqref{q2pkq2M200}, while the corresponding scattering-potential poles contain no such term. Since the leading $Q_0^-$ contribution of Eq.~\eqref{eq:polecounting} does not cancel in any pole difference involving $\delta Q^-$, its effect on the residues remains beyond first-subeikonal accuracy and is therefore not promoted in the final result. Consequently, no mass-dependent residue-denominator correction of the type identified in Eq.~\eqref{eq:flow_massresidue} arises in this contribution.

Thus, substituting the summed residue contributions into Eq.~\eqref{eq:M0M200} and retaining terms through first-subeikonal order and linear in $\mathbf{u}$, we obtain
\begin{widetext}
\begin{align}
\frac{2}{d_T}
\operatorname{Re}\operatorname{Tr}
\left\langle
\mathcal{M}_0^\dagger
\mathcal{M}_{2,0,0}^{c}
\right\rangle
{}&=
-\frac{4  g_s^2 C_2(T) N }{ A_\perp d_A}
\operatorname{Tr}(c^2 a_1^2)
\int_0^L dz \rho(z)
\int\frac{d^2\mathbf{q}}{(2\pi)^2}
\left|v(\mathbf{q})\right|^2
\left|J(E, \mathbf{p})\right|^2
\frac{\mathbf{k}^2}{\left(\mathbf{k}^2+\chi\right)^2}
\nonumber \\
&\times
\Bigg\{
1-
\cos\left(
\frac{\mathbf{k}^2+\chi}{2xE} z\right)
+
\frac{3}{16}
\frac{R_0}{xE}
\frac{\left(\mathbf{k}^2+\chi\right)}{v^2(\mathbf{q})}
\frac{\partial v^2(\mathbf{q})}{\partial \mathbf{q}^2}
\sin\left(
\frac{\mathbf{k}^2+\chi}{2xE}z
\right)
\Bigg\}.
\label{eq:M0M200final}
\end{align}
\end{widetext}
In Eq.~\eqref{eq:M0M200final}, terms odd under $\mathbf{q}\to-\mathbf{q}$ have been omitted, since they vanish upon transverse-momentum integration.

The first-subeikonal sine term originates from the residues of the scattering-potential poles in $\mathcal{J}^1_{2,0,0}$. The associated derivative structure, given by Eq.~\eqref{vt2}, reflects the shifted momentum dependence of the potentials induced by the $\mathcal{I}^1_{2,0,0}$ pole. No phase-induced $z\sin$ correction arises in this contribution, while corrections from $\delta Q^-$ to the Lorentz-contraction prefactors remain beyond first-subeikonal accuracy. 
The sine term thus belongs to the scattering-potential-pole sine contribution, as classified in Sec.~\ref{FullSpectrumClassification}.

In the massless limit, $M=m_g=0$, our result, Eq.~\eqref{eq:M0M200final}, agrees to linear order in $\mathbf{u}$ with Eq.~(38) of Ref.~\cite{KML}.

%%%%%%%%%%%%%%M201

\begin{figure*}[!t]
\centering

% ============================================================
% PANEL (a): M_{2,0,1}, non-CL
% ============================================================
\begin{minipage}[t]{0.43\textwidth}
\centering
\resizebox{\linewidth}{!}{%
\begin{tikzpicture}[
    line cap=round,
    line join=round,
    >=latex
]

\path[use as bounding box] (-0.8,-2.65) rectangle (8.25,1.65);

\node at (-0.55,1.55) {\Large (a)};
\node at (0.95,1.45) {\Large $\mathcal{M}_{2,0,1}$};

\draw[
    pattern=north east lines,
    pattern color=black,
    line width=0.8pt
] (0,0) circle (0.28);

\node[left=4pt] at (-0.28,0) {$J$};

\coordinate (V1) at (2.20,0.00);

\coordinate (G1) at (4.15,-0.85);

\coordinate (V2) at (6.05,0.00);

\coordinate (QinStart) at (0.28,0.00);
\coordinate (Qout)     at (8.10,0.00);

\coordinate (Gout) at (8.10,-0.85);

\coordinate (C1) at (4.15,-2.20);
\coordinate (C2) at (6.05,-2.20);

\coordinate (B1) at (4.15,-2.02);
\coordinate (B2) at (6.05,-2.02);

\draw[line width=0.85pt] (QinStart) -- (V1);
\draw[line width=0.85pt] (V1) -- (V2);
\draw[line width=0.85pt] (V2) -- (Qout);

\draw[
    line width=0.85pt,
    decorate,
    decoration={
        coil,
        amplitude=2.3pt,
        segment length=5.2pt
    }
] (V1) -- (G1);

\draw[
    line width=0.85pt,
    decorate,
    decoration={
        coil,
        amplitude=2.3pt,
        segment length=5.2pt
    }
] (G1) -- (Gout);

\draw[
    line width=0.85pt,
    decorate,
    decoration={
        coil,
        amplitude=2.1pt,
        segment length=4.8pt
    }
] (G1) -- (B1);

\draw[
    line width=0.85pt,
    decorate,
    decoration={
        coil,
        amplitude=2.1pt,
        segment length=4.8pt
    }
] (V2) -- (B2);

\fill[black] (V1) circle (1.9pt);
\fill[black] (G1) circle (1.9pt);
\fill[black] (V2) circle (1.9pt);

\draw[line width=0.8pt] (C1) circle (0.18);

\draw[line width=0.8pt]
    ($(C1)+(-0.12,-0.12)$) --
    ($(C1)+(0.12,0.12)$);

\draw[line width=0.8pt]
    ($(C1)+(-0.12,0.12)$) --
    ($(C1)+(0.12,-0.12)$);

\draw[line width=0.8pt] (C2) circle (0.18);

\draw[line width=0.8pt]
    ($(C2)+(-0.12,-0.12)$) --
    ($(C2)+(0.12,0.12)$);

\draw[line width=0.8pt]
    ($(C2)+(-0.12,0.12)$) --
    ($(C2)+(0.12,-0.12)$);

\node[below=7pt] at (C1) {$a_1,i,\mu$};
\node[below=7pt] at (C2) {$a_2,j,\rho$};

\draw[line width=0.75pt,->]
    (0.95,0.22) -- (1.75,0.22);

\node[above=1pt] at (1.35,0.22)
    {$p-q_1-q_2$};

\draw[line width=0.75pt,->]
    (3.75,0.22) -- (4.55,0.22);

\node[above=1pt] at (4.15,0.22)
    {$p-k-q_2$};

\draw[line width=0.75pt,->]
    (6.72,0.22) -- (7.52,0.22);

\node[above=1pt] at (7.12,0.22)
    {$p-k$};

\draw[line width=0.75pt,->]
    (2.78,-0.50) -- (3.52,-0.83)
    node[midway, below=2pt, sloped]
    {$k-q_1$};

\draw[line width=0.75pt,->]
    (6.95,-1.07) -- (7.75,-1.07);

\node[below=1pt] at (7.35,-1.07)
    {$k,c,\nu$};

\draw[line width=0.75pt,->]
    (3.93,-1.82) -- (3.93,-1.20);

\node[left=2pt] at (3.93,-1.60)
    {$q_1$};

\draw[line width=0.75pt,->]
    (5.83,-1.62) -- (5.83,-1.00);

\node[left=2pt] at (5.83,-1.40)
    {$q_2$};

\end{tikzpicture}%
}
\end{minipage}%
\hspace{0.04\textwidth}%
%
% ============================================================
% PANEL (b): M^c_{2,0,1}, CL
% ============================================================
\begin{minipage}[t]{0.43\textwidth}
\centering
\resizebox{\linewidth}{!}{%
\begin{tikzpicture}[
    line cap=round,
    line join=round,
    >=latex
]

\path[use as bounding box] (-0.8,-2.75) rectangle (8.25,1.65);

\node at (-0.55,1.55) {\Large (b)};
\node at (0.95,1.45) {\Large $\mathcal{M}^{c}_{2,0,1}$};

\draw[
    pattern=north east lines,
    pattern color=black,
    line width=0.8pt
] (0,0) circle (0.28);

\node[left=4pt] at (-0.28,0) {$J$};

\coordinate (V1) at (2.20,0.00);

\coordinate (G1) at (4.15,-0.85);

\coordinate (V2) at (6.05,0.00);

\coordinate (QinStart) at (0.28,0.00);
\coordinate (Qout)     at (8.10,0.00);

\coordinate (Gout) at (8.10,-0.85);

\coordinate (B1) at (4.15,-1.991);
\coordinate (B2) at (6.05,-1.991);

\draw[line width=0.85pt] (QinStart) -- (V1);
\draw[line width=0.85pt] (V1) -- (V2);
\draw[line width=0.85pt] (V2) -- (Qout);

\draw[
    line width=0.85pt,
    decorate,
    decoration={
        coil,
        amplitude=2.3pt,
        segment length=5.2pt
    }
] (V1) -- (G1);

\draw[
    line width=0.85pt,
    decorate,
    decoration={
        coil,
        amplitude=2.3pt,
        segment length=5.2pt
    }
] (G1) -- (Gout);

\draw[
    line width=0.85pt,
    decorate,
    decoration={
        coil,
        amplitude=2.1pt,
        segment length=4.8pt
    }
] (G1) -- (B1);

\draw[
    line width=0.85pt,
    decorate,
    decoration={
        coil,
        amplitude=2.1pt,
        segment length=4.8pt
    }
] (V2) -- (B2);

\fill[black] (V1) circle (1.9pt);
\fill[black] (G1) circle (1.9pt);
\fill[black] (V2) circle (1.9pt);

\draw[
    line width=0.8pt,
    pattern=grid,
    pattern color=black
] (5.10,-2.18) ellipse (1.45 and 0.25);

\node at (5.10,-2.72) {$a,\mu$};

\draw[line width=0.75pt,->]
    (0.95,0.22) -- (1.75,0.22);

\node[above=1pt] at (1.35,0.22)
    {$p-q_1-q_2$};

\draw[line width=0.75pt,->]
    (3.75,0.22) -- (4.55,0.22);

\node[above=1pt] at (4.15,0.22)
    {$p-k-q_2$};

\draw[line width=0.75pt,->]
    (6.72,0.22) -- (7.52,0.22);

\node[above=1pt] at (7.12,0.22)
    {$p-k$};

\draw[line width=0.75pt,->]
    (2.78,-0.50) -- (3.52,-0.83)
    node[midway, below=2pt, sloped]
    {$k-q_1$};

\draw[line width=0.75pt,->]
    (6.95,-1.07) -- (7.75,-1.07);

\node[below=1pt] at (7.35,-1.07)
    {$k,c,\nu$};

\draw[line width=0.75pt,->]
    (3.93,-1.82) -- (3.93,-1.20);

\node[left=2pt] at (3.93,-1.60)
    {$q_1$};

\draw[line width=0.75pt,->]
    (5.83,-1.62) -- (5.83,-1.00);

\node[left=2pt] at (5.83,-1.40)
    {$q_2$};

\end{tikzpicture}%
}
\end{minipage}

\caption{
Diagram $\mathcal{M}_{2,0,1}$ contributing at first order in opacity  $L/{\lambda}$:
(a) one medium interaction on the radiated-gluon line and one on the outgoing heavy-quark line, with momentum transfers $q_1$ and $q_2$, respectively, and
(b) the corresponding contact-limit configuration.
}
\label{fig:M201}

\end{figure*}
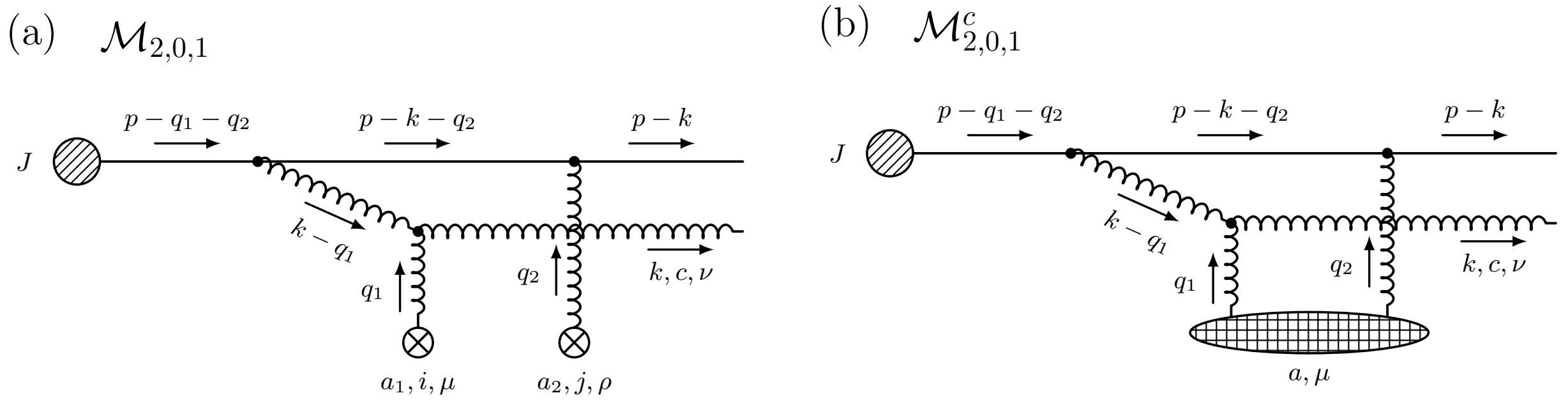

\section{\label{sec:M201}Two-scattering contributions $\mathcal{M}_{2,0,1}$ and $\mathcal{M}_{2,0,2}$ }

The next two-scattering contribution is $\mathcal{M}_{2,0,1}$, shown in Fig.~\ref{fig:M201}, in which, following the gluon emission, the radiated gluon and the outgoing heavy quark each undergo one interaction with the medium, with momentum transfers $q_1$ and $q_2$, and emerge with momenta $k$ and $p-k$, respectively.

With the momentum and index assignments of Fig.~\ref{fig:M201}, the amplitude takes the form
\begin{widetext}
\begin{align}
\mathcal{M}_{2,0,1}
={}&
\int \frac{d^4q_1}{(2\pi)^4}
\int \frac{d^4q_2}{(2\pi)^4}
 \left[
i g_s(2p-2k-q_2)_\rho \, A_{ext}^{\rho, a_2}(q_2) T_{jet}^{a_2}
\right]\,
\frac{i}{(p-k-q_2)^2-M^2+i\epsilon} \,
\varepsilon^{*}_{\nu}(k) \,
\Gamma_{\xi \nu\mu}^{a c a_1} (k-q_1, -k, q_1)\,
\nonumber \\
&\times
A_{ext}^{\mu, a_1}(q_1) \,
\frac{(-i)P^{\xi \xi'}(k-q_1)\delta_{aa'}}
{(k-q_1)^2-m_g^2+i\epsilon} \,
 \left[
i g_s(2p-k-q_1-2q_2)_{\xi'} T_{jet}^{a'}
\right]
\frac{i}{(p-q_1-q_2)^2-M^2+i\epsilon} \,
iJ(p-q_1-q_2) \, .
\label{m201_p}
\end{align}
\end{widetext}

Inserting the external fields and performing the $q_1^0$ and $q_2^0$ integrations, we obtain, after interference with $\mathcal{M}_0^\dagger$ and the same CL averaging procedure as above,
%%%scalar--gluon and DA LI DA DODAJEM TAJ VERTEX
\begin{widetext}
\begin{align}
\frac{2}{d_T}
\operatorname{Re}\operatorname{Tr}
\left\langle
\mathcal{M}_0^\dagger
\mathcal{M}_{2,0,1}^{c}
\right\rangle
{}&=
\frac{4 g_s^2 C_2(T)  N}{A_\perp d_A} \alpha \,
\operatorname{Re}
\Bigg\{
\int_0^L dz\,\rho(z)
\int\frac{d^2\mathbf{q}}{(2\pi)^2}\, \left|J(E, \mathbf{p})\right|^2
\frac{
\boldsymbol{\varepsilon} \cdot \mathbf{k}}{
\mathbf{k}^2+\chi
}  \, I_{2,0,1}
\int\frac{dq_{1z}}{2\pi}
\int\frac{dq_{2z}}{2\pi}\, e^{-iq_{1z}z}
\nonumber\\
&\times
e^{-iq_{2z}z}\,
\frac{
 v(q_1) v(q_2)
}{
\left[
(p-q_1-q_2)^2-M^2+i\epsilon
\right]
\left[
(k-q_1)^2-m_g^2+i\epsilon
\right]
}\,  \frac{1}{
(p-k-q_2)^2-M^2+i\epsilon
}
\Bigg\} \, ,
\label{eq:M0M201}
\end{align}
\end{widetext}
Here, $q_1$ and $q_2$ are specified in Eqs.~\eqref{kin1M2} and~\eqref{kin2M2}. As established in Sec.~\ref{sec:M203}, the Lorentz contraction $I_{2,0,1}$ introduces no additional longitudinal poles. For the present diagram, its evaluation yields
\begin{widetext}
\begin{align}
I_{2,0,1} &{}=-8 E^2 \left[ \boldsymbol{\varepsilon} \cdot (\mathbf{k}- \mathbf{q}) - \boldsymbol{\varepsilon} \cdot \mathbf{k} \frac{\mathbf{u} \cdot  (\mathbf{k}-\mathbf{q})}{x E} -\mathbf{q} \cdot (\mathbf{q} - \mathbf{k}) \,  \frac{ \boldsymbol{\varepsilon} \cdot \mathbf{u}}{x E}\right] \, .
\label{I201}
\end{align}
\end{widetext}

The longitudinal momentum integrations are performed successively, starting with the \(q_{1z}\) contour integral,
\begin{widetext}
\begin{align}
\mathcal{I}_{2,0,1}
{}&=
\int\frac{dq_{1z}}{2\pi}
\frac{
e^{-iq_{1z}z}
v(q_1)
}{
\left[
(p-q_1-q_2)^2-M^2+i\epsilon
\right]
\left[
(k-q_1)^2-m_g^2+i\epsilon
\right]
} \, .
\label{IM201p}
\end{align}
\end{widetext}
Since the contour is closed in the lower half-plane, the scattering-potential pole is exponentially suppressed according to Eq.~\eqref{eq:separation}. The remaining singularities arise from the heavy-quark and gluon propagators. The heavy-quark pole coincides with that in Eq.~\eqref{m220_pol},
$q^-_{1z,p-q_1-q_2}=-\frac{\mathbf{k}^2+\chi}{2xE}-q_{2z}-i\epsilon$ , 
and, as discussed below Eq.~\eqref{m220_pol}, it receives no correction at \(\mathcal{O}\left(\perp^3/(xE)^2\right)\). The gluon propagator instead gives
\begin{widetext}
\begin{align}
q^-_{1z,k-q_1}
={}&\mathbf{u}\cdot\mathbf{q}
+\frac{(\mathbf{k}-\mathbf{q})^2-\mathbf{k}^2}{2xE}
+\frac{\mathbf{u}\cdot\mathbf{q}}{xE}
\frac{(\mathbf{k}-\mathbf{q})^2+m_g^2}{2x E}
-i\epsilon \, .
\label{q1kq1}
\end{align}
\end{widetext}
Equation~\eqref{q1kq1}, obtained explicitly, also follows from Eq.~\eqref{q2kq2} under $\mathbf{q}\to-\mathbf{q}$, consistently with Eq.~\eqref{q1q2}. Its flow-dependent $\mathcal{O}\!\left(\perp^3/(xE)^2\right)$ correction is retained in the $q_{1z}$ residue denominators and the longitudinal phase, according to Eqs.~\eqref{eq:polecounting} and~\eqref{deltaQ}. Its contribution at first-subeikonal accuracy is determined by the subsequent $q_{2z}$ integration, with the corresponding phase correction governed by Eq.~\eqref{eq:deltaphicounting}.

We denote the residues of the heavy-quark and gluon poles by $\mathcal{I}^1_{2,0,1}$ and $\mathcal{I}^2_{2,0,1}$, respectively. The remaining longitudinal integrations are then
\begin{align}
\mathcal{J}^{1,2}_{2,0,1}
={}&
\int\frac{dq_{2z}}{2\pi} \,
\frac{
e^{-iq_{2z}z}
v(q_2)
}{
(p-k-q_2)^2-M^2+i\epsilon
} \,
\mathcal{I}^{1,2}_{2,0,1} \, .
\end{align}

For $\mathcal{J}^{1}_{2,0,1}$, the factor $e^{iq_{2z}z}$ generated by the heavy-quark $q_{1z}$ pole cancels the explicit $e^{-iq_{2z}z}$, leaving no $q_{2z}$-dependent phase. Thus, subleading pole corrections may affect the residue denominators but generate no phase-induced correction on this branch.

The $q_{2z}$ integration in $\mathcal{J}^{1}_{2,0,1}$ (in the lower half-plane) contains the heavy-quark propagator pole $q^-_{2z,p-k-q_2}$, given in Eq.~\eqref{q2pkq2M200}, the two scattering-potential poles specified in Eqs.~\eqref{q2v1} and~\eqref{q2v2}, and an additional singularity on the real axis arising from the difference between the two $q^-_{1z}$ propagator poles, $q^-_{1z,p-q_1-q_2}$ and $q^-_{1z, k-q_1}$. The latter is treated in the Cauchy principal-value sense and is located at
\begin{align}
q^-_{2z,\Delta q_1^-}
={}&
-\mathbf{u}\cdot\mathbf{q}
-\frac{(\mathbf{k}-\mathbf{q})^2+\chi}{2xE}
-\frac{\mathbf{u}\cdot\mathbf{q}}{xE}
\frac{(\mathbf{k}-\mathbf{q})^2+m_g^2}{2xE} \, .
\label{q2Deltaq1M201}
\end{align}

The common $\mathcal{O}(\perp)$ contributions to $q^-_{2z,p-k-q_2}$ and $q^-_{2z,\Delta q_1^-}$ cancel in their difference, whose leading term is therefore of order $\mathcal{O}(\perp^2/(xE))$. Consequently, the surviving $\mathcal{O}\!\left(\perp^3/(xE)^2\right)$ mismatch contributes to the residue denominators at relative first-subeikonal order, in accordance with Eq.~\eqref{deltaQ}.
\begin{widetext}
\begin{align}
q^-_{2z,p-k-q_2}-q^-_{2z,\Delta q_1^-}
={}& \frac{(\mathbf{k}-\mathbf{q})^2+\chi}{2xE} \left[1+ \frac{\mathbf{u}\cdot\mathbf{q}}{xE} -\frac{\mathbf{u}\cdot\mathbf{q}}{xE} \,  \frac{2x^2M^2}{(\mathbf{k}-\mathbf{q})^2+\chi} \right] \, .
\label{q2zpkq2q2zdeltaq1}
\end{align}
\end{widetext}
Among the two-scattering contributions, this correction survives in $\mathcal{M}_{2,0,1}$ and its counterpart $\mathcal{M}_{2,0,2}$, whereas in $\mathcal{M}_{2,1,1}$ analogous terms enter the individual residue denominators but cancel upon summing the corresponding residues. A similar mechanism occurs for $\mathcal{M}_{1,0,1}$ through the quark-gluon pole difference in Eq.~\eqref{razlikapolova}. By contrast, differences between propagator poles and scattering-potential poles retain their leading $\mathcal{O}(\perp)$ separation, so the $\mathcal{O}\left(\perp^3/(xE)^2\right)$ shifts remain beyond first-subeikonal accuracy in the corresponding residue denominators and regular prefactors.

For $\mathcal{J}^{2}_{2,0,1}$, the gluon $q^-_{1z, k-q_1}$ residue leaves the $q_{2z}$ dependence of the exponential unchanged. Since $q^-_{1z,k-q_1}$ is independent of $q_{2z}$, the first scattering potential generates no additional $q_{2z}$ pole. The pole of $v(q_2)$ at $q^-_{2z,v_2}=-iR$  is exponentially suppressed according to Eq.~\eqref{eq:separation} and is therefore discarded. Thus, no scattering-potential pole contributes to $\mathcal{J}^{2}_{2,0,1}$.

The contributing propagator poles are $q^-_{2z,p-k-q_2}$, originating from the $(p-k-q_2)$ heavy-quark propagator and given in Eq.~\eqref{q2pkq2M200}, and $q^-_{2z,\Delta q_1^-}$ in Eq.~\eqref{q2Deltaq1M201}, generated by the $(p-q_1-q_2)$ propagator after evaluating $\mathcal{I}^{2}_{2,0,1}$ at the gluon-propagator pole $q^-_{1z,k-q_1}$ (and treated in the principal-value sense).
%%%%%%%%%%%%%%%%%%%%%%%%%

As shown in Eq.~\eqref{q2zpkq2q2zdeltaq1}, the same pole difference enters the residue denominators of $\mathcal{J}^{2}_{2,0,1}$, retaining the $\mathcal{O}\!\left(\perp^3/(xE)^2\right)$ correction at first-subeikonal accuracy. Its effect on the formation phase, however, depends on the contributing pole.

For the residue at $q^-_{2z,\Delta q_1^-}$, the $\mathcal{O}\!\left(\perp^3/(xE)^2\right)$ corrections in $q^-_{1z,k-q_1}$ and $q^-_{2z,\Delta q_1^-}$ cancel in their sum, which determines the longitudinal phase. Consequently, no subleading phase correction $\delta\phi$, as defined in Eqs.~\eqref{eq:phidecomposition} and~\eqref{eq:deltaphicounting}, survives on this branch.

For the residue at $q^-_{2z,p-k-q_2}$, however, the cancellation is incomplete, leaving the first-subeikonal formation-phase correction $\delta\phi_{2,0,1}$ given in Eq.~\eqref{eq:flow_deltaphi}.

With the notation of Eq.~\eqref{eq:phidecomposition}, the leading part of the corresponding formation phase is
\begin{align}
\phi_0
={}&
\frac{\mathbf{k}^2-(\mathbf{k}-\mathbf{q})^2}{2xE}z \, .
\end{align}
The corresponding cosine therefore contains $\phi_0-\delta\phi_{2,0,1}$. Expanding it according to Eq.~\eqref{eq:cosphaseexpansion} generates the first-subeikonal term $+\delta\phi_{2,0,1} \sin\phi_0$. The overall minus sign preceding this cosine then gives the $-\delta\phi_{2,0,1} \sin\phi_0$ contribution in Eq.~\eqref{eq:M0M201final}.

Combining the longitudinal contributions and substituting Eq.~\eqref{I201} into Eq.~\eqref{eq:M0M201}, we obtain, through first-subeikonal order and linear in $\mathbf{u}$,
\begin{widetext}
\begin{align}
\frac{2}{d_T}
\operatorname{Re}\operatorname{Tr}
\left\langle
\mathcal{M}_0^\dagger
\mathcal{M}^{c}_{2,0,1}
\right\rangle
{}&=
-\frac{8g_s^2 C_2(T)N}{A_\perp d_A} \, \alpha
\int_0^L dz \, \rho(z)
\int\frac{d^2\mathbf{q}}{(2\pi)^2} \,
\left|v(\mathbf{q})\right|^2
\left|J(E,\mathbf{p})\right|^2
\frac{1}{
\left(\mathbf{k}^2+\chi\right)
\left[(\mathbf{k}-\mathbf{q})^2+\chi\right]
}
\nonumber \\
&\times
\Bigg\{
\Bigg[
\mathbf{k}\cdot(\mathbf{k}-\mathbf{q})
-\mathbf{k}^2
\frac{\mathbf{u}\cdot(\mathbf{k}-\mathbf{q})}{xE}
-\mathbf{q}\cdot(\mathbf{q}-\mathbf{k})\, \frac{
(\mathbf{u}\cdot\mathbf{k})
}{xE}
\nonumber \\
&\qquad
-\mathbf{k}\cdot(\mathbf{k}-\mathbf{q})
\frac{\mathbf{u}\cdot\mathbf{q}}{2xE}
\frac{
\mathbf{k}^2-(\mathbf{k}-\mathbf{q})^2
}{v^2}
\frac{\partial v^2}{\partial\mathbf{q}^2}
+\mathbf{k}\cdot(\mathbf{k}-\mathbf{q}) \,\frac{\mathbf{u}\cdot\mathbf{q}}{xE}
\frac{2x^2M^2}{
(\mathbf{k}-\mathbf{q})^2+\chi
}
\Bigg]
\nonumber \\
&\qquad\times
\Bigg[
\cos\left(
\frac{\mathbf{k}^2+\chi}{2xE} \, z
\right)
-\cos\left(
\frac{
\mathbf{k}^2-(\mathbf{k}-\mathbf{q})^2
}{2xE} \, z
\right) \Bigg]
\nonumber \\
&\qquad
-\mathbf{k}\cdot(\mathbf{k}-\mathbf{q}) \, \frac{\mathbf{u}\cdot\mathbf{q}}{xE}
\frac{
(\mathbf{k}-\mathbf{q})^2+m_g^2-x^2M^2
}{2xE} \,
z \,
\sin\left(
\frac{
\mathbf{k}^2-(\mathbf{k}-\mathbf{q})^2
}{2xE} \,z
\right)
\nonumber \\
&\qquad
-\frac{3}{16} \,
\left[(\mathbf{k}-\mathbf{q})^2+\chi\right]
\frac{R_0
}{
xE
} \, 
\frac{\mathbf{k}^2}{v^2} \,
\frac{\partial v^2}{\partial\mathbf{q}^2}
\sin\left(
\frac{\mathbf{k}^2+\chi}{2xE} \,z
\right)
\Bigg\} \, ,
\label{eq:M0M201final}
\end{align}
\end{widetext}
where terms odd under $\mathbf{q}\to-\mathbf{q}$ have been omitted, since they vanish upon transverse-momentum integration.

The terms in Eq.~\eqref{eq:M0M201final} have distinct origins. The first term in the first square bracket is the leading-eikonal contribution, while the next two arise from subleading Lorentz contractions. The term proportional to $\partial v^2/\partial\mathbf{q}^2$ in the same bracket is the flow-induced potential-derivative correction generated by expanding the scattering potentials at the corresponding propagator poles.

The mass-dependent residue-denominator correction and the phase-induced $z\sin$ term can both be traced to the same underlying $\mathcal{O}\!\left(\perp^3/(xE)^2\right)$ flow-dependent gluon–heavy-quark pole mismatch, although they involve different combinations of longitudinal poles. The former enters through the difference $q^-_{2z,p-k-q_2}-q^-_{2z,\Delta q_1^-}$ in the residue denominators, whereas the latter arises from the sum $q^-_{1z,k-q_1}+q^-_{2z,p-k-q_2}$ in the longitudinal phase of $\mathcal{J}^{2}_{2,0,1}$. Indeed, the combination of longitudinal poles entering the formation phase can be decomposed as
\begin{align*}
&q^-_{1z,k-q_1}+q^-_{2z,p-k-q_2}
\\
&\quad=
\left(q^-_{1z,k-q_1}+q^-_{2z,\Delta q_1^-}\right)
\\
&\quad+
\left(q^-_{2z,p-k-q_2}-q^-_{2z,\Delta q_1^-}\right).
\end{align*}
The subleading flow-dependent shifts cancel in the first parentheses, leaving only $-(\mathbf{k}^2+\chi)/(2xE)$ in the real part. Consequently, the surviving phase correction is governed by the same pole difference that enters the residue denominators and generates the $z\sin$ term upon expansion according to Eq.~\eqref{eq:cosphaseexpansion}.

Finally, the last sine term, proportional to $R_0\partial v^2/\partial\mathbf{q}^2$, is the scattering-potential-pole sine contribution originating from the corresponding residues of $\mathcal{J}^{1}_{2,0,1}$. All these first-subeikonal corrections are identified according to the classification established in Sec.~\ref{FullSpectrumClassification}.

To linear order in $\mathbf{u}$, Eq.~\eqref{eq:M0M201final} agrees in the massless limit, $M=m_g=0$, with Eq.~(48) of Ref.~\cite{KML}, apart from the phase-induced $z\sin$ contribution. While Ref.~\cite{KML} retains the $\mathcal{O}\left(\perp^3/(xE)^2\right)$ pole corrections in the residue denominators, the formation phases are evaluated at leading-pole accuracy. In our calculation, the surviving subleading correction is retained in the formation phase as well, generating the additional $z\sin$ term through Eqs.~\eqref{eq:flow_deltaphi} and~\eqref{eq:cosphaseexpansion}.

 %%%%%%%%%%%%%%%%%%%%
%%%%M202
\begin{figure*}[!t]
\centering

% ============================================================
% PANEL (a): M_{2,0,2}, non-CL
% ============================================================
\begin{minipage}[t]{0.43\textwidth}
\centering
\resizebox{\linewidth}{!}{%
\begin{tikzpicture}[
    line cap=round,
    line join=round,
    >=latex
]

\path[use as bounding box] (-0.8,-2.65) rectangle (8.25,1.65);

\node at (-0.55,1.55) {\Large (a)};
\node at (0.95,1.45) {\Large $\mathcal{M}_{2,0,2}$};

\draw[
    pattern=north east lines,
    pattern color=black,
    line width=0.8pt
] (0,0) circle (0.28);

\node[left=4pt] at (-0.28,0) {$J$};

\coordinate (V1) at (2.20,0.00);

\coordinate (V2) at (4.15,0.00);

\coordinate (G1) at (4.15,-0.85);

\coordinate (G2) at (6.05,-0.85);

\coordinate (QinStart) at (0.28,0.00);
\coordinate (Qout)     at (8.10,0.00);

\coordinate (Gout) at (8.10,-0.85);

\coordinate (C1) at (4.15,-2.20);
\coordinate (C2) at (6.05,-2.20);

\coordinate (B1) at (4.15,-2.02);
\coordinate (B2) at (6.05,-2.02);

\draw[line width=0.85pt] (QinStart) -- (V1);
\draw[line width=0.85pt] (V1) -- (V2);
\draw[line width=0.85pt] (V2) -- (Qout);

\draw[
    line width=0.85pt,
    decorate,
    decoration={
        coil,
        amplitude=2.3pt,
        segment length=5.2pt
    }
] (V1) -- (G1);

\draw[
    line width=0.85pt,
    decorate,
    decoration={
        coil,
        amplitude=2.3pt,
        segment length=5.2pt
    }
] (G1) -- (G2);

\draw[
    line width=0.85pt,
    decorate,
    decoration={
        coil,
        amplitude=2.3pt,
        segment length=5.2pt
    }
] (G2) -- (Gout);

\draw[
    line width=0.85pt,
    decorate,
    decoration={
        coil,
        amplitude=2.1pt,
        segment length=4.8pt
    }
] (V2) -- (B1);

\draw[
    line width=0.85pt,
    decorate,
    decoration={
        coil,
        amplitude=2.1pt,
        segment length=4.8pt
    }
] (G2) -- (B2);

\fill[black] (V1) circle (1.9pt);
\fill[black] (V2) circle (1.9pt);
\fill[black] (G2) circle (1.9pt);

\draw[line width=0.8pt] (C1) circle (0.18);

\draw[line width=0.8pt]
    ($(C1)+(-0.12,-0.12)$) --
    ($(C1)+(0.12,0.12)$);

\draw[line width=0.8pt]
    ($(C1)+(-0.12,0.12)$) --
    ($(C1)+(0.12,-0.12)$);

\draw[line width=0.8pt] (C2) circle (0.18);

\draw[line width=0.8pt]
    ($(C2)+(-0.12,-0.12)$) --
    ($(C2)+(0.12,0.12)$);

\draw[line width=0.8pt]
    ($(C2)+(-0.12,0.12)$) --
    ($(C2)+(0.12,-0.12)$);

\node[below=7pt] at (C1) {$a_1,i,\mu$};
\node[below=7pt] at (C2) {$a_2,j,\rho$};

\draw[line width=0.75pt,->]
    (0.95,0.22) -- (1.75,0.22);

\node[above=1pt] at (1.35,0.22)
    {$p-q_1-q_2$};

\draw[line width=0.75pt,->]
    (2.70,0.22) -- (3.50,0.22);

\node[above=1pt] at (3.10,0.22)
    {$p-k-q_1$};

\draw[line width=0.75pt,->]
    (5.72,0.22) -- (6.52,0.22);

\node[above=1pt] at (6.12,0.22)
    {$p-k$};

\draw[line width=0.75pt,->]
    (2.78,-0.50) -- (3.52,-0.83)
    node[midway, below=2pt, sloped]
    {$k-q_2$};

\draw[line width=0.75pt,->]
    (6.95,-1.07) -- (7.75,-1.07);

\node[below=1pt] at (7.35,-1.07)
    {$k,c,\nu$};

\draw[line width=0.75pt,->]
    (3.93,-1.62) -- (3.93,-1.00);

\node[left=2pt] at (3.93,-1.40)
    {$q_1$};

\draw[line width=0.75pt,->]
    (5.83,-1.82) -- (5.83,-1.20);

\node[left=2pt] at (5.83,-1.60)
    {$q_2$};

\end{tikzpicture}%
}
\end{minipage}%
\hspace{0.04\textwidth}%
%
% ============================================================
% PANEL (b): M^c_{2,0,2}, CL
% ============================================================
\begin{minipage}[t]{0.43\textwidth}
\centering
\resizebox{\linewidth}{!}{%
\begin{tikzpicture}[
    line cap=round,
    line join=round,
    >=latex
]

\path[use as bounding box] (-0.8,-2.75) rectangle (8.25,1.65);

\node at (-0.55,1.55) {\Large (b)};
\node at (0.95,1.45) {\Large $\mathcal{M}^{c}_{2,0,2}$};

\draw[
    pattern=north east lines,
    pattern color=black,
    line width=0.8pt
] (0,0) circle (0.28);

\node[left=4pt] at (-0.28,0) {$J$};

\coordinate (V1) at (2.20,0.00);

\coordinate (V2) at (4.15,0.00);

\coordinate (G1) at (4.15,-0.85);

\coordinate (G2) at (6.05,-0.85);

\coordinate (QinStart) at (0.28,0.00);
\coordinate (Qout)     at (8.10,0.00);

\coordinate (Gout) at (8.10,-0.85);

\coordinate (B1) at (4.15,-1.991);
\coordinate (B2) at (6.05,-1.991);

\draw[line width=0.85pt] (QinStart) -- (V1);
\draw[line width=0.85pt] (V1) -- (V2);
\draw[line width=0.85pt] (V2) -- (Qout);

\draw[
    line width=0.85pt,
    decorate,
    decoration={
        coil,
        amplitude=2.3pt,
        segment length=5.2pt
    }
] (V1) -- (G1);

\draw[
    line width=0.85pt,
    decorate,
    decoration={
        coil,
        amplitude=2.3pt,
        segment length=5.2pt
    }
] (G1) -- (G2);

\draw[
    line width=0.85pt,
    decorate,
    decoration={
        coil,
        amplitude=2.3pt,
        segment length=5.2pt
    }
] (G2) -- (Gout);

\draw[
    line width=0.85pt,
    decorate,
    decoration={
        coil,
        amplitude=2.1pt,
        segment length=4.8pt
    }
] (V2) -- (B1);

\draw[
    line width=0.85pt,
    decorate,
    decoration={
        coil,
        amplitude=2.1pt,
        segment length=4.8pt
    }
] (G2) -- (B2);

\fill[black] (V1) circle (1.9pt);
\fill[black] (V2) circle (1.9pt);
\fill[black] (G2) circle (1.9pt);

\draw[
    line width=0.8pt,
    pattern=grid,
    pattern color=black
] (5.10,-2.18) ellipse (1.45 and 0.25);

\node at (5.10,-2.72) {$a,\mu$};

\draw[line width=0.75pt,->]
    (0.95,0.22) -- (1.75,0.22);

\node[above=1pt] at (1.35,0.22)
    {$p-q_1-q_2$};

\draw[line width=0.75pt,->]
    (2.70,0.22) -- (3.50,0.22);

\node[above=1pt] at (3.10,0.22)
    {$p-k-q_1$};

\draw[line width=0.75pt,->]
    (5.72,0.22) -- (6.52,0.22);

\node[above=1pt] at (6.12,0.22)
    {$p-k$};

\draw[line width=0.75pt,->]
    (2.78,-0.50) -- (3.52,-0.83)
    node[midway, below=2pt, sloped]
    {$k-q_2$};

\draw[line width=0.75pt,->]
    (6.95,-1.07) -- (7.75,-1.07);

\node[below=1pt] at (7.35,-1.07)
    {$k,c,\nu$};

\draw[line width=0.75pt,->]
    (3.93,-1.62) -- (3.93,-1.00);

\node[left=2pt] at (3.93,-1.40)
    {$q_1$};

\draw[line width=0.75pt,->]
    (5.83,-1.82) -- (5.83,-1.20);

\node[left=2pt] at (5.83,-1.60)
    {$q_2$};

\end{tikzpicture}%
}
\end{minipage}

\caption{
Diagram $\mathcal{M}_{2,0,2}$ contributing at first order in opacity  $L/{\lambda}$:
(a) one interaction with the medium on the heavy-quark line, with momentum transfer $q_1$, and one on the radiated gluon, with momentum transfer $q_2$, and
(b) the corresponding contact-limit configuration.
}
\label{fig:M202}

\end{figure*}
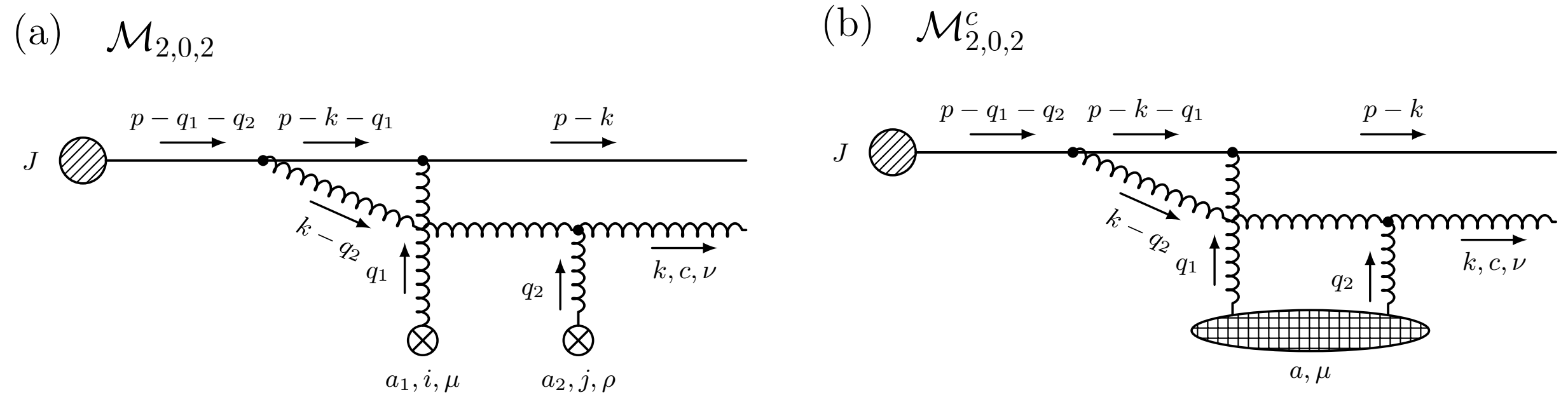

The corresponding counterpart, $\mathcal{M}_{2,0,2}$, is shown in Fig.~\ref{fig:M202}. Here, $q_1$ is transferred to the final heavy quark, whereas $q_2$ is exchanged with the emitted gluon. We evaluate this diagram independently, using its relation to $\mathcal{M}_{2,0,1}$ only as a consistency check. The amplitude becomes
\begin{widetext}
\begin{align}
\mathcal{M}_{2,0,2}
={}&
\int \frac{d^4q_1}{(2\pi)^4}
\int \frac{d^4q_2}{(2\pi)^4} \,
\varepsilon^{* \nu}(k) \,
\Gamma_{\xi\nu\rho}^{a c a_2}(k-q_2,-k,q_2) \,
A_{ext}^{\rho, a_2}(q_2) \,
\frac{(-i)P^{\xi\xi'}(k-q_2)\delta_{aa'}}
{(k-q_2)^2-m_g^2+i\epsilon}
\left[
i g_s (2p-2k-q_1)_\mu
A_{ext}^{\mu, a_1}(q_1)
T_{jet}^{a_1}
\right]
\nonumber \\
&\times
\frac{i}{(p-k-q_1)^2-M^2+i\epsilon}
\left[
i g_s (2p-k-2q_1-q_2)_{\xi'}
T_{jet}^{a'}
\right]
\frac{i}{(p-q_1-q_2)^2-M^2+i\epsilon} \,
iJ(p-q_1-q_2) \, .
\label{m202_p}
\end{align}
\end{widetext}

Performing the energy integrations and the CL medium average a for $\mathcal{M}_{2,0,1}$ again yields the momentum constraint in Eq.~\eqref{q1q2}. The resulting interference contribution has the same overall structure as Eq.~\eqref{eq:M0M201}, with the scattering labels interchanged, $q_1\leftrightarrow q_2$, and reads

\begin{widetext}
\begin{align}
\frac{2}{d_T}
\operatorname{Re}\operatorname{Tr}
\left\langle
\mathcal{M}_0^\dagger
\mathcal{M}_{2,0,2}^{c}
\right\rangle
{}&=
\frac{4 g_s^2 C_2(T)  N}{A_\perp d_A} \alpha \,
\operatorname{Re}
\Bigg\{
\int_0^L dz\,\rho(z)
\int\frac{d^2\mathbf{q}}{(2\pi)^2}\, \left|J(E, \mathbf{p})\right|^2
\frac{
\boldsymbol{\varepsilon} \cdot \mathbf{k}}{
\mathbf{k}^2+\chi
}  \, I_{2,0,2}
\int\frac{dq_{1z}}{2\pi}
\int\frac{dq_{2z}}{2\pi}\, e^{-iq_{1z}z}
\nonumber\\
&\times
e^{-iq_{2z}z}\,
\frac{
 v(q_1) v(q_2)
}{
\left[
(p-q_1-q_2)^2-M^2+i\epsilon
\right]
\left[
(p-k-q_1)^2-M^2+i\epsilon
\right]
}\,  \frac{1}{
(k-q_2)^2-m_g^2+i\epsilon
}
\Bigg\} \, .
\label{eq:M0M202}
\end{align}
\end{widetext}

At first-subeikonal accuracy, the Lorentz contraction receives no contribution from the $\mathcal{O}\!\left(\perp^3/(xE)^2\right)$ corrections to the longitudinal propagator poles and reads
\begin{widetext}
\begin{align}
I_{2,0,2}
={}&
-8 E^2 \left[ \boldsymbol{\varepsilon} \cdot (\mathbf{k}+ \mathbf{q}) - \boldsymbol{\varepsilon} \cdot \mathbf{k} \frac{\mathbf{u} \cdot  (\mathbf{k}+\mathbf{q})}{x E} -\mathbf{q} \cdot (\mathbf{q} + \mathbf{k}) \,  \frac{ \boldsymbol{\varepsilon} \cdot \mathbf{u}}{x E}\right] \, .
\label{I202}
\end{align}
\end{widetext}
As for Eq.~\eqref{I201}, the result is independent of the longitudinal momentum transfers and introduces no additional poles. We begin with the $q_{1z}$ integral,
\begin{widetext}
\begin{align}
\mathcal{I}_{2,0,2}
={}&
\int\frac{dq_{1z}}{2\pi} \,
\frac{e^{-iq_{1z}z}\,v(q_1)}
{\left[(p-q_1-q_2)^2-M^2+i\epsilon\right]
\left[(p-k-q_1)^2-M^2+i\epsilon\right]} \,.
\label{IM202p}
\end{align}
\end{widetext}
Closing the contour in the lower half-plane, we neglect the exponentially suppressed scattering-potential pole according to Eq.~\eqref{eq:separation}. The two contributing heavy-quark poles are given by Eq.~\eqref{m220_pol} and 
\begin{align}
q^-_{1z,p-k-q_1}
={}&
\mathbf{u}\cdot\mathbf{q}
+\frac{\mathbf{u}\cdot\mathbf{q}}{xE} \, \frac{x^2M^2}{2 x E}
-i\epsilon \,.
\label{q1pkq1M202}
\end{align}
We denote the corresponding residues by $\mathcal{I}^{1}_{2,0,2}$ and $\mathcal{I}^{2}_{2,0,2}$, respectively. 

The remaining longitudinal integrations are
\begin{equation}
\mathcal{J}^{1,2}_{2,0,2}
=
\int\frac{dq_{2z}}{2\pi}\,
\frac{e^{-iq_{2z}z}\, v(q_2)}
{(k-q_2)^2-m_g^2+i\epsilon} \,
\mathcal{I}^{1,2}_{2,0,2}\,.
\label{JM202}
\end{equation}

For $\mathcal{J}^{1}_{2,0,2}$, the residue at the heavy-quark pole in Eq.~\eqref{m220_pol} supplies a factor $e^{iq_{2z}z}$, canceling the explicit $e^{-iq_{2z}z}$ in Eq.~\eqref{JM202}. The remaining singularities include the gluon-propagator pole in Eq.~\eqref{q2kq2} (it can be straightforwardly checked that it reduces to Eq.~\eqref{q1kq1} under $\mathbf{q}\to-\mathbf{q}$), the scattering-potential poles in Eqs.~\eqref{q2v1} and~\eqref{q2v2}, and a real-axis singularity associated with the difference of the two $q_{1z}$ propagator poles. The latter is treated in the Cauchy principal-value sense and is located at
\begin{align}
q^{'-}_{2z,\Delta q_1^-}
={}&
-\mathbf{u}\cdot\mathbf{q}
-\frac{\mathbf{k}^2+\chi}{2xE}
-\frac{\mathbf{u}\cdot\mathbf{q}}{xE} \,
\frac{x^2M^2}{2xE} \,.
\label{q2Deltaq1M202}
\end{align}

The leading $\mathcal{O}(\perp)$ contributions cancel in the difference between the gluon pole in Eq.~\eqref{q2kq2} and the effective pole in Eq.~\eqref{q2Deltaq1M202}, yielding
\begin{widetext}
\begin{align}
q^-_{2z,k-q_2}-q^{'-}_{2z,\Delta q_1^-}
=\frac{(\mathbf{k}+\mathbf{q})^2+\chi}{2xE}
\left[
1-\frac{\mathbf{u}\cdot\mathbf{q}}{xE}
+\frac{\mathbf{u}\cdot\mathbf{q}}{xE}\,
\frac{2x^2M^2}{(\mathbf{k}+\mathbf{q})^2+\chi}
\right]\,.
\label{deltaq2J202}
\end{align}
\end{widetext}
Consequently, the surviving $\mathcal{O}\!\left(\perp^3/(xE)^2\right)$ correction contributes to the residue denominators of $\mathcal{J}^{1}_{2,0,2}$ at relative first-subeikonal accuracy. By contrast, the differences between either of these poles and the scattering-potential poles retain their leading $\mathcal{O}(\perp)$ separation. Therefore, the $\mathcal{O}\!\left(\perp^3/(xE)^2\right)$ corrections remain beyond first-subeikonal accuracy in the corresponding residue denominators. The $q_{2z}$-dependent phases cancel on this branch, so no phase-induced correction arises.

For $\mathcal{J}^{2}_{2,0,2}$, the residue at $q^-_{1z,p-k-q_1}$ retains the $q_{2z}$ dependence of the exponential. The scattering-potential pole is exponentially suppressed, leaving only the gluon pole in Eq.~\eqref{q2kq2} and the principal-value singularity in Eq.~\eqref{q2Deltaq1M202}. The pole difference in Eq.~\eqref{deltaq2J202} yields the same first-subeikonal residue denominators correction for this branch. However, at the principal-value singularity, the subleading shifts cancel in $q_{1z}+q_{2z}$, so no phase-induced correction survives.

At the gluon-propagator pole, the formation phase is determined by the sum $q^-_{1z,p-k-q_1}+q^-_{2z,k-q_2}$, in which the $\mathcal{O}\!\left(\perp^3/(xE)^2\right)$ corrections do not cancel. The resulting first-subeikonal phase shift is
\begin{align}
\delta\phi_{2,0,2}^{(2)}
={}&
\frac{\mathbf{u}\cdot\mathbf{q}\,
\left[
(\mathbf{k}+\mathbf{q})^2+m_g^2-x^2M^2
\right]}
{2(xE)^2}\,z\,.
\label{dphiM202}
\end{align}

The leading phase is $\phi_0=[\mathbf{k}^2-(\mathbf{k}+\mathbf{q})^2]z/(2xE)$, so the corresponding cosine contains $\phi_0+\delta\phi_{2,0,2}^{(2)}$. Its expansion according to Eq.~\eqref{eq:cosphaseexpansion}, together with the relative minus sign between the leading cosine contributions, generates the positive $z\sin\phi_0$ term displayed below. Combining the two longitudinal contributions and substituting Eq.~\eqref{I202} into Eq.~\eqref{eq:M0M202}, we obtain, through first-subeikonal order and linear in $\mathbf{u}$,
\begin{widetext}
\begin{align}
\frac{2}{d_T}
\operatorname{Re}\operatorname{Tr}
\left\langle
\mathcal{M}_0^\dagger
\mathcal{M}_{2,0,2}^{c}
\right\rangle
{}&=
-\frac{8g_s^2 C_2(T)N}{A_\perp d_A} \,
\alpha
\int_0^L dz\,\rho(z)
\int\frac{d^2\mathbf{q}}{(2\pi)^2} \,
|v(\mathbf{q})|^2
|J(E,\mathbf{p})|^2
\frac{1}
{(\mathbf{k}^2+\chi)
\left[(\mathbf{k}+\mathbf{q})^2+\chi\right]}
\nonumber \\
&\times
\Bigg\{
\Bigg[
\mathbf{k}\cdot(\mathbf{k}+\mathbf{q})
-\mathbf{k}^2
\frac{\mathbf{u}\cdot(\mathbf{k}+\mathbf{q})}{xE}
-\mathbf{q}\cdot(\mathbf{q}+\mathbf{k})
\frac{\mathbf{u}\cdot\mathbf{k}}{xE}
\nonumber \\
&\hspace{1.1cm}
+\mathbf{k}\cdot(\mathbf{k}+\mathbf{q})
\frac{\mathbf{u}\cdot\mathbf{q}}{2xE} \,
\frac{\mathbf{k}^2-(\mathbf{k}+\mathbf{q})^2}
{v^2(\mathbf{q})}\,
\frac{\partial v^2(\mathbf{q})}{\partial\mathbf{q}^2} 
-\mathbf{k}\cdot(\mathbf{k}+\mathbf{q})
\frac{\mathbf{u}\cdot\mathbf{q}}{xE}\,
\frac{2x^2M^2}
{(\mathbf{k}+\mathbf{q})^2+\chi}
\Bigg]
\nonumber \\
&\hspace{0.6cm}\times
\Bigg[
\cos\left(
\frac{\mathbf{k}^2+\chi}{2xE} \,z
\right)
-\cos\left(
\frac{\mathbf{k}^2-(\mathbf{k}+\mathbf{q})^2}{2xE} \,z
\right)
\Bigg]
\nonumber \\
&\quad
+\mathbf{k}\cdot(\mathbf{k}+\mathbf{q})
\frac{\mathbf{u}\cdot\mathbf{q}}{xE} \,
\frac{(\mathbf{k}+\mathbf{q})^2+m_g^2-x^2M^2}{2xE} \,
z \,
\sin\left(
\frac{\mathbf{k}^2-(\mathbf{k}+\mathbf{q})^2}{2xE} \,z
\right)
\nonumber \\
&\quad
-\frac{3}{16} \,
\left[(\mathbf{k}+\mathbf{q})^2+\chi\right]
\frac{R_0}{xE} \,
\frac{\mathbf{k}^2}{v^2(\mathbf{q})}\,
\frac{\partial v^2(\mathbf{q})}{\partial\mathbf{q}^2} \,
\sin\left(
\frac{\mathbf{k}^2+\chi}{2xE} \,z
\right)
\Bigg\} \,,
\label{eq:M0M202final}
\end{align}
\end{widetext}
where terms odd in $\mathbf{q}$ have been omitted, as they vanish upon transverse-momentum integration.

The first-subeikonal terms in Eq.~\eqref{eq:M0M202final} have the same physical origins as those in Eq.~\eqref{eq:M0M201final}: subleading Lorentz contractions, flow-induced potential-derivative corrections, a mass-dependent  residue-denominator correction, and two distinct sine contributions. The phase-induced $z\sin$ term arises from the surviving subleading correction to the longitudinal phase of $\mathcal{J}^{2}_{2,0,2}$, whereas the ordinary sine term originates from the scattering-potential-pole residues of $\mathcal{J}^{1}_{2,0,2}$. These mechanisms follow the classification in Sec.~\ref{FullSpectrumClassification}.

Although evaluated independently, Eq.~\eqref{eq:M0M202final} maps onto Eq.~\eqref{eq:M0M201final} under the interchange $(q_1,a_1)\leftrightarrow(q_2,a_2)$, corresponding to $\mathbf{q}\to-\mathbf{q}$ in the CL configuration. This provides a direct consistency check of the two calculations and confirms the equivalence of their CL contributions, consistent with Ref.~\cite{DGLV}. To avoid double counting, only one contribution is retained in the final two-scattering result, or equivalently, each may be included with a factor of $1/2$.

Among the two-scattering contributions, phase-induced $z\sin\phi_0$ terms arise only in the interferences of $\mathcal{M}_0^\dagger$ with $\mathcal{M}_{2,0,1}^{c}$ and $\mathcal{M}_{2,0,2}^{c}$, where subleading gluon-heavy-quark pole shifts survive in the longitudinal combinations determining the formation phases. For the remaining diagrams, these shifts either cancel or do not contribute at first-subeikonal accuracy. This mechanism is distinct from the ordinary sine contributions generated by scattering-potential-pole residues.

\section{\label{sec:M210} Two-scattering contributions $\mathcal{M}_{2, 1, 0}$ and $\mathcal{M}_{2, 1, 1}$}

We next consider the CL contributions of $\mathcal{M}_{2,1,0}$ and $\mathcal{M}_{2,1,1}$, in which one scattering occurs before and the other after gluon emission, with momentum transfers $q_1$ and $q_2$, emitted gluon momentum $k$, and final heavy-quark momentum $p-k$. These contributions vanish at leading-eikonal order in the (D)GLV treatment~\cite{GLV2, GLV1,DGLV,timeordered1}, but were shown to survive at first-subeikonal order in the massless case~\cite{KML}. Here, we independently examine them at the same accuracy, retaining finite heavy-quark and gluon masses and transverse medium flow.

%%%%M210
%%%%%%%%%%%%%%%%%%%%%%%%%%%%%%%%%%%%%%%%%%%%
%%%%%%%%%%%%%%%%%%%%%%%%%%%%%%%%%%%%%%%%%%%%%%%
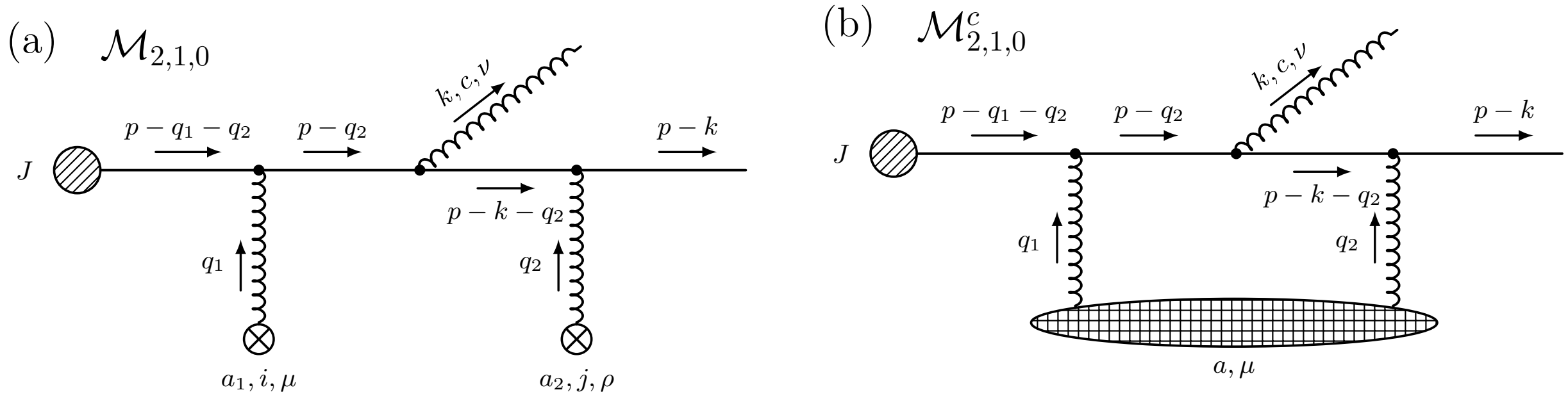
\begin{figure*}[t]
\centering

% ============================================================
% PANEL (a): M_{2,1,0}, non-CL
% ============================================================
\begin{minipage}[t]{0.43\textwidth}
\centering
\resizebox{\linewidth}{!}{%
\begin{tikzpicture}[
    line cap=round,
    line join=round,
    >=latex
]

\path[use as bounding box] (-0.8,-2.55) rectangle (8.25,1.65);

\node at (-0.55,1.55) {\Large (a)};
\node at (0.95,1.45) {\Large $\mathcal{M}_{2,1,0}$};

\draw[
    pattern=north east lines,
    pattern color=black,
    line width=0.8pt
] (0,0) circle (0.28);

\node[left=4pt] at (-0.28,0) {$J$};

\coordinate (V1) at (2.20,0.00);   % prva interakcija
\coordinate (V2) at (4.15,0.00);   % emisija gluona
\coordinate (V3) at (6.05,0.00);   % druga interakcija

\coordinate (QinStart) at (0.28,0.00);
\coordinate (Qout)     at (8.10,0.00);

\coordinate (Gout) at (6.10,1.50);

\coordinate (C1) at (2.20,-2.05);
\coordinate (C2) at (6.05,-2.05);

\coordinate (B1) at (2.20,-1.87);
\coordinate (B2) at (6.05,-1.87);

\draw[line width=0.85pt] (QinStart) -- (V1);
\draw[line width=0.85pt] (V1) -- (V2);
\draw[line width=0.85pt] (V2) -- (V3);
\draw[line width=0.85pt] (V3) -- (Qout);

\draw[
    line width=0.85pt,
    decorate,
    decoration={
        coil,
        amplitude=2.1pt,
        segment length=4.8pt
    }
] (V1) -- (B1);

\draw[
    line width=0.85pt,
    decorate,
    decoration={
        coil,
        amplitude=2.1pt,
        segment length=4.8pt
    }
] (V3) -- (B2);

\draw[
    line width=0.85pt,
    decorate,
    decoration={
        coil,
        amplitude=2.3pt,
        segment length=5.2pt
    }
] (V2) -- (Gout);

\fill[black] (V1) circle (1.9pt);
\fill[black] (V2) circle (1.9pt);
\fill[black] (V3) circle (1.9pt);

\draw[line width=0.8pt] (C1) circle (0.18);

\draw[line width=0.8pt]
    ($(C1)+(-0.12,-0.12)$) --
    ($(C1)+(0.12,0.12)$);

\draw[line width=0.8pt]
    ($(C1)+(-0.12,0.12)$) --
    ($(C1)+(0.12,-0.12)$);

\draw[line width=0.8pt] (C2) circle (0.18);

\draw[line width=0.8pt]
    ($(C2)+(-0.12,-0.12)$) --
    ($(C2)+(0.12,0.12)$);

\draw[line width=0.8pt]
    ($(C2)+(-0.12,0.12)$) --
    ($(C2)+(0.12,-0.12)$);

\node[below=7pt] at (C1) {$a_1,i,\mu$};
\node[below=7pt] at (C2) {$a_2,j,\rho$};

\draw[line width=0.75pt,->]
    (0.95,0.22) -- (1.75,0.22);

\node[above=1pt] at (1.35,0.22)
    {$p-q_1-q_2$};

\draw[line width=0.75pt,->]
    (2.75,0.22) -- (3.45,0.22);

\node[above=1pt] at (3.10,0.22)
    {$p-q_2$};

\draw[line width=0.75pt,->]
    (4.85,-0.22) -- (5.55,-0.22);

\node[below=1pt] at (5.20,-0.22)
    {$p-k-q_2$};

\draw[line width=0.75pt,->]
    (7.05,0.22) -- (7.75,0.22);

\node[above=1pt] at (7.40,0.22)
    {$p-k$};

\draw[line width=0.75pt,->]
    (1.98,-1.46) -- (1.98,-0.84);

\node[left=2pt] at (1.98,-1.15)
    {$q_1$};

\draw[line width=0.75pt,->]
    (5.83,-1.46) -- (5.83,-0.84);

\node[left=2pt] at (5.83,-1.15)
    {$q_2$};

\draw[line width=0.75pt,->]
    (4.57,0.59) -- (5.19,1.07)
    node[midway, above=1pt, sloped] {$k,c,\nu$};

\end{tikzpicture}%
}
\end{minipage}%
\hspace{0.04\textwidth}%
%
% ============================================================
% PANEL (b): M^c_{2,1,0}, CL
% ============================================================
\begin{minipage}[t]{0.43\textwidth}
\centering
\resizebox{\linewidth}{!}{%
\begin{tikzpicture}[
    line cap=round,
    line join=round,
    >=latex
]

\path[use as bounding box] (-0.8,-2.75) rectangle (8.25,1.65);

\node at (-0.55,1.55) {\Large (b)};
\node at (0.95,1.45) {\Large $\mathcal{M}^{c}_{2,1,0}$};

\draw[
    pattern=north east lines,
    pattern color=black,
    line width=0.8pt
] (0,0) circle (0.28);

\node[left=4pt] at (-0.28,0) {$J$};

\coordinate (V1) at (2.20,0.00);   % prva interakcija
\coordinate (V2) at (4.15,0.00);   % emisija gluona
\coordinate (V3) at (6.05,0.00);   % druga interakcija

\coordinate (QinStart) at (0.28,0.00);
\coordinate (Qout)     at (8.10,0.00);

\coordinate (Gout) at (6.10,1.50);

\coordinate (B1) at (2.20,-1.87);
\coordinate (B2) at (6.05,-1.87);

\draw[line width=0.85pt] (QinStart) -- (V1);
\draw[line width=0.85pt] (V1) -- (V2);
\draw[line width=0.85pt] (V2) -- (V3);
\draw[line width=0.85pt] (V3) -- (Qout);

\draw[
    line width=0.85pt,
    decorate,
    decoration={
        coil,
        amplitude=2.1pt,
        segment length=4.8pt
    }
] (V1) -- (B1);

\draw[
    line width=0.85pt,
    decorate,
    decoration={
        coil,
        amplitude=2.1pt,
        segment length=4.8pt
    }
] (V3) -- (B2);

\draw[
    line width=0.85pt,
    decorate,
    decoration={
        coil,
        amplitude=2.3pt,
        segment length=5.2pt
    }
] (V2) -- (Gout);

\fill[black] (V1) circle (1.9pt);
\fill[black] (V2) circle (1.9pt);
\fill[black] (V3) circle (1.9pt);

\draw[
    line width=0.8pt,
    pattern=grid,
    pattern color=black
] (4.125,-2.05) ellipse (2.46 and 0.29);
\node at (4.125,-2.62) {$a,\mu$};

\draw[line width=0.75pt,->]
    (0.95,0.22) -- (1.75,0.22);

\node[above=1pt] at (1.35,0.22)
    {$p-q_1-q_2$};

\draw[line width=0.75pt,->]
    (2.75,0.22) -- (3.45,0.22);

\node[above=1pt] at (3.10,0.22)
    {$p-q_2$};

\draw[line width=0.75pt,->]
    (4.85,-0.22) -- (5.55,-0.22);

\node[below=1pt] at (5.20,-0.22)
    {$p-k-q_2$};

\draw[line width=0.75pt,->]
    (7.05,0.22) -- (7.75,0.22);

\node[above=1pt] at (7.40,0.22)
    {$p-k$};

\draw[line width=0.75pt,->]
    (1.98,-1.32) -- (1.98,-0.7);

\node[left=2pt] at (1.98,-1.1)
    {$q_1$};

\draw[line width=0.75pt,->]
    (5.83,-1.32) -- (5.83,-0.7);

\node[left=2pt] at (5.83,-1.1)
    {$q_2$};

\draw[line width=0.75pt,->]
    (4.57,0.59) -- (5.19,1.07)
    node[midway, above=1pt, sloped] {$k,c,\nu$};

\end{tikzpicture}%
}
\end{minipage}
\caption{
Diagram $\mathcal{M}_{2,1,0}$ contributing at first order in opacity  $L/{\lambda}$:
(a) two interactions with the medium, occurring before and after gluon emission, respectively, and
(b) the corresponding contact-limit configuration.
}
\label{fig:M210}

\end{figure*}

For the momentum assignment shown in Fig.~\ref{fig:M210}, the amplitude is
\begin{widetext}
\begin{align}
\mathcal{M}_{2,1,0}
{}&=
\int\frac{d^4q_1}{(2\pi)^4}
\int\frac{d^4q_2}{(2\pi)^4}
\left[
i g_s
(2p-2k-q_2)_{\rho}
A_{ext}^{\rho, a_2}(q_2)
T_{ jet}^{a_2}
\right]
\frac{i}{(p-k-q_2)^2-M^2+i\epsilon}
\nonumber \\
&\times
\left[
i g_s
(2p-k-2q_2)^{\nu}
\varepsilon_{\nu}^{*}(k)
T_{ jet}^{c}
\right]
\frac{i}{(p-q_2)^2-M^2+i\epsilon}
\nonumber \\
&\times
\left[
i g_s
(2p-q_1-2q_2)_{\mu}
A_{ext}^{\mu, a_1}(q_1)
T_{jet}^{a_1}
\right]
\frac{i}{(p-q_1-q_2)^2-M^2+i\epsilon} \,
iJ(p-q_1-q_2) \, .
\label{m210_p}
\end{align}
\end{widetext}

Inserting the external fields and using the CL constraint obtained from the target color trace in Eq.~\eqref{tr2} and the medium average, the interference with $\mathcal{M}_0^\dagger$ takes the form
\begin{widetext}
\begin{align}
\frac{2}{d_T}
\operatorname{Re}\operatorname{Tr}
\left\langle
\mathcal{M}_0^\dagger
\mathcal{M}_{2,1,0}^{c}
\right\rangle
{}&=
\frac{32 g_s^2 C_2(T)  N}{A_\perp d_A} \operatorname{Tr}(c a_1 c a_1)\,
\operatorname{Re}
\Bigg\{
\int_0^L dz\,\rho(z)
\int\frac{d^2\mathbf{q}}{(2\pi)^2}\, \left|J(E, \mathbf{p})\right|^2
\frac{
E^2 \,  \mathbf{k}^2
}{
x\left(
\mathbf{k}^2+\chi
\right)
} \int\frac{dq_{1z}}{2\pi}
\int\frac{dq_{2z}}{2\pi}\,
\nonumber\\
&\times
e^{-iq_{1z}z}
e^{-iq_{2z}z}\,
\frac{
 v(q_1) v(q_2)
}{
\left[
(p-q_1-q_2)^2-M^2+i\epsilon
\right]
\left[
(p-k-q_2)^2-M^2+i\epsilon
\right]
}
\frac{1}{
(p-q_2)^2-M^2+i\epsilon
}
\Bigg\} \, ,
\label{eq:M0M210}
\end{align}
\end{widetext}
where, $q_1$ and $q_2$ are specified in Eqs.~\eqref{kin1M2} and~\eqref{kin2M2}. 

We first perform the $q_{1z}$ integration,
\begin{align}
\mathcal{I}_{2,1,0}
{}&=
\int\frac{dq_{1z}}{2\pi}
\frac{
e^{-iq_{1z}z} \,
v(q_1)
}{
(p-q_1-q_2)^2-M^2+i\epsilon
} \, .
\label{IM210p}
\end{align}
Closing the contour in the lower half-plane, the scattering-potential pole is exponentially suppressed according to Eq.~\eqref{eq:separation}. The only contributing pole is therefore that of the heavy-quark propagator, $q^-_{1z,p-q_1-q_2}$, given in Eq.~\eqref{m220_pol}. As established there, it receives no $\mathcal{O}\!\left(\perp^3/(xE)^2\right)$ correction.

Having evaluated the residue at $q^-_{1z,p-q_1-q_2}$, the remaining longitudinal integration becomes
\begin{widetext}
\begin{align}
\mathcal{J}_{2,1,0}
={}&
\int\frac{dq_{2z}}{2\pi}\,
\frac{
e^{-iq_{2z}z} \,
v(q_2)
}{\left[
(p-k-q_2)^2-M^2+i\epsilon
\right] \left[  (p-q_2)^2-M^2+i\epsilon \right]} \,
\mathcal{I}_{2,1,0} \, .
\label{JM210}
\end{align}
\end{widetext}
The factor $e^{iq_{2z}z}$ from the $q_{1z}$ residue cancels the explicit $e^{-iq_{2z}z}$ in Eq.~\eqref{JM210}, leaving a longitudinal phase independent of $q_{2z}$. Since the $q_{1z}$ pole receives no $\mathcal{O}\!\left(\perp^3/(xE)^2\right)$ correction, no first-subeikonal formation-phase shift arises.

The integral $\mathcal{J}_{2,1,0}$ contains four poles in the lower half-plane: the heavy-quark propagator poles in Eqs.~\eqref{q2pkq2M200} and~\eqref{q2pq2}, and the scattering-potential poles in Eqs.~\eqref{q2v1} and~\eqref{q2v2}. The two heavy-quark poles have identical $\mathcal{O}\!\left(\perp^3/(xE)^2\right)$ corrections, which cancel in their mutual pole difference. Their separations from either scattering-potential pole, however, retain leading $\mathcal{O}(\perp)$ contributions, so the subleading shifts remain beyond first-subeikonal accuracy in the corresponding residue denominators. No additional contributions arise from these shifts through the Lorentz-contraction prefactors,  as we checked.

Summing the residues of $\mathcal{J}_{2,1,0}$ and substituting the result into Eq.~\eqref{eq:M0M210}, we obtain, through first-subeikonal order and linear in $\mathbf{u}$,
\begin{widetext}
\begin{align}
\frac{2}{d_T}
\operatorname{Re}\operatorname{Tr}
\left\langle
\mathcal{M}_0^\dagger
\mathcal{M}_{2,1,0}^{c}
\right\rangle
{}&=
\frac{8  g_s^2 C_2(T) N }{ A_\perp d_A}
\operatorname{Tr}(c a_1 c a_1)
\int_0^L dz \rho(z)
\int\frac{d^2\mathbf{q}}{(2\pi)^2}
\left|v(\mathbf{q})\right|^2
\left|J(E, \mathbf{p})\right|^2
\frac{\mathbf{k}^2}{\left(\mathbf{k}^2+\chi\right)^2}
\nonumber \\
&\times
\frac{3}{16}
\frac{R_0}{xE}
\frac{\left(\mathbf{k}^2+\chi\right)}{v^2(\mathbf{q})}
\frac{\partial v^2(\mathbf{q})}{\partial \mathbf{q}^2}
\sin\left(
\frac{\mathbf{k}^2+\chi}{2xE}z
\right) \, .
\label{eq:M0M210final}
\end{align}
\end{widetext}
The term odd under $\mathbf{q}\to-\mathbf{q}$ has been omitted, since it vanishes upon transverse-momentum integration.

Unlike the vanishing leading-eikonal CL contribution of Ref.~\cite{DGLV}, Eq.~\eqref{eq:M0M210final} contains a nonzero first-subeikonal term originating from the scattering-potential-pole residues of $\mathcal{J}_{2,1,0}$. According to the classification in Sec.~\ref{FullSpectrumClassification}, this is a scattering-potential-pole sine contribution, rather than a phase-induced first-subeikonal correction. 

In the massless limit, $M=m_g=0$, our result reduces to Eq.~(53) of Ref.~\cite{KML} upon expansion to linear order in $\mathbf{u}$, consistently with Eq.~\eqref{eq:flowexpansion}.

%%%%%%%%%%%%%%%%%%%%%%%%%%%%%%%%%%%%
%%%%%%%%%%%%%%%%%%%%%%%%%%%%%%%%%%%%%%
%%%M211

\begin{figure*}[!t]
\centering

% ============================================================
% PANEL (a): M_{2,1,1}, non-CL
% ============================================================
\begin{minipage}[t]{0.43\textwidth}
\centering
\resizebox{\linewidth}{!}{%
\begin{tikzpicture}[
    line cap=round,
    line join=round,
    >=latex
]

\path[use as bounding box] (-0.8,-2.65) rectangle (8.25,1.65);

\node at (-0.55,1.55) {\Large (a)};
\node at (0.95,1.45) {\Large $\mathcal{M}_{2,1,1}$};

\draw[
    pattern=north east lines,
    pattern color=black,
    line width=0.8pt
] (0,0) circle (0.28);

\node[left=4pt] at (-0.28,0) {$J$};

\coordinate (V1) at (2.20,0.00);

\coordinate (V2) at (4.15,0.00);

\coordinate (G2) at (6.05,-0.85);

\coordinate (QinStart) at (0.28,0.00);
\coordinate (Qout)     at (8.10,0.00);

\coordinate (Gout) at (8.10,-0.85);

\coordinate (C1) at (2.20,-2.20);
\coordinate (C2) at (6.05,-2.20);

\coordinate (B1) at (2.20,-2.02);
\coordinate (B2) at (6.05,-2.02);

\draw[line width=0.85pt] (QinStart) -- (V1);
\draw[line width=0.85pt] (V1) -- (V2);
\draw[line width=0.85pt] (V2) -- (Qout);

\draw[
    line width=0.85pt,
    decorate,
    decoration={
        coil,
        amplitude=2.3pt,
        segment length=5.2pt
    }
] (V2) -- (G2);

\draw[
    line width=0.85pt,
    decorate,
    decoration={
        coil,
        amplitude=2.3pt,
        segment length=5.2pt
    }
] (G2) -- (Gout);

\draw[
    line width=0.85pt,
    decorate,
    decoration={
        coil,
        amplitude=2.1pt,
        segment length=4.8pt
    }
] (V1) -- (B1);

\draw[
    line width=0.85pt,
    decorate,
    decoration={
        coil,
        amplitude=2.1pt,
        segment length=4.8pt
    }
] (G2) -- (B2);

\fill[black] (V1) circle (1.9pt);
\fill[black] (V2) circle (1.9pt);
\fill[black] (G2) circle (1.9pt);

\draw[line width=0.8pt] (C1) circle (0.18);

\draw[line width=0.8pt]
    ($(C1)+(-0.12,-0.12)$) --
    ($(C1)+(0.12,0.12)$);

\draw[line width=0.8pt]
    ($(C1)+(-0.12,0.12)$) --
    ($(C1)+(0.12,-0.12)$);

\draw[line width=0.8pt] (C2) circle (0.18);

\draw[line width=0.8pt]
    ($(C2)+(-0.12,-0.12)$) --
    ($(C2)+(0.12,0.12)$);

\draw[line width=0.8pt]
    ($(C2)+(-0.12,0.12)$) --
    ($(C2)+(0.12,-0.12)$);

\node[below=7pt] at (C1) {$a_1,i,\mu$};
\node[below=7pt] at (C2) {$a_2,j,\rho$};

\draw[line width=0.75pt,->]
    (0.95,0.22) -- (1.75,0.22);

\node[above=1pt] at (1.35,0.22)
    {$p-q_1-q_2$};

\draw[line width=0.75pt,->]
    (2.75,0.22) -- (3.45,0.22);

\node[above=1pt] at (3.10,0.22)
    {$p-q_2$};

\draw[line width=0.75pt,->]
    (5.72,0.22) -- (6.52,0.22);

\node[above=1pt] at (6.12,0.22)
    {$p-k$};

\draw[line width=0.75pt,->]
    (4.78,-0.50) -- (5.52,-0.83)
    node[midway, below=2pt, sloped]
    {$k-q_2$};

\draw[line width=0.75pt,->]
    (6.75,-0.63) -- (7.55,-0.63);

\node[above=1pt] at (7.15,-0.63)
    {$k,c,\nu$};

\draw[line width=0.75pt,->]
    (1.98,-1.62) -- (1.98,-1.0);

\node[left=2pt] at (1.98,-1.40)
    {$q_1$};

\draw[line width=0.75pt,->]
    (5.83,-1.82) -- (5.83,-1.20);

\node[left=2pt] at (5.83,-1.60)
    {$q_2$};

\end{tikzpicture}%
}
\end{minipage}%
\hspace{0.04\textwidth}%
%
% ============================================================
% PANEL (b): M^c_{2,1,1}, CL
% ============================================================
\begin{minipage}[t]{0.43\textwidth}
\centering
\resizebox{\linewidth}{!}{%
\begin{tikzpicture}[
    line cap=round,
    line join=round,
    >=latex
]

\path[use as bounding box] (-0.8,-2.75) rectangle (8.25,1.65);

\node at (-0.55,1.55) {\Large (b)};
\node at (0.95,1.45) {\Large $\mathcal{M}^{c}_{2,1,1}$};

\draw[
    pattern=north east lines,
    pattern color=black,
    line width=0.8pt
] (0,0) circle (0.28);

\node[left=4pt] at (-0.28,0) {$J$};

\coordinate (V1) at (2.20,0.00);

\coordinate (V2) at (4.15,0.00);

\coordinate (G2) at (6.05,-0.85);

\coordinate (QinStart) at (0.28,0.00);
\coordinate (Qout)     at (8.10,0.00);

\coordinate (Gout) at (8.10,-0.85);

\coordinate (B1) at (2.20,-2.024);
\coordinate (B2) at (6.05,-2.024);

\draw[line width=0.85pt] (QinStart) -- (V1);
\draw[line width=0.85pt] (V1) -- (V2);
\draw[line width=0.85pt] (V2) -- (Qout);

\draw[
    line width=0.85pt,
    decorate,
    decoration={
        coil,
        amplitude=2.3pt,
        segment length=5.2pt
    }
] (V2) -- (G2);

\draw[
    line width=0.85pt,
    decorate,
    decoration={
        coil,
        amplitude=2.3pt,
        segment length=5.2pt
    }
] (G2) -- (Gout);

\draw[
    line width=0.85pt,
    decorate,
    decoration={
        coil,
        amplitude=2.1pt,
        segment length=4.8pt
    }
] (V1) -- (B1);

\draw[
    line width=0.85pt,
    decorate,
    decoration={
        coil,
        amplitude=2.1pt,
        segment length=4.8pt
    }
] (G2) -- (B2);

\fill[black] (V1) circle (1.9pt);
\fill[black] (V2) circle (1.9pt);
\fill[black] (G2) circle (1.9pt);

\draw[
    line width=0.8pt,
    pattern=grid,
    pattern color=black
] (4.125,-2.18) ellipse (2.46 and 0.25);

\node at (4.125,-2.72) {$a,\mu$};

\draw[line width=0.75pt,->]
    (0.95,0.22) -- (1.75,0.22);

\node[above=1pt] at (1.35,0.22)
    {$p-q_1-q_2$};

\draw[line width=0.75pt,->]
    (2.75,0.22) -- (3.45,0.22);

\node[above=1pt] at (3.10,0.22)
    {$p-q_2$};

\draw[line width=0.75pt,->]
    (5.72,0.22) -- (6.52,0.22);

\node[above=1pt] at (6.12,0.22)
    {$p-k$};

\draw[line width=0.75pt,->]
    (4.78,-0.50) -- (5.52,-0.83)
    node[midway, below=2pt, sloped]
    {$k-q_2$};

\draw[line width=0.75pt,->]
    (6.75,-0.63) -- (7.55,-0.63);

\node[above=1pt] at (7.15,-0.63)
    {$k,c,\nu$};

\draw[line width=0.75pt,->]
    (1.98,-1.62) -- (1.98,-1.0);

\node[left=2pt] at (1.98,-1.40)
    {$q_1$};

\draw[line width=0.75pt,->]
    (5.83,-1.82) -- (5.83,-1.20);

\node[left=2pt] at (5.83,-1.60)
    {$q_2$};

\end{tikzpicture}%
}
\end{minipage}
\caption{
Diagram $\mathcal{M}_{2,1,1}$ contributing at first order in opacity  $L/{\lambda}$:
(a) the heavy quark interacts with the medium before gluon emission, while the emitted gluon undergoes the second interaction, and
(b) the corresponding contact-limit configuration.
} 
\label{fig:M211}
\end{figure*}
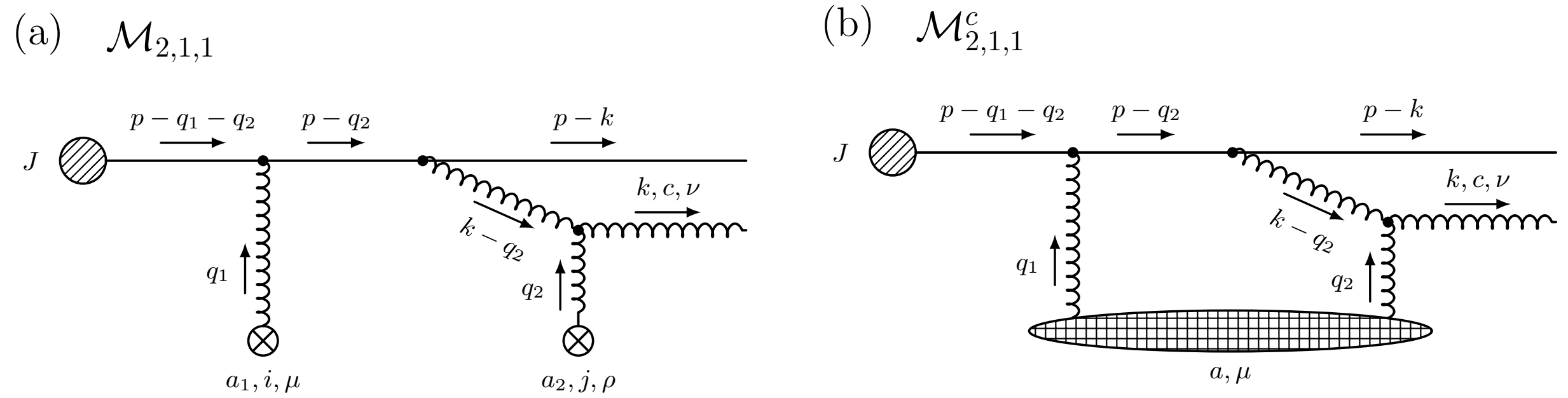
%%%%%%%%%%%%%%%%%%%%

The related diagram $\mathcal{M}_{2,1,1}$, shown in Fig.~\ref{fig:M211}, involves a heavy-quark scattering with momentum transfer $q_1$ before gluon emission, followed by an interaction of the emitted gluon with the medium through momentum transfer $q_2$. The final heavy quark and gluon carry momenta $p-k$ and $k$, respectively. Its amplitude reads
\begin{widetext}
\begin{align}
\mathcal{M}_{2,1,1}
={}&
\int \frac{d^4q_1}{(2\pi)^4}
\int \frac{d^4q_2}{(2\pi)^4}
\varepsilon^{*\nu}(k) \,
\Gamma_{\delta \nu\rho}^{a c a_2} (k-q_2, -k, q_2)\,
A_{ext}^{\rho, a_2}(q_2) \,
\frac{(-i)P^{\delta \delta'}(k-q_2)\delta_{aa'}}
{(k-q_2)^2-m_g^2+i\epsilon} \,
 \left[
i g_s(2p-k-q_2)_{\delta'} T_{jet}^{a'}
\right]
\nonumber \\
&\times
\frac{i}{(p-q_2)^2-M^2+i\epsilon} \,
 \left[
i g_s(2p-q_1-2q_2)_\mu \, A_{ext}^{\mu, a_1}(q_1) T_{jet}^{a_1}
\right]\,
\frac{i}{(p-q_1-q_2)^2-M^2+i\epsilon} \,
iJ(p-q_1-q_2) \, .
\label{m211_p}
\end{align}
\end{widetext}
As established for the analogous three-gluon structure in Sec.~\ref{sec:M203}, the fully contracted Lorentz structure is independent of $q_{1z}$ and $q_{2z}$ at first-subeikonal accuracy and introduces no additional longitudinal poles. Using the CL momentum constraint in Eq.~\eqref{q1q2}, the interference with $\mathcal{M}_0^\dagger$ is given by
\begin{widetext}
\begin{align}
\frac{2}{d_T}
\operatorname{Re}\operatorname{Tr}
\left\langle
\mathcal{M}_0^\dagger
\mathcal{M}_{2,1,1}^{c}
\right\rangle
{}&=
\frac{32 g_s^2 C_2(T)  N}{A_\perp d_A} \alpha \,
\operatorname{Re}
\Bigg\{
\int_0^L dz\,\rho(z)
\int\frac{d^2\mathbf{q}}{(2\pi)^2}\, \left|J(E, \mathbf{p})\right|^2
\frac{
E^2 }{
\mathbf{k}^2+\chi
} 
\int\frac{dq_{1z}}{2\pi}
\int\frac{dq_{2z}}{2\pi}\,
e^{-iq_{1z}z}
e^{-iq_{2z}z}\,
\nonumber\\
&\times
\left[ \mathbf{k} \cdot (\mathbf{k}+ \mathbf{q}) - \mathbf{k}^2 \frac{\mathbf{u} \cdot  (\mathbf{k}+ \mathbf{q})}{x E} -\mathbf{q} \cdot (\mathbf{k} + \mathbf{q}) \,  \frac{\mathbf{u} \cdot \mathbf{k}}{x E}\right] \, 
\frac{
 v(q_1) v(q_2)
}{
\left[
(p-q_1-q_2)^2-M^2+i\epsilon
\right]
\left[
(p-q_2)^2-M^2+i\epsilon
\right]
}
\nonumber \\
&\times \frac{1}{
(k-q_2)^2-m_g^2+i\epsilon
}
\Bigg\} \, .
\label{eq:M0M211}
\end{align}
\end{widetext}

The $q_{1z}$ integration coincides with Eq.~\eqref{IM210p} and therefore has the same heavy-quark pole and residue. Hence, $\mathcal{I}_{2,1,1}=\mathcal{I}_{2,1,0}$, and the remaining longitudinal integral becomes
\begin{widetext}
\begin{align}
\mathcal{J}_{2,1,1}
={}&
\int\frac{dq_{2z}}{2\pi}\,
\frac{
e^{-iq_{2z}z} \,
v(q_2)
}{\left[
(p-q_2)^2-M^2+i\epsilon
\right] \left[  (k-q_2)^2-m_g^2+i\epsilon \right]} \,
\mathcal{I}_{2,1,1} \, .
\label{JM211}
\end{align}
\end{widetext}
The $q_{1z}$ residue removes the explicit $q_{2z}$ dependence of the exponential in Eq.~\eqref{JM211}. Consequently, the remaining formation phase receives no first-subeikonal correction.

The integral $\mathcal{J}_{2,1,1}$ contains four poles in the lower half-plane: the gluon-propagator pole in Eq.~\eqref{q2kq2}, the heavy-quark pole in Eq.~\eqref{q2pq2}, and the scattering-potential poles in Eqs.~\eqref{q2v1} and~\eqref{q2v2}. In the difference between the gluon and heavy-quark propagator poles, the leading $\mathcal{O}(\perp)$ terms cancel, so the surviving $\mathcal{O}\!\left(\perp^3/(xE)^2\right)$ mismatch contributes to their individual residue denominators at relative first-subeikonal order. However, these corrections enter the two propagator residues with opposite signs and cancel upon summation. By contrast, differences between either propagator pole and a scattering-potential pole retain their leading $\mathcal{O}(\perp)$ separation, leaving the corresponding subleading shifts beyond the working accuracy. Thus, the first-subeikonal corrections to the individual residue denominators cancel upon summation, leaving no net contribution to $\mathcal{J}_{2,1,1}$.

Combining all four longitudinal residues, the interference contribution reduces to
\begin{widetext}
\begin{align}
\frac{2}{d_T}
\operatorname{Re}\operatorname{Tr}
\left\langle
\mathcal{M}_0^\dagger
\mathcal{M}_{2,1,1}^{c}
\right\rangle
{}&=
\frac{4  g_s^2 C_2(T) N }{ A_\perp d_A}
\alpha
\int_0^L dz \rho(z)
\int\frac{d^2\mathbf{q}}{(2\pi)^2}
\left|v(\mathbf{q})\right|^2
\left|J(E, \mathbf{p})\right|^2
\frac{1}{\mathbf{k}^2+\chi}
\nonumber \\
&\times
\Bigg\{
\frac{\mathbf{u} \cdot \mathbf{k}}{2 xE}
\frac{\mathbf{q}^2}{v^2(\mathbf{q})}
\frac{\partial v^2(\mathbf{q})}{\partial \mathbf{q}^2}
\cos\left(
\frac{\mathbf{k}^2+\chi}{2xE}z
\right) \, 
+\frac{3}{8}
\frac{R_0}{xE}
\frac{\mathbf{k}^2}{v^2(\mathbf{q})}
\frac{\partial v^2(\mathbf{q})}{\partial \mathbf{q}^2}
\sin\left(
\frac{\mathbf{k}^2+\chi}{2xE}z
\right) 
\Bigg\} \, .
\label{eq:M0M211final}
\end{align}
\end{widetext}
Terms odd under $\mathbf{q}\to-\mathbf{q}$ have been omitted, since they vanish upon transverse-momentum integration. Of the three terms in the Lorentz contraction of Eq.~\eqref{eq:M0M211}, only the leading $\mathcal{O}(\perp^2)$ term contributes at the working accuracy; the remaining two enter at higher order.

Unlike the vanishing leading-eikonal CL contribution of Ref.~\cite{DGLV}, Eq.~\eqref{eq:M0M211final} yields a nonzero first-subeikonal result. Following the classification in Sec.~\ref{FullSpectrumClassification}, the cosine term represents a flow-induced potential-derivative correction, receiving contributions from all four longitudinal residues, whereas the ordinary sine term originates solely from the scattering-potential-pole residues. Neither a mass-dependent  residue-denominator correction nor a phase-induced $z\sin$ contribution survives.

In the massless limit, $M=m_g=0$, Eq.~\eqref{eq:M0M211final} reduces to Eq.~(58) of Ref.~\cite{KML} upon expansion to linear order in $\mathbf{u}$, consistently with Eq.~\eqref{eq:flowexpansion}.

\begin{figure*}[!t]
\centering

% ============================================================
% PANEL (a): 4-gluon vertex, non-CL
% ============================================================
\begin{minipage}[t]{0.43\textwidth}
\centering
\resizebox{\linewidth}{!}{%
\begin{tikzpicture}[
    line cap=round,
    line join=round,
    >=latex
]

\path[use as bounding box] (-0.8,-2.65) rectangle (8.25,1.65);

\node at (-0.55,1.55) {\Large (a)};
\node at (1.10,1.45) {\Large $\mathcal{M}^{(4g)}_{2,0,3}$};

\draw[
    pattern=north east lines,
    pattern color=black,
    line width=0.8pt
] (0,0) circle (0.28);

\node[left=4pt] at (-0.28,0) {$J$};

\coordinate (Vq) at (2.20,0.00);

\coordinate (V4) at (5.10,-0.95);

\coordinate (QinStart) at (0.28,0.00);
\coordinate (Qout)     at (8.10,0.00);

\coordinate (Gout) at (8.10,-0.95);

\coordinate (C1) at (4.15,-2.20);
\coordinate (C2) at (6.05,-2.20);

\coordinate (B1) at (4.15,-2.02);
\coordinate (B2) at (6.05,-2.02);

\draw[line width=0.85pt] (QinStart) -- (Vq);
\draw[line width=0.85pt] (Vq) -- (Qout);

\draw[
    line width=0.85pt,
    decorate,
    decoration={
        coil,
        amplitude=2.3pt,
        segment length=5.2pt
    }
] (Vq) -- (V4);

\draw[
    line width=0.85pt,
    decorate,
    decoration={
        coil,
        amplitude=2.3pt,
        segment length=5.2pt
    }
] (V4) -- (Gout);

\draw[
    line width=0.85pt,
    decorate,
    decoration={
        coil,
        amplitude=2.1pt,
        segment length=4.8pt
    }
] (B1) -- (V4);

\draw[
    line width=0.85pt,
    decorate,
    decoration={
        coil,
        amplitude=2.1pt,
        segment length=4.8pt
    }
] (B2) -- (V4);

\fill[black] (Vq) circle (1.9pt);

\fill[black] (V4) circle (2.3pt);

\draw[line width=0.8pt] (C1) circle (0.18);

\draw[line width=0.8pt]
    ($(C1)+(-0.12,-0.12)$) --
    ($(C1)+(0.12,0.12)$);

\draw[line width=0.8pt]
    ($(C1)+(-0.12,0.12)$) --
    ($(C1)+(0.12,-0.12)$);

\draw[line width=0.8pt] (C2) circle (0.18);

\draw[line width=0.8pt]
    ($(C2)+(-0.12,-0.12)$) --
    ($(C2)+(0.12,0.12)$);

\draw[line width=0.8pt]
    ($(C2)+(-0.12,0.12)$) --
    ($(C2)+(0.12,-0.12)$);

\node[below=7pt] at (C1) {$a_1,i,\mu$};
\node[below=7pt] at (C2) {$a_2,j,\rho$};

\draw[line width=0.75pt,->]
    (0.95,0.22) -- (1.75,0.22);

\node[above=1pt] at (1.35,0.22)
    {$p-q_1-q_2$};

\draw[line width=0.75pt,->]
    (5.95,0.22) -- (6.75,0.22);

\node[above=1pt] at (6.35,0.22)
    {$p-k$};

\draw[line width=0.75pt,->]
    (2.65,-0.36) -- (3.55,-0.66)
    node[midway, below=1.5pt, sloped]
    {$k-q_1-q_2$};

\draw[line width=0.75pt,->]
    (6.95,-1.17) -- (7.75,-1.17);

\node[below=1pt] at (7.35,-1.17)
    {$k,c,\nu$};

\draw[line width=0.75pt,->]
    (4.03,-1.78) -- (4.45,-1.30)
    node[midway, left=2pt]
    {$q_1$};

\draw[line width=0.75pt,->]
    (6.17,-1.78) -- (5.75,-1.30)
    node[midway, right=2pt]
    {$q_2$};

\end{tikzpicture}%
}
\end{minipage}%
\hspace{0.04\textwidth}%
%
% ============================================================
% PANEL (b): 4-gluon vertex, CL
% ============================================================
\begin{minipage}[t]{0.43\textwidth}
\centering
\resizebox{\linewidth}{!}{%
\begin{tikzpicture}[
    line cap=round,
    line join=round,
    >=latex
]

\path[use as bounding box] (-0.8,-2.75) rectangle (8.25,1.65);

\node at (-0.55,1.55) {\Large (b)};
\node at (1.10,1.45) {\Large $\mathcal{M}^{c,(4g)}_{2,0,3}$};

\draw[
    pattern=north east lines,
    pattern color=black,
    line width=0.8pt
] (0,0) circle (0.28);

\node[left=4pt] at (-0.28,0) {$J$};

\coordinate (Vq) at (2.20,0.00);

\coordinate (V4) at (5.10,-0.95);

\coordinate (QinStart) at (0.28,0.00);
\coordinate (Qout)     at (8.10,0.00);

\coordinate (Gout) at (8.10,-0.95);

\coordinate (B1) at (4.15,-1.991);
\coordinate (B2) at (6.05,-1.991);

\draw[line width=0.85pt] (QinStart) -- (Vq);
\draw[line width=0.85pt] (Vq) -- (Qout);

\draw[
    line width=0.85pt,
    decorate,
    decoration={
        coil,
        amplitude=2.3pt,
        segment length=5.2pt
    }
] (Vq) -- (V4);

\draw[
    line width=0.85pt,
    decorate,
    decoration={
        coil,
        amplitude=2.3pt,
        segment length=5.2pt
    }
] (V4) -- (Gout);

\draw[
    line width=0.85pt,
    decorate,
    decoration={
        coil,
        amplitude=2.1pt,
        segment length=4.8pt
    }
] (B1) -- (V4);

\draw[
    line width=0.85pt,
    decorate,
    decoration={
        coil,
        amplitude=2.1pt,
        segment length=4.8pt
    }
] (B2) -- (V4);

\fill[black] (Vq) circle (1.9pt);
\fill[black] (V4) circle (2.3pt);

\draw[
    line width=0.8pt,
    pattern=grid,
    pattern color=black
] (5.10,-2.18) ellipse (1.45 and 0.25);

\node at (5.10,-2.72) {$a,\mu$};

\draw[line width=0.75pt,->]
    (0.95,0.22) -- (1.75,0.22);

\node[above=1pt] at (1.35,0.22)
    {$p-q_1-q_2$};

\draw[line width=0.75pt,->]
    (5.95,0.22) -- (6.75,0.22);

\node[above=1pt] at (6.35,0.22)
    {$p-k$};

\draw[line width=0.75pt,->]
    (2.65,-0.36) -- (3.55,-0.66)
    node[midway, below=1.5pt, sloped]
    {$k-q_1-q_2$};

\draw[line width=0.75pt,->]
    (6.95,-1.17) -- (7.75,-1.17);

\node[below=1pt] at (7.35,-1.17)
    {$k,c,\nu$};

\draw[line width=0.75pt,->]
    (4.03,-1.76) -- (4.45,-1.28)
    node[midway, left=2pt]
    {$q_1$};

\draw[line width=0.75pt,->]
    (6.17,-1.76) -- (5.75,-1.28)
    node[midway, right=2pt]
    {$q_2$};

\end{tikzpicture}%
}
\end{minipage}

\caption{
Diagram $\mathcal{M}_{2,0,3}^{(4g)}$ contributing at first order in opacity  $L/{\lambda}$:
(a) the non-contact-limit contribution containing the four-gluon vertex and
(b) the corresponding contact-limit configuration.
}
\label{fig:M2034g}

\end{figure*}
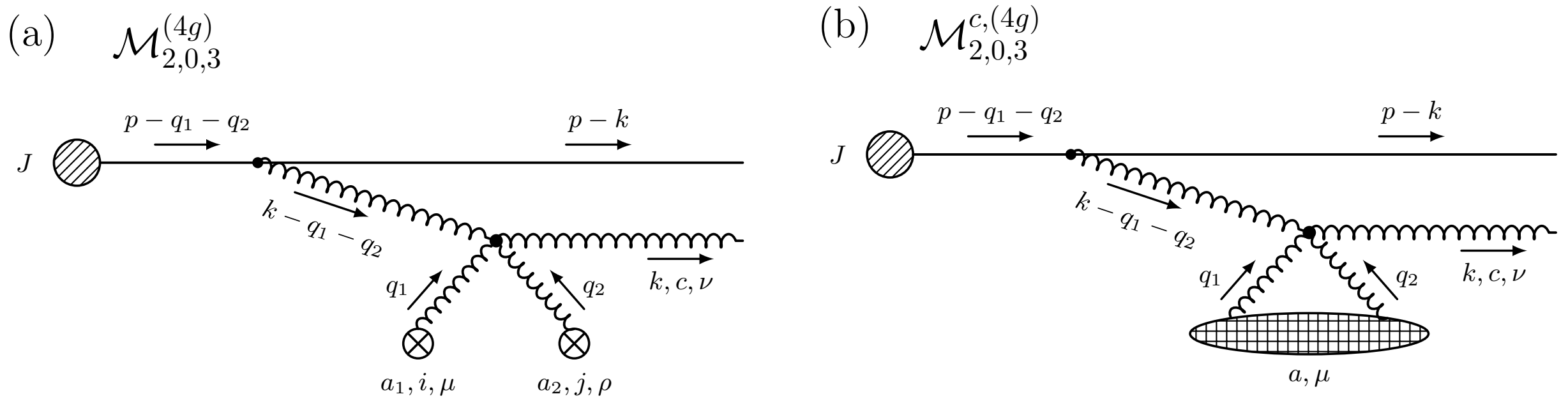
 
  \section{\label{sec:M4g203}Two-scattering contribution $\mathcal{M}_{2,0,3}^{(4g)}$}

The four-gluon-vertex contribution $\mathcal{M}_{2,0,3}^{(4g)}$ vanishes at leading-eikonal accuracy and is therefore absent from the standard (D)GLV calculation~\cite{GLV1,GLV2,DGLV}. This topology first contributes at first-subeikonal order in Ref.~\cite{KML}. As shown in Fig.~\ref{fig:M2034g}, the heavy quark emits an off-shell gluon with momentum $k-q_1-q_2$, which interacts with two medium gluons carrying momenta $q_1$ and $q_2$ through the four-gluon vertex. The outgoing gluon and heavy quark carry momenta $k$ and $p-k$, respectively.

Applying the Feynman rules of Sec.~\ref{sec:FeynmanRules} and the four-gluon vertex in Eq.~\eqref{eq:4gluon_vertex}, we obtain the amplitude
\begin{widetext}
\begin{align}
\mathcal{M}_{2,0,3}^{(4g)}
{}&=
\int\frac{d^4q_1}{(2\pi)^4}
\int\frac{d^4q_2}{(2\pi)^4}
\varepsilon^{*\nu}(k) \,
V_{\mu\rho\nu\sigma'}^{a_1a_2ca'} \,
\frac{(-i)P^{\sigma\sigma'}(k-q_1-q_2)\delta_{aa'}}
{(k-q_1-q_2)^2-m_g^2+i\epsilon} \,
\left[
i g_s(2p-k-q_1-q_2)_{\sigma}T_{jet}^{a}
\right]
\nonumber \\
&\times
A_{ext}^{\rho,a_2}(q_2) \,
A_{ext}^{\mu,a_1}(q_1) \,
\frac{i}{(p-q_1-q_2)^2-M^2+i\epsilon} \,
iJ(p-q_1-q_2)\, .
\end{align}
\end{widetext}

As in the preceding two-scattering calculations, the CL medium average and the target color trace in Eq.~\eqref{tr2} impose $a_1=a_2$ and the momentum constraint in Eq.~\eqref{q1q2}. The resulting interference with $\mathcal{M}_0^\dagger$ reads
\begin{widetext}
\begin{align}
\frac{2}{d_T}
\operatorname{Re}\operatorname{Tr}
\left\langle
\mathcal{M}_0^\dagger
\mathcal{M}_{2,0,3}^{c,(4g)}
\right\rangle
{}&=
-\frac{4g_s^2 C_2(T)N}{A_\perp d_A}
\operatorname{Re}
\Bigg\{
\int_0^L dz \,\rho(z)
\int\frac{d^2\mathbf{q}}{(2\pi)^2}\,
\left|J(E,\mathbf{p})\right|^2
\frac{\boldsymbol{\varepsilon}\cdot\mathbf{k}}
{\mathbf{k}^2+\chi}
\int\frac{dq_{1z}}{2\pi}
\int\frac{dq_{2z}}{2\pi}\,
\nonumber \\
&\times
e^{-iq_{1z}z}
e^{-iq_{2z}z}\,
v(q_1)v(q_2)
\frac{I_{2,0,3}^{(4g)}}
{\left[(p-q_1-q_2)^2-M^2+i\epsilon\right]
\left[(k-q_1-q_2)^2-m_g^2+i\epsilon\right]}
\Bigg\}\, .
\label{EM4g}
\end{align}
\end{widetext}
Here, $q_1$ and $q_2$ are specified in Eqs.~\eqref{kin1M2} and~\eqref{kin2M2}, respectively. Using the color relations in Eqs.~\eqref{tr2} and~\eqref{gen1}, the projector in Eq.~\eqref{Proj}, and the transversality conditions in Eqs.~\eqref{kin0M1} and~\eqref{vektori}, together with the CL identity $\varepsilon(k)\cdot(q_1+q_2)=0$, the Lorentz-contracted structure simplifies considerably, and becomes independent of $q_{1z}$ and $q_{2z}$ at first-subeikonal accuracy. It therefore introduces no additional longitudinal poles and yields
\begin{align}
I_{2,0,3}^{(4g)}
\approx{}&
8\alpha\,
\frac{\boldsymbol{\varepsilon}\cdot\mathbf{k}}{x}
\left[
1-\frac{\mathbf{u}\cdot\mathbf{k}}{xE}
\right] \approx 8\alpha\,
\frac{\boldsymbol{\varepsilon}\cdot\mathbf{k}}{x} .
\label{I2034g}
\end{align}
Since this diagram starts at first-subeikonal order,  the
omitted relative correction contributes only beyond working accuracy to the final interference term.

The first longitudinal integration is
\begin{widetext}
\begin{align}
\mathcal{I}_{2,0,3}^{(4g)}
={}&
\int\frac{dq_{1z}}{2\pi}\,
\frac{
e^{-iq_{1z}z}\,v(q_1)
}{
\left[(p-q_1-q_2)^2-M^2+i\epsilon\right]
\left[(k-q_1-q_2)^2-m_g^2+i\epsilon\right]
}\, .
\end{align}
\end{widetext}
Closing the contour in the lower half-plane, the scattering-potential pole is exponentially suppressed according to Eq.~\eqref{eq:separation}. The contributing propagator poles are the heavy-quark pole $q^-_{1z,p-q_1-q_2}$ given by Eq.~\eqref{m220_pol} and the gluon pole $q^-_{1z,k-q_1-q_2}$ presented in Eq.~\eqref{q1kq1q2}. Neither contains an $\mathcal{O}\!\left(\perp^3/(xE)^2\right)$ correction, so their difference generates no mass-dependent  residue-denominator correction at first-subeikonal accuracy.

Denoting the heavy-quark- and gluon-pole residues by $\mathcal{I}_{2,0,3}^{1,(4g)}$ and $\mathcal{I}_{2,0,3}^{2,(4g)}$, respectively, the remaining longitudinal integrations are
\begin{align}
\mathcal{J}_{2,0,3}^{1,2,(4g)}
={}&
\int\frac{dq_{2z}}{2\pi}\,
e^{-iq_{2z}z}\,
v(q_2)\,
\mathcal{I}_{2,0,3}^{1,2,(4g)}\, .
\end{align}
Both $q_{1z}$ propagator poles contain the term $-q_{2z}$, so their residues generate a factor $e^{iq_{2z}z}$ that cancels the explicit $e^{-iq_{2z}z}$ in the second longitudinal integration. The heavy-quark-pole branch retains the formation factor $\exp[i(\mathbf{k}^2+\chi)z/(2xE)]$, whereas the gluon-pole branch carries no additional formation phase. Since neither pole contains a subleading correction of order $\mathcal{O}\!\left(\perp^3/(xE)^2\right)$, no phase-induced first-subeikonal contribution arises.

For $\mathcal{J}_{2,0,3}^{1,(4g)}$, the contributing singularities are the scattering-potential poles $q^-_{2z,v_1}$ and $q^-_{2z,v_2}$, given in Eqs.~\eqref{q2v1} and~\eqref{q2v2}, respectively.

For $\mathcal{J}_{2,0,3}^{2,(4g)}$, the two scattering-potential poles coincide at $q^-_{2z,v_2}$, producing a second-order pole. The corresponding contribution to the interference is purely imaginary and therefore vanishes upon taking the real part. Consequently, the final result arises entirely from $\mathcal{J}_{2,0,3}^{1,(4g)}$.

%For $\mathcal{J}_{2,0,3}^{2,(4g)}$, these poles coincide at $q^-_{2z,v_2}$, producing a second-order pole. Its residue is purely imaginary, and the corresponding contribution vanishes upon taking the real part of the interference. Consequently, only the heavy-quark-pole branch contributes to the final result.

Substituting the surviving longitudinal contribution and Eq.~\eqref{I2034g} into Eq.~\eqref{EM4g}, we obtain, through first-subeikonal order and linear in $\mathbf{u}$,
\begin{widetext}
\begin{align}
\frac{2}{d_T}
\operatorname{Re}\operatorname{Tr}
\left\langle
\mathcal{M}_0^\dagger
\mathcal{M}_{2,0,3}^{c,(4g)}
\right\rangle
{}&=
\frac{4g_s^2 C_2(T)N}{A_\perp d_A} \, \alpha 
\int_0^L dz\,\rho(z)
\int\frac{d^2\mathbf{q}}{(2\pi)^2}\,
\left|v(\mathbf{q})\right|^2
\left|J(E,\mathbf{p})\right|^2
\nonumber \\
&\times
\frac{
\mathbf{k}^2
}{
\left(\mathbf{k}^2+\chi\right)^2
}\, 
\frac{R_0}{x E}
\sin\left(
\frac{\mathbf{k}^2+\chi}{2xE}\,z
\right)\, .
\label{eq:M0M4g203final}
\end{align}
\end{widetext}
The sine term in Eq.~\eqref{eq:M0M4g203final} originates from the scattering-potential-pole residues, and is identified as a scattering-potential-pole sine contribution in the classification of Sec.~\ref{FullSpectrumClassification}. No mass-dependent residue-denominator correction arises, since the second longitudinal integration involves only scattering-potential poles, and the absence of a surviving subleading formation-phase shift excludes any phase-induced $z\sin$ contribution.

In the massless limit, $M=m_g=0$, Eq.~\eqref{eq:M0M4g203final} reduces to the four-gluon-vertex contribution in Eq.~(63) of Ref.~\cite{KML} upon expansion to linear order in $\mathbf{u}$, consistently with Eq.~\eqref{eq:flowexpansion}.

\section{\label{sec:M2sum}Sum of the two-scattering contributions}

The complete CL two-scattering contribution at first order in opacity is obtained by summing the interferences with $\mathcal{M}_0^\dagger$ derived in the preceding sections. To avoid double counting, the equivalent $\mathcal{M}_{2,0,1}^{c}$ and $\mathcal{M}_{2,0,2}^{c}$ contributions are included with a factor of $1/2$ each. Thus,
\begin{widetext}
\begin{align}
\frac{2}{d_T}\,
\operatorname{Re}\operatorname{Tr}
\left\langle
\mathcal{M}_0^\dagger \mathcal{M}_2
\right\rangle
{}&=
\frac{2}{d_T}\,
\operatorname{Re}\operatorname{Tr}
\left\langle
\mathcal{M}_0^\dagger \mathcal{M}^{c}_{2,2,0}
\right\rangle
+
\frac{2}{d_T}\,
\operatorname{Re}\operatorname{Tr}
\left\langle
\mathcal{M}_0^\dagger \mathcal{M}^{c}_{2,0,3}
\right\rangle
+
\frac{2}{d_T}\,
\operatorname{Re}\operatorname{Tr}
\left\langle
\mathcal{M}_0^\dagger \mathcal{M}^{c}_{2,0,0}
\right\rangle
\nonumber \\
&+
\frac{1}{2}
\left[
\frac{2}{d_T}\,
\operatorname{Re}\operatorname{Tr}
\left\langle
\mathcal{M}_0^\dagger \mathcal{M}^{c}_{2,0,1}
\right\rangle
+
\frac{2}{d_T}\,
\operatorname{Re}\operatorname{Tr}
\left\langle
\mathcal{M}_0^\dagger \mathcal{M}^{c}_{2,0,2}
\right\rangle
\right]
\nonumber \\
&+
\frac{2}{d_T}\,
\operatorname{Re}\operatorname{Tr}
\left\langle
\mathcal{M}_0^\dagger \mathcal{M}^{c}_{2,1,0}
\right\rangle
+
\frac{2}{d_T}\,
\operatorname{Re}\operatorname{Tr}
\left\langle
\mathcal{M}_0^\dagger \mathcal{M}^{c}_{2,1,1}
\right\rangle
\nonumber \\
&+
\frac{2}{d_T}\,
\operatorname{Re}\operatorname{Tr}
\left\langle
\mathcal{M}_0^\dagger \mathcal{M}^{c,(4g)}_{2,0,3}
\right\rangle \, .
\label{M2sum}
\end{align}
\end{widetext}

Substituting Eqs.~\eqref{eq:M0M220final}, \eqref{eq:M0M203final}, \eqref{eq:M0M200final}, \eqref{eq:M0M201final}, ~\eqref{eq:M0M202final}, \eqref{eq:M0M210final}, \eqref{eq:M0M211final}, and~\eqref{eq:M0M4g203final} into Eq.~\eqref{M2sum}, we obtain the complete two-scattering interference contribution through first-subeikonal order and linear in $\mathbf{u}$,
\begin{widetext}
\begin{align}
\frac{2}{d_T}\,
\operatorname{Re}\operatorname{Tr}
\left\langle
\mathcal{M}_0^\dagger\mathcal{M}_2
\right\rangle
{}&=
-\frac{4g_s^2 C_2(T)N}{A_\perp d_A}
\int_0^L dz\,\rho(z)
\int\frac{d^2\mathbf{q}}{(2\pi)^2}\,
\frac{|v(\mathbf{q})|^2
|J(E,\mathbf{p})|^2}{\mathbf{k}^2+\chi}
\Bigg\{
\operatorname{Tr}(c^2a_1^2)\,
\frac{\mathbf{k}^2}{\mathbf{k}^2+\chi}
\nonumber \\
&\quad
+2\alpha\,
\frac{\mathbf{k}^2}{\mathbf{k}^2+\chi}
\left[
1-\frac{2\,\mathbf{u}\cdot\mathbf{k}}{xE}
-\frac{\mathbf{u}\cdot\mathbf{k}}{xE}\,
\frac{\mathbf{q}^2}{v^2(\mathbf{q})}\,
\frac{\partial v^2(\mathbf{q})}{\partial\mathbf{q}^2}
\right]
\left[
1-\cos\left(
\frac{\mathbf{k}^2+\chi}{2xE}\,z
\right)
\right]
\nonumber \\
&\quad
+\frac{2\alpha}
{(\mathbf{k}-\mathbf{q})^2+\chi}
\Bigg[
\mathbf{k}\cdot(\mathbf{k}-\mathbf{q})
-\mathbf{k}^2 \, \frac{\mathbf{u}\cdot(\mathbf{k}-\mathbf{q})}{xE}
-\mathbf{q}\cdot(\mathbf{q}-\mathbf{k}) \, \frac{\mathbf{u}\cdot\mathbf{k}}{xE}
\nonumber \\
&\qquad
-\mathbf{k}\cdot(\mathbf{k}-\mathbf{q})\,
\frac{\mathbf{u}\cdot\mathbf{q}}{2xE}\,
\frac{\mathbf{k}^2-(\mathbf{k}-\mathbf{q})^2}{v^2(\mathbf{q})}\,
\frac{\partial v^2(\mathbf{q})}{\partial\mathbf{q}^2}
+\mathbf{k}\cdot(\mathbf{k}-\mathbf{q})\,
\frac{\mathbf{u}\cdot\mathbf{q}}{xE}\,
\frac{2x^2M^2}{(\mathbf{k}-\mathbf{q})^2+\chi}
\Bigg]
\nonumber \\
&\qquad\times
\left[
\cos\left(
\frac{\mathbf{k}^2+\chi}{2xE}\,z
\right)
-\cos\left(
\frac{\mathbf{k}^2-(\mathbf{k}-\mathbf{q})^2}{2xE}\,z
\right)
\right]
\nonumber \\
&\quad
-\alpha\,
\frac{\mathbf{u}\cdot\mathbf{k}}{2xE}\,
\frac{\mathbf{q}^2}{v^2(\mathbf{q})}\,
\frac{\partial v^2(\mathbf{q})}{\partial\mathbf{q}^2}\,
\cos\left(
\frac{\mathbf{k}^2+\chi}{2xE}\,z
\right)
\nonumber \\
&\quad
-\alpha\,
\frac{\mathbf{k}^2}{\mathbf{k}^2+\chi}\,
\frac{R_0}{xE}
\left[
1+\frac{\mu^2-2\mathbf{q}^2}{4\,v^2(\mathbf{q})}\,
\frac{\partial v^2(\mathbf{q})}{\partial\mathbf{q}^2}
\right]
\sin\left(
\frac{\mathbf{k}^2+\chi}{2xE}\,z
\right)
\nonumber \\
&\quad
-\alpha\, \frac{\mathbf{k}\cdot(\mathbf{k}-\mathbf{q})}
{(\mathbf{k}-\mathbf{q})^2+\chi}\,
\frac{\mathbf{u}\cdot\mathbf{q}}{xE}\,
\frac{(\mathbf{k}-\mathbf{q})^2+m_g^2-x^2M^2}{xE}\,
z\,
\sin\left(
\frac{\mathbf{k}^2-(\mathbf{k}-\mathbf{q})^2}{2xE}\,z
\right)
\Bigg\} \, .
\label{M2sumfinal}
\end{align}
\end{widetext}

\section{\label{M1M2NE}First-order-in-opacity gluon spectrum and radiative energy loss}

Combining the single- and two-scattering contributions obtained in the preceding sections, the squared-amplitude combination entering the medium-induced gluon spectrum at first order in opacity is
\begin{widetext}
\begin{align}
\frac{1}{d_T}\Tr\mean{|\mathcal{M}_1|^2}+
\frac{2}{d_T}\operatorname{Re}\Tr\mean{\mathcal{M}_0^\dagger\mathcal{M}_2}
{}&=
\frac{8g_s^2 C_2(T)N}{A_\perp d_A} \alpha \,
\int_0^L dz\,\rho(z)
\int\frac{d^2\mathbf{q}}{(2\pi)^2}\,
|v(\mathbf{q})|^2
|J(E,\mathbf{p})|^2
\nonumber \\
\quad&
\times \Bigg\{2 \left[
1-
\cos\left(
\frac{(\mathbf{k}-\mathbf{q})^2+\chi}{2xE}\,z
\right)
\right]
\Bigg\{
\left[
\frac{(\mathbf{k}-\mathbf{q})^2}
{\left[(\mathbf{k}-\mathbf{q})^2+\chi\right]^2}
-\frac{\mathbf{k}\cdot(\mathbf{k}-\mathbf{q})}
{(\mathbf{k}^2+\chi)\left[(\mathbf{k}-\mathbf{q})^2+\chi\right]}
\right] \nonumber \\
&\quad \times\Bigg[
1
-\frac{\mathbf{u}\cdot\mathbf{q}}{2xE} \, 
\frac{2\mathbf{k}^2-(\mathbf{k}-\mathbf{q})^2+\chi}{v^2(\mathbf{q})} \, 
\frac{\partial v^2(\mathbf{q})}{\partial\mathbf{q}^2}
\Bigg]
\nonumber \\
&\quad+
\frac{\mathbf{u}\cdot(\mathbf{k}-\mathbf{q})}{xE}
\left[
\frac{\mathbf{k}^2}
{(\mathbf{k}^2+\chi)\left[(\mathbf{k}-\mathbf{q})^2+\chi\right]}
-\frac{2(\mathbf{k}-\mathbf{q})^2}
{\left[(\mathbf{k}-\mathbf{q})^2+\chi\right]^2}
\right]
\nonumber \\
&\quad-
\frac{\mathbf{u}\cdot\mathbf{k}}{xE}
\frac{\mathbf{q}\cdot(\mathbf{k}-\mathbf{q})}
{(\mathbf{k}^2+\chi)\left[(\mathbf{k}-\mathbf{q})^2+\chi\right]}
\nonumber \\
&\quad+
\frac{\mathbf{u}\cdot\mathbf{q}}{xE}
\frac{2x^2M^2}
{(\mathbf{k}-\mathbf{q})^2+\chi}
\left[
\frac{2(\mathbf{k}-\mathbf{q})^2}
{\left[(\mathbf{k}-\mathbf{q})^2+\chi\right]^2}
-\frac{\mathbf{k}\cdot(\mathbf{k}-\mathbf{q})}
{(\mathbf{k}^2+\chi)\left[(\mathbf{k}-\mathbf{q})^2+\chi\right]}
\right]
\Bigg\}
\nonumber \\
&\quad
+
\left[
1-
\cos\left(
\frac{\mathbf{k}^2+\chi}{2xE}\,z
\right)
\right]
\frac{\mathbf{u}\cdot\mathbf{k}}{2xE}
\frac{1}{(\mathbf{k}^2+\chi)^2}
\Bigg[
4\mathbf{k}^2
+
(\mathbf{k}^2-\chi)
\frac{\mathbf{q}^2}{v^2(\mathbf{q})}
\frac{\partial v^2(\mathbf{q})}{\partial\mathbf{q}^2}
\Bigg]
\nonumber \\
&\quad
+
\sin\left(
\frac{\mathbf{k}^2+\chi}{2xE}\,z
\right) \, 
\frac{\mathbf{k}^2}{(\mathbf{k}^2+\chi)^2}
\frac{R_0}{2xE}
\Bigg[
1+
\frac{\mu^2-2\mathbf{q}^2}{4v^2(\mathbf{q})}
\frac{\partial v^2(\mathbf{q})}{\partial\mathbf{q}^2}
\Bigg]
\nonumber \\
&\quad
+
\frac{1}{(\mathbf{k}-\mathbf{q})^2+\chi}
\frac{\mathbf{u}\cdot\mathbf{q}}{xE}
\frac{
(\mathbf{k}-\mathbf{q})^2+m_g^2-x^2M^2
}{xE}\,
z
\Bigg[
\frac{(\mathbf{k}-\mathbf{q})^2}
{(\mathbf{k}-\mathbf{q})^2+\chi}
-\frac{\mathbf{k}\cdot(\mathbf{k}-\mathbf{q})}
{\mathbf{k}^2+\chi}
\Bigg]
\nonumber \\
&\qquad\times
\sin\left(
\frac{(\mathbf{k}-\mathbf{q})^2+\chi}{2xE}\,z
\right)
\Bigg\} \, .
\label{M1M2}
\end{align}
\end{widetext}
Retaining only the leading-eikonal terms, Eq.~\eqref{M1M2} reproduces the massive DGLV kernel in Eq.~(82) of Ref.~\cite{DGLV}, including its characteristic LPM interference factor.

The first-subeikonal terms in Eq.~\eqref{M1M2} comprise the distinct contributions identified by their physical origin in Sec.~\ref{FullSpectrumClassification}: subleading Lorentz contractions, flow-induced potential derivatives, mass-dependent residue-denominator corrections, phase-induced $z\sin$ terms, and scattering-potential-pole sine contributions.

In the massless limit, $M=m_g=0$, and to linear order in $\mathbf{u}$, the subleading Lorentz-contraction and potential-derivative corrections agree with Eq.~(65) of Ref.~\cite{KML}. The ordinary sine contribution, generated by the scattering-potential-pole residues in two-scattering sector, is also recovered. The phase-induced $z\sin$ term, however, is absent from that result, since Ref.~\cite{KML} evaluates the formation phases at leading-pole accuracy, whereas we retain the surviving subleading longitudinal-pole corrections in their arguments.

%%%%%%%%%%%%%%%%%%%%%%%%%%%%%%%%%%%%%%

Starting from the joint gluon-heavy-quark distribution in Eq.~\eqref{spectum}, we apply the broad-source approximation in Eq.~\eqref{eq:broadsource}, which allows the hard-source factor to be expressed through the parent-quark distribution defined in Eq.~\eqref{NJ}. The final-state phase space is reduced according to Eq.~\eqref{M0phase}, with the soft-gluon approximation $1/(1-x)\simeq 1$ and the Jacobian correction omitted beyond first-subeikonal accuracy.

For compactness, we denote the expression enclosed by the outermost braces in Eq.~\eqref{M1M2} by $\mathcal{K}^{(1)}(\mathbf{k},\mathbf{q},z;x,M,m_g)$, with the dependence on the medium parameters and transverse flow implicit. For a heavy quark in the fundamental representation, we use the color
identity $\alpha/d_F=C_A C_2(F)/2$ derived above. Thus, we obtain
\begin{align}
\frac{dN_g^{(1)}}
{dx\,d^2\mathbf{k}}
={}&
\frac{\alpha_s}{x \, \pi^2}\,
C_A \, C_2(F)\,
\frac{C_2(T)N}{A_\perp d_A}
\int_0^L dz\,\rho(z)
\int\frac{d^2\mathbf{q}}{(2\pi)^2}\,
\nonumber \\
&\times
|v(\mathbf{q})|^2 \, 
\mathcal{K}^{(1)}(\mathbf{k},\mathbf{q},z;x,M,m_g)\,.
\label{gluonspectrum}
\end{align}

Weighting the gluon spectrum by the radiated energy $\omega=xE$ and integrating over the transverse gluon momentum gives
\begin{equation}
\frac{dE_{rad}^{(1)}}{dx}=
E
\int d^2\mathbf{k}\,
x\frac{dN_g^{(1)}}
{dx\,d^2\mathbf{k}}\,.
\end{equation}
Substituting Eq.~\eqref{gluonspectrum}, we find
\begin{align}
\frac{dE_{rad}^{(1)}}{dx}
={}&
\frac{\alpha_s E}{\pi^2}\,
C_A \, C_2(F)\,
\frac{C_2(T)N}{A_\perp d_A}
\int d^2\mathbf{k}
\int_0^L dz\,\rho(z)
\nonumber \\
\times&
\int\frac{d^2\mathbf{q}}{(2\pi)^2}\,
|v(\mathbf{q})|^2\,
\mathcal{K}^{(1)}(\mathbf{k},\mathbf{q},z;x,M,m_g)\,.
\label{diffradEloss}
\end{align}
The total radiative energy loss at first order in opacity is then
\begin{equation}
\Delta E_{rad}^{(1)}
=
\int dx\,
\frac{dE_{rad}^{(1)}}{dx}\,.
\label{radEloss}
\end{equation}

For the numerical evaluation, we use the normalized exponential longitudinal profile of Refs.~\cite{GLV2,DGLV,KML}, as specified in the main text,
\begin{equation}
\rho(z)
=
\frac{2}{L}\,
e^{-2z/L},\
\qquad
z\geq 0\, ,
\qquad
\int_0^\infty dz\,\rho(z)=1 .
\end{equation}
The corresponding longitudinal averages, including the phase-induced first-subeikonal contribution, are evaluated analytically over $z\in[0,\infty)$.

We further express the medium average in terms of the standard opacity normalization~\cite{GW,GLV2,DGLV}.  In the DGLV convention, the opacity relevant for inclusive medium-induced radiation is defined by the gluon-target elastic cross section
\begin{equation}
\frac{d\sigma_g}{d^2\mathbf{q}}
=
\frac{C_A C_2(T)}{d_A}\,
\frac{\left|v(\mathbf{q})\right|^2}{(2\pi)^2} \, .
\end{equation}
The flow dependence of the scattering potential begins at $\mathcal{O}(u^2)$. At linear order in $\mathbf{u}$, the elastic cross section is therefore evaluated with
\begin{equation}
v(\mathbf{q})
=
-\frac{g_s^2}{\mathbf{q}^2+\mu^2} \, ,
\end{equation}
which gives
\begin{equation}
\sigma_{ g}
=
\int d^2\mathbf{q}\,
\frac{d\sigma_{g}}{d^2\mathbf{q}}
=
\frac{C_A C_2(T)}{d_A}\,
\frac{4\pi\alpha_s^2}{\mu^2}.
\end{equation}
Defining the opacity as
\begin{equation}
\frac{L}{\lambda_g}
=
\frac{N\sigma_{g}}{A_\perp},
\end{equation}
we obtain the normalized transverse-momentum-transfer distribution
\begin{equation}
\frac{1}{\sigma_{g}}\,
\frac{d\sigma_{g}}{d^2\mathbf{q}}
=
\frac{1}{\pi}\,
\frac{\mu^2}
{\left(\mathbf{q}^2+\mu^2\right)^2},
\qquad
\int d^2\mathbf{q}\,
\frac{1}{\pi}\,
\frac{\mu^2}
{\left(\mathbf{q}^2+\mu^2\right)^2}
=
1 .
\end{equation}

The medium-dependent factor in Eq.~\eqref{gluonspectrum} can therefore be expressed as
\begin{equation}
\frac{C_A C_2(F)C_2(T)N}{A_\perp d_A}\,
\frac{d^2\mathbf{q}}{(2\pi)^2}\,
\left|v(\mathbf{q})\right|^2
=
C_2(F) \frac{L}{\lambda_g}\,
\frac{d^2\mathbf{q}}{\pi}\,
\frac{\mu^2}
{\left(\mathbf{q}^2+\mu^2\right)^2}.
\end{equation}

Combining the opacity normalization with the analytically evaluated longitudinal contributions, we obtain the double-differential medium-induced gluon spectrum in Eq.~\eqref{specExp}, retaining all corrections through first-subeikonal order and linear in $\mathbf{u}$.
%%%%%%%%%%%%%%%%%%%%%%%%%%%%%%%%%%%%%%%%%%%%

\end{document}